\documentclass[aps,pra,reprint,superscriptaddress,twocolumn]{revtex4}
\usepackage[dvipdfmx]{graphicx}
\usepackage{graphicx}
\usepackage{color}
\usepackage[T1]{fontenc}
\usepackage{textcomp}
\usepackage{bm}
\usepackage{amsmath,amssymb,amsthm}
\usepackage{comment}
\usepackage{ulem}
\usepackage{grffile}
\usepackage{algorithm}
\usepackage{algpseudocode}

\newcommand{\cket}[1]{\left|#1\right\rangle}
\newcommand{\bra}[1]{\left\langle#1\right|}
\newcommand{\bracket}[2]{\left\langle#1|#2\right\rangle}

\begin{document}


\title{Proposal for simulating quantum spin models with the Dzyaloshinskii-Moriya interaction using Rydberg atoms and the construction of asymptotic quantum many-body scar states}

\author{Masaya Kunimi}
\email{kunimi@rs.tus.ac.jp}
\affiliation{Department of Physics, Tokyo University of Science, 1-3 Kagurazaka, Tokyo 162-8601,  Japan}
\author{Takafumi Tomita}
\email{tomita@ims.ac.jp}
\affiliation{Department of Photo-Molecular Science, Institute for Molecular Science, National Institutes of Natural Sciences, 38 Nishigo-Naka, Myodaiji, Okazaki, Aichi 444-8585, Japan}
\author{Hosho Katsura}
\email{katsura@phys.s.u-tokyo.ac.jp}
\affiliation{Department of Physics, Graduate School of Science, The University of Tokyo, 7-3-1 Hongo, Tokyo 113-0033, Japan}
\affiliation{Institute for Physics of Intelligence, The University of Tokyo, 7-3-1 Hongo, Tokyo 113-0033, Japan}
\affiliation{Trans-scale Quantum Science Institute, The University of Tokyo, 7-3-1 Hongo, Tokyo 113-0033, Japan}
\author{Yusuke Kato}
\email{yusuke@phys.c.u-tokyo.ac.jp}
\affiliation{Department of Basic Science, The University of Tokyo, 3-8-1 Komaba, Tokyo 153-8902, Japan}
\affiliation{Quantum Research Center for Chirality, Institute for Molecular Science, Okazaki, Aichi 444-8585, Japan}


\date{\today}

\begin{abstract}
We have developed a method to simulate quantum spin models with the Dzyaloshinskii-Moriya interaction (DMI) using Rydberg atom quantum simulators. Our approach involves a two-photon Raman transition and a transformation to the spin-rotating frame, both of which are feasible with current experimental techniques. As a model that can be simulated in our setup but not in solid-state systems, we consider an $S=\frac{1}{2}$ spin chain with a Hamiltonian consisting of the DMI and Zeeman energy. We study the magnetization curve in the ground state of this model and quench dynamics. Further, we show the existence of quantum many-body scar states and asymptotic quantum many-body scar states. The observed nonergodicity in this model demonstrates the importance of the highly tunable DMI that can be realized by the proposed quantum simulator.
\end{abstract}
\maketitle
\section{Introduction}\label{sec:Introduction}
In recent years, quantum simulators have become powerful tools for exploring various quantum many-body phenomena. These simulators, which include ultracold gases \cite{Bloch2008,Gross2017,Schafer2020}, Rydberg atoms \cite{Browaeys2020,Morgado2021}, trapped ions \cite{Blatt2012,Monroe2021}, and superconducting qubits \cite{Wendin2017,Kjaergaard2020,Altman2022}, offer high controllability, allowing researchers to realize different types of Hamiltonians in experiments. Rydberg atom quantum simulators, in particular, have attracted much attention due to their ability to create quantum spin models. By leveraging the strong dipole-dipole interaction between highly excited atoms and optical tweezers in a programmable array, these simulators have successfully realized a range of models, such as 
Ising \cite{Labuhn2016,Zeiher2017,Bernien2017,Leseleuc2018,Lienhard2018,Guardado-Sanchez2018,Keesling2019,Ebadi2021,Semeghini2021,Bluvstein2021,Scholl2021,Bharti2022a}, {\it XY}  \cite{Orioli2018,Leseleuc2019,Chen2023}, {\it XXZ}  \cite{Signoles2021,Geier2021,Scholl2022,Franz2022a}, and {\it XYZ} \cite{Geier2021,Steinert2023} models. This versatility makes them valuable for investigating issues in statistical mechanics, condensed matter physics, and quantum computing. Notable achievements in Rydberg atom quantum simulators include the discovery of quantum many-body scar states \cite{Bernien2017,Turner2018,Turner2018_2}, observations of topological edge states \cite{Leseleuc2019}, and realizations of quantum spin-liquid states \cite{Semeghini2021}. Furthermore, researchers have successfully implemented several gate operations \cite{Levine2018,Levine2019,Madjarov2020,Bluvstein2022,Graham2022,Jenkins2022,Ma2022,Wu2022,Chew2022}, demonstrating the potential of these simulators for advancing quantum technologies. These developments highlight the promising role of quantum simulators in advancing our understanding of complex quantum systems and their applications in various scientific and technological domains. 

In solid-state physics, the Dzyaloshinskii-Moriya interaction (DMI) \cite{Dzyaloshinsky1958,Moriya1960} is crucial in understanding magnetism in crystals without space-inversion symmetry. This interaction leads to nontrivial magnetic structures in materials with structural chirality, known as chiral magnets. The classical effects of DMI include the formation of chiral soliton lattices \cite{Togawa2012,Kishine2015,Togawa2016} and skyrmions \cite{Muhlbauer2009,Yu2010,Nagaosa2013}. Moreover, the role of DMI in quantum phenomena has been extensively studied \cite{Oshikawa1997,Affleck1999,Takashima2016,Kodama_thesis,Kodama2023}. One of the most conspicuous quantum effects in magnets is the spin-parity effect. This effect refers to the phenomenon in which properties differ significantly depending on whether the spin is half-integer or integer. Recent theoretical work by Kodama et al. has shed light on the spin-parity effects in one-dimensional chiral magnets, particularly when the DMI exceeds the exchange interaction~\cite{Kodama_thesis,Kodama2023}. However, experimental observation of these effects is challenging due to the relatively small magnitude of the DMI compared to the exchange interaction in typical crystal systems. A promising candidate for quantum simulations of the DMI is neutral cold atoms in optical lattices with synthetic spin-orbit coupling (SOC) \cite{Dalibard2011,Goldman2014,Zhai2015}. Previous theoretical studies have shown that an effective Hamiltonian in the strongly correlated regimes becomes the spin Hamiltonian with the DMI \cite{Cole2012,Radic2012,Cai2012,Peotta2014,Gong2015,Goldman2023a}. However, this synthetic SOC induces heating, which makes the low-energy physics and long-time evolution inaccessible \cite{Wang2012,Huang2016,Wu2016,Sun2018}.

The purpose of this paper is two-fold. First, we propose a method to simulate quantum spin models with DMI in Rydberg atom quantum simulators, in a way accessible with current experimental techniques. Second, we explore the nontrivial properties of a quantum spin model that exemplifies the importance of our quantum simulator.

We can obtain the effective DMI in the spin-rotating frame using two-photon Raman transition and unitary transformation. In our scheme, we can tune the ratio between the DMI and the exchange interaction from zero to infinity. This tunability allows us to realize the Hamiltonian with only DMI and magnetic field term, called the DH model \cite{Kodama_thesis,Kodama2023}. We investigate the ground-state properties and nonequilibrium dynamics of the DH model using matrix-product state methods \cite{Schollwock2011,Paeckel2019}. We thereby show that the time-averaged magnetization continuously changes as a function of the magnetic field in the fully quantum case, whereas it is discontinuous in the classical case. We show that the DH model has QMBS states under periodic and open boundary conditions and argue that they are responsible for the slow thermalization after a quench from a particular initial state. We also show that the DH model has asymptotic quantum many-body scar (AQMBS) states, which are not eigenstates of the Hamiltonian but whose energy variance vanishes in the thermodynamic limit \cite{Gotta2023}. We analytically construct a series of AQMBS states and find the slow thermalization after a quench from such states. 

This paper is organized as follows: In Sec.~\ref{sec:setup}, we explain our proposal for simulating spin models with DMI. In Secs.~\ref{sec:Ground_state} and \ref{sec:sweep_dynamics}, we analyze the ground-state and sweep dynamics of the DH model, respectively. In Secs.~\ref{sec:quantum_many-body_scar} and \ref{sec:asymptotic_scar}, we show the existence of the QMBS states and AQMBS states in the DH model, respectively. In Sec.~\ref{sec:summary}, we summarize our results.

\section{Experimental proposal for simulating Dzyaloshinskii-Moriya interaction}\label{sec:setup}

\begin{figure}[t]
\centering
\includegraphics[width=8.6cm,clip]{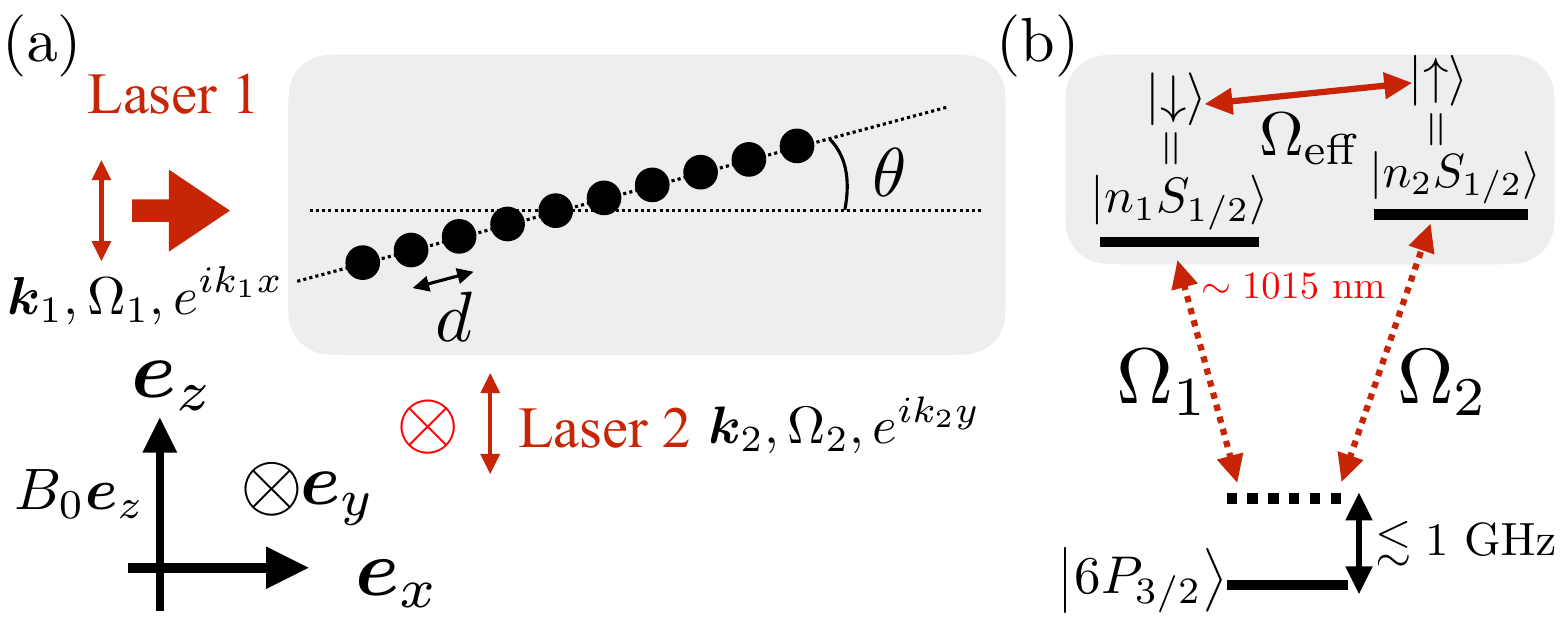}
\caption{(a) Schematic of the experimental setup for realization of the DMI. The filled black circles represent the position of the Rydberg atoms. $\bm{e}_{x,y,z}$ is the unit vector in each direction. The magnetic field is parallel to the $z$ axis. (b) Level diagram of ${}^{87}$Rb atom. Using the two-photon Raman scheme, we obtain an effective two-level system consisting of $\cket{\uparrow}$ and $\cket{\downarrow}$.}
\label{fig:schematic_figure}
\vspace{-0.75em}
\end{figure}

We consider ${}^{87}$Rb atoms arranged in a one-dimensional open chain with lattice spacing $d$ in the $xz$ plane [see Fig.~\ref{fig:schematic_figure} (a)]. To construct spin-$\frac{1}{2}$ systems, we assign the $\cket{\downarrow}$ state as a Rydberg state $|n_1S_{1/2}, m_J=+\frac{1}{2}\rangle$ and the $\cket{\uparrow}$ state as another Rydberg state $|n_2S_{1/2},m_J=+\frac{1}{2}\rangle$, where $n_i\;(i=1,2)$ is the principal quantum number. Here, we assume that the magnetic quantum numbers of both Rydberg states are $m_J=\frac{1}{2}$. According to Refs.~\cite{Whitlock2017,Signoles2021}, the interaction Hamiltonian is of the {\it XXZ} type:
\begin{align}
\hat{H}_{XXZ}&=J\sum_{j=1}^{M-1}(\hat{S}_j^x\hat{S}_{j+1}^x+\hat{S}_j^y\hat{S}_{j+1}^y+\delta\hat{S}_j^z\hat{S}_{j+1}^z),\label{eq:XXZ_Hamiltonian}
\end{align}
where $M$ is the number of spins, $\hat{S}_j^{\mu}$\;$(\mu=x,y,z)$ is the spin-$\frac{1}{2}$ operator at site $j$, $J$ is the exchange interaction energy, which originates from the van der Waals interaction between the Rydberg atoms, and $\delta$ is the anisotropy parameter. We consider only the nearest-neighbor (NN) part of the interaction because the magnitude of the next NN interaction is $\frac{1}{64}$ that of the NN interaction \cite{note_interaction}. See Appendix~\ref{app:derivation_XXZ} for a detailed derivation of the {\it XXZ} Hamiltonian from the dipole-dipole interaction between the Rydberg atoms.

To generate the DMI in this system, we irradiate two linearly polarized Raman lasers to the Rydberg atoms [See Fig.~\ref{fig:schematic_figure} (a)]. One laser propagates along the $x$ axis with Rabi frequency $\Omega_1$ and wave number $\bm{k}_1=k_1\bm{e}_x$, and the other propagates along the $y$ axis with Rabi frequency $\Omega_2$ and wave number $\bm{k}_2=k_2\bm{e}_y$. We denote the position of the $j$th atom as $\bm{R}_j\equiv d j(\cos\theta, 0, \sin\theta)$
, where $\theta$ is the angle between the chain and the $x$ axis shown in Fig.~\ref{fig:schematic_figure} (a). We can express the effective Hamiltonian for the $j$th atom as
\begin{align}
\hat{h}_j&=-\hbar\Omega_{\rm eff}[\cos(q j)\hat{S}_j^x+\sin(q j)\hat{S}_j^y]-\hbar\tilde{\Delta}\hat{S}_j^z.\label{eq:effective_Hamiltonian_for_j-th_atom}
\end{align}
Here, $\Omega_{\rm eff}\equiv \Omega_1\Omega_2/(2\Delta)$ is the effective Rabi frequency with $\Delta\equiv (\Delta_1+\Delta_2)/2$. The one-photon detuning $\hbar\Delta_i\equiv\hbar\omega_i-(E_{n_iS}-E_P)\;(i=1,2)$ is written in terms of the frequency of the laser $i$ ($\omega_i$) and the energies of $\cket{n_iS_{1/2}}$ and $\cket{6P_{3/2}}$ states ($E_{n_iS_{1/2}}$ and $E_P$). We also define $q\equiv k_1d\cos\theta$ and the two-photon detuning including the ac-Stark shift $\tilde{\Delta}\equiv -[\Delta_1-\Delta_2+(\Omega_1^2-\Omega_2^2)/(4\Delta)]$. See Appendix~\ref{app:rotating_transverse_field} for a detailed derivation of the rotating transverse field term. The total Hamiltonian in the spin-laboratory frame \cite{note1} is given by $\hat{H}_{\text{s-lab}}\equiv \hat{H}_{XXZ}+\sum_j\hat{h}_j$.

We then move to the spin-rotating frame. The unitary transformation with the operator \cite{Perk1976,Calvo1981,Oshikawa1997,Affleck1999,Shekhtman1992,Nikuni1993}
\begin{align}
\hat{U}_{\text{s-rot}}\equiv \prod_{j=1}^Me^{-i q j\hat{S}_j^z},\label{eq:unitary_operator_rotating_frame}
\end{align}
yields the Hamiltonian in the spin-rotating frame:
\begin{align}
\hat{H}_{\text{s-rot}}&\equiv\hat{U}^{\dagger}_{\text{s-rot}}\hat{H}_{\text{s-lab}}\hat{U}_{\text{s-rot}}\notag \\
&=J\cos q\sum_{j=1}^{M-1}(\hat{S}_j^x\hat{S}_{j+1}^x+\hat{S}_j^y\hat{S}_{j+1}^y)\notag \\
&-J\sin q\sum_{j=1}^{M-1}(\hat{S}_j^x\hat{S}_{j+1}^y-\hat{S}_j^y\hat{S}_{j+1}^x)\notag \\
&+J\delta\sum_{j=1}^{M-1}\hat{S}_j^z\hat{S}_{j+1}^z-\hbar\Omega_{\rm eff}\sum_{j=1}^M\hat{S}_j^x-\hbar\tilde{\Delta}\sum_{j=1}^{M}\hat{S}_j^z.\label{eq:Hamiltonian_in_rotating_frame}
\end{align}
In the new frame, the DMI appears in the third line of Eq.~(\ref{eq:Hamiltonian_in_rotating_frame}). This expression implies that we can simulate the Hamiltonian with the {\it XXZ} and DMI terms under the uniform magnetic field by the {\it XXZ} Hamiltonian with the rotating transverse field. In Appendices~\ref{app:relation_lab_and_rot_frame} and \ref{app:two-dimension}, we discuss the relation between the spin-laboratory and spin-rotating frames and extension to two-dimensional systems, respectively.

In the following,  we discuss the tunability of the Hamiltonian. The ratio between the {\it XY} interaction and DMI is given by $\tan(k_1d\cos\theta)$. We can tune $\theta$ by arranging the Rydberg atoms in appropriate positions using the optical tweezers [see Fig.~\ref{fig:schematic_figure} (a)]. Thereby the strong DMI regime   [$|\tan(k_1d\cos\theta)|>1$] can be realized, which is practically inaccessible in solid-state systems. The {\it XY} interaction term vanishes under the condition $k_1d\cos\theta=\pi/2$,  which can be realized when $k_1=2\pi/(1015~{\rm nm})$, $d=5~\mu{\rm m}$, and $\theta\simeq 87.1^{\circ}$. We can also vary the anisotropy parameter $\delta$ and the detunings  by choosing different combinations of Rydberg states and applying additional static electric and magnetic fields \cite{Whitlock2017,Leseleuc2018,Weber2017,Sibalic2017,Robertson2021}. This tunability offers the potential to realize the DH model \cite{Kodama_thesis,Kodama2023} in the Rydberg atom quantum simulators:
\begin{align}
\hat{H}_{\rm DH}=D\sum_{j=1}^{M-1}(\hat{S}_j^x\hat{S}_{j+1}^y-\hat{S}_j^y\hat{S}_{j+1}^x)-h^x\sum_{j=1}^M\hat{S}_j^x,\label{eq:Hamiltonian_DH_model}
\end{align}
where $D\equiv -J$ is the magnitude of the DMI and $h^x\equiv \hbar\Omega_{\rm eff}$ is the magnetic field.

We remark that the DMI can be expressed as the Peierls phase because the first two terms in Eq.~(\ref{eq:Hamiltonian_in_rotating_frame}) reduce to $e^{-iq}\hat{S}_j^+\hat{S}_{j+1}^-+e^{+iq}\hat{S}_j^-\hat{S}_{j+1}^+$. Recently, the methods for controlling the Peierls phase in the Rydberg atom quantum simulators have been studied experimentally \cite{Lienhard2020} and theoretically \cite{Yang2022,Wu2022_2,Perciavalle2023,Poon2024}.

\section{Ground state properties of the DH model}\label{sec:Ground_state}
In this section, we show the ground-state properties of the DH model based on the density matrix renormalization group (DMRG) calculations \cite{White1992,White1993}. Some of the results have been reported in Ref.~\cite{Kodama_thesis}, but we reproduce them here for completeness.

\begin{figure*}[t]
\centering
\includegraphics[width=17cm,clip]{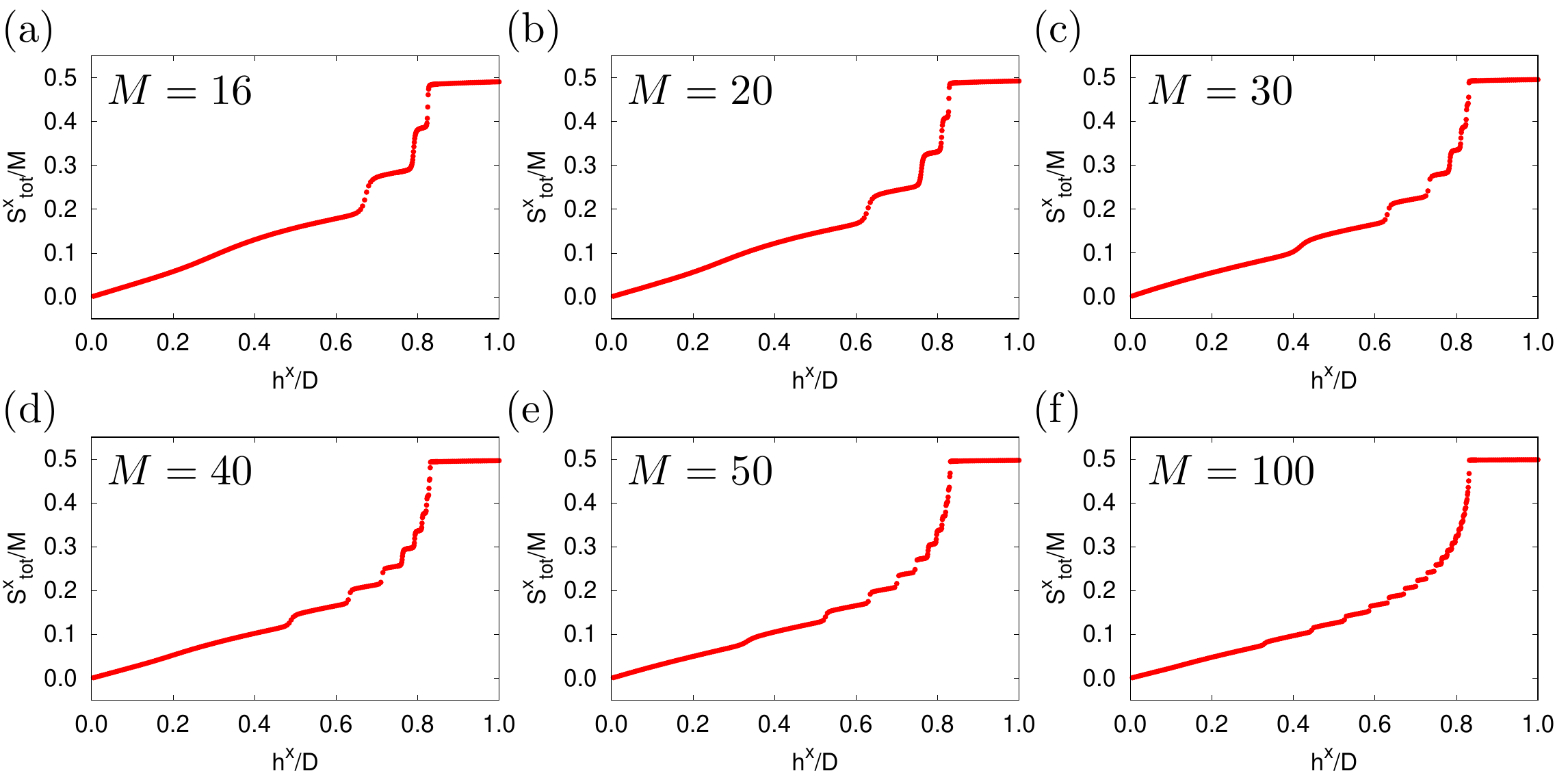}
\caption{Magnetization curve of the DH model (\ref{eq:Hamiltonian_DH_model}) in the ground states for various system sizes. (a) $M=16$, (b) $M=20$, (c) $M=30$, (d) $M=40$, (e) $M=50$, (f) $M=100$.}
\label{fig:m-h_curve_dmrg}
\vspace{-0.75em}
\end{figure*}

\begin{figure*}[t]
\centering
\includegraphics[width=17cm,clip]{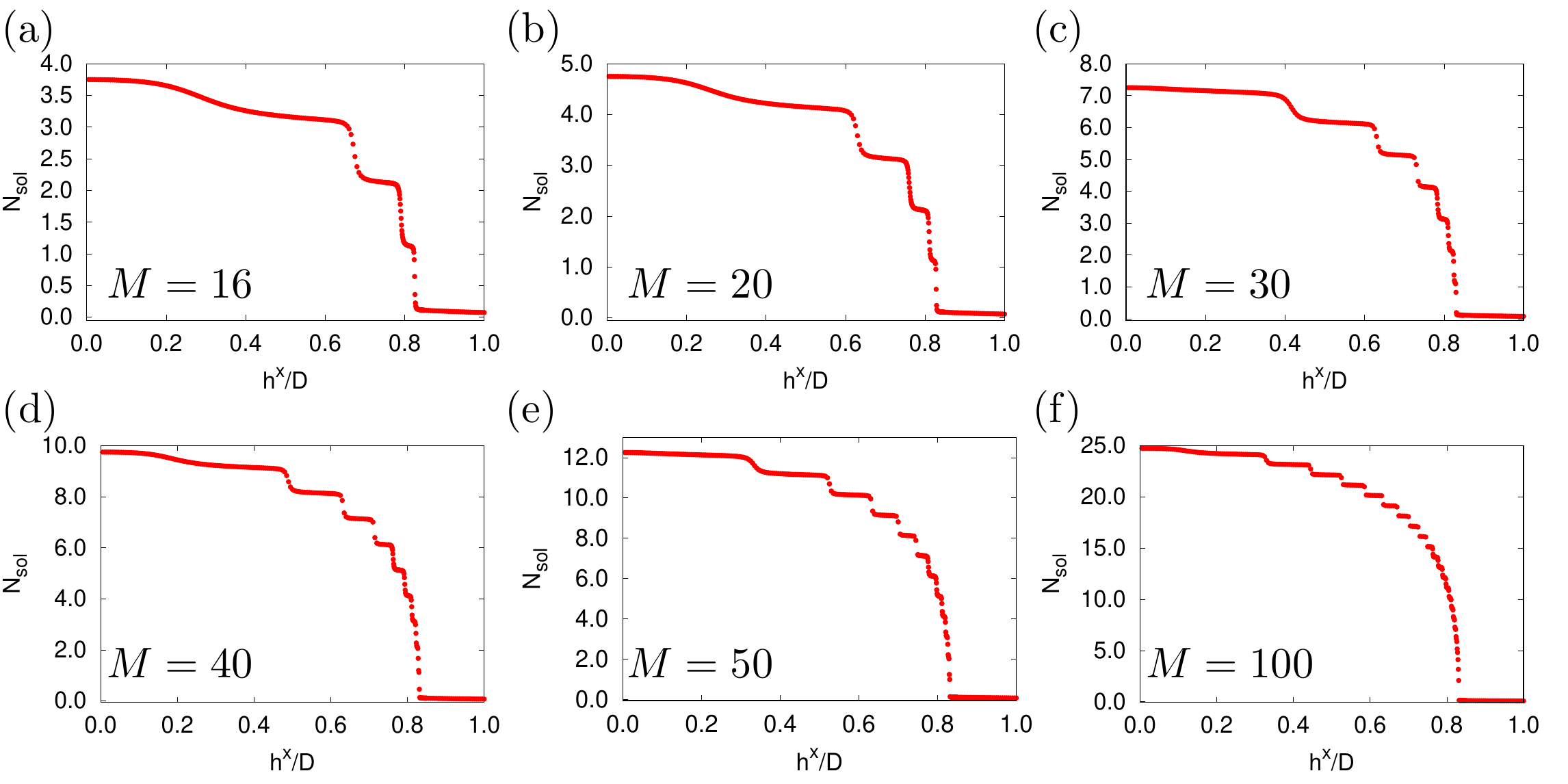}
\caption{Ground-state soliton number of the DH model (\ref{eq:Hamiltonian_DH_model}) as a function of $h^x$ for various system sizes. (a) $M=16$, (b) $M=20$, (c) $M=30$, (d) $M=40$, (e) $M=50$, (f) $M=100$.}
\label{fig:soliton_number_dmrg}
\vspace{-0.75em}
\end{figure*}

\begin{figure*}[t]
\centering
\includegraphics[width=17cm,clip]{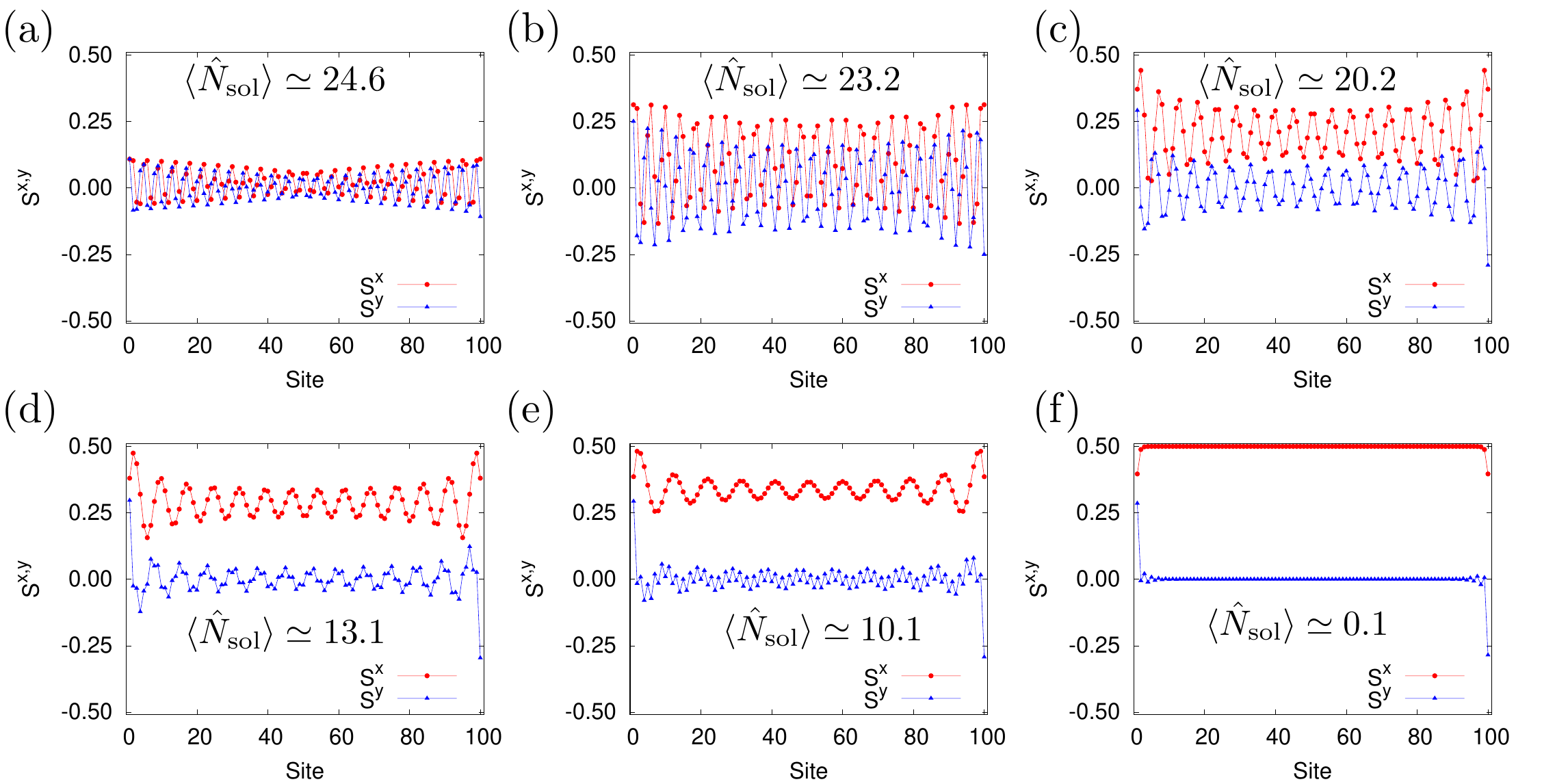}
\caption{Expectation value of the local spins $S_j^x$ (red circle) and $S_j^y$ (blue triangle) of the DH model (\ref{eq:Hamiltonian_DH_model}) in the ground states for $M=100$. (a) $h^x=0.1D$, (b) $h^x=0.4D$, (c) $h^x=0.6D$, (d) $h^x=0.78D$, (e) $h^x=0.81D$, (f) $h^x=0.85D$. We also show the expectation value of the soliton number operator for each parameter.}
\label{fig:spin_density_dmrg}
\vspace{-0.75em}
\end{figure*}

Figure~\ref{fig:m-h_curve_dmrg} shows the magnetization curve of the DH model (\ref{eq:Hamiltonian_DH_model}) for various system sizes. We see some steep changes in the magnetization curve. Magnetization curves in classical chiral magnets exhibit similar behavior, which we attribute to level crossings between states with different winding numbers~\cite{Kishine2014}. Although the winding number is not well defined in quantum systems, the soliton number operator
\begin{align}
\hat{N}_{\rm sol}\equiv \sum_{j=1}^{M-1}\left(\frac{1}{4}-\hat{S}_j^x\hat{S}_{j+1}^x\right),\label{eq:definition_of_soliton_number_in_quantum}
\end{align}
plays a similar role \cite{Kishine2014,Kodama_thesis,Kodama2023}. We plot the ground state expectation value of $\hat{N}_{\rm sol}$ as a function of the magnetic field in Fig.~\ref{fig:soliton_number_dmrg}. The behavior is consistent with the magnetization curve shown in Fig.~\ref{fig:m-h_curve_dmrg}. The changes in the expectation value of the soliton number $\langle\hat{N}_{\rm sol}\rangle$ by $1$ correspond to the sharp changes in the magnetization. The local spin density exhibits similar behavior as shown in Fig.~\ref{fig:spin_density_dmrg}, where we find a helical structure of the spin in low-field regions. In high-field regions, the spin density is almost uniform except near the edges of the system. When we lower the magnetic field, the solitons enter the system, and the expectation value of $\hat{S}_{\rm tot}^x$ reduces.

\section{Sweep dynamics of the DH model}\label{sec:sweep_dynamics}
In this section, we calculate the sweep dynamics of the DH model to investigate the quantum nature of the chiral magnets. We consider the DH Hamiltonian with a time-dependent field:
\begin{align}
\hat{H}(t)&=D\sum_{j=1}^{M-1}(\hat{S}_j^x\hat{S}_{j+1}^y-\hat{S}_j^y\hat{S}_{j+1}^x)-h^x(t)\sum_{j=1}^M\hat{S}_j^x.\label{eq:time-dependent_DH_Hamiltonian}
\end{align}
We will discuss the time dependence of $h^x(t)$ in the next paragraph. We use the time-evolving block decimation (TEBD) method \cite{Vidal2003,Vidal2004} with the second-order Trotter decomposition for real-time evolution. We treat the explicit time dependence of the Hamiltonian by the technique proposed in Ref.~\cite{Hatano2005}. We set the discrete time step to $\Delta t=0.01\hbar/D$, and take the bond dimension to be sufficiently large to avoid artifact of truncation. For comparison, we also calculate the dynamics of the classical DH model by replacing the spin operators with {\it c}-numbers and using the standard fourth-order Runge-Kutta method for the time evolution.

\begin{figure}[t]
\centering
\includegraphics[width=8.6cm,clip]{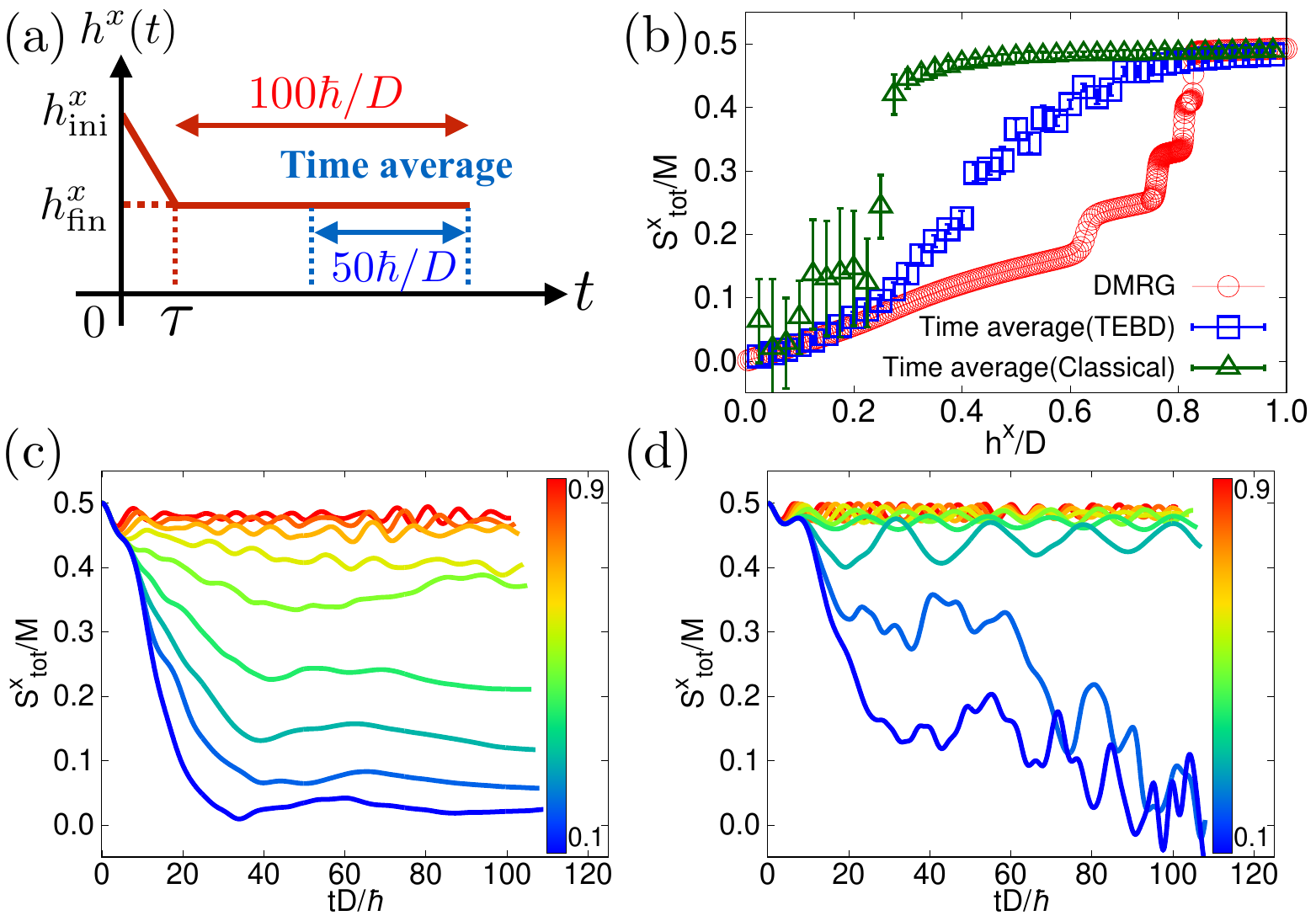}
\caption{(a) Time sequence of the magnetic field $h^x(t)$. We calculate the time averages in the range $\tau+50\hbar/D\le t\le \tau+100\hbar/D$. (b) The $x$ component of the time-averaged magnetization per site for TEBD (blue square) and classical (green triangle) results. The error bars represent the standard deviation of the time average. For reference, we plot the ground-state magnetization calculated by the DMRG method (red circle). (c),~(d) Magnetization of $x$ component vs time for TEBD (c) and classical (d) results. The color bars represent the results for $h^x_{\rm fin}=0.9D$ (red) to $h^x_{\rm fin}=0.1D$ (blue). 
}
\label{fig:sweep}
\vspace{-0.75em}
\end{figure}

We set the initial condition as $\cket{\psi(0)}=\cket{+_1+_2\ldots +_M}$, where $\cket{\pm_j}\equiv (\cket{\uparrow_j}\pm\cket{\downarrow_j})/\sqrt{2}$ are the eigenstates of $\hat{S}_j^x$ with eigenvalues $\pm\frac{1}{2}$. This state is almost the ground state of the DH model for large magnetic-field regimes, as shown in the previous section. We set the magnetic field to $h_{\rm ini}^x=1D$ at $t=0$ and start to ramp down the magnetic field $h^x$ linearly to the final value $h_{\rm fin}^x$ during $0\le t\le \tau$, then fix it until $t=\tau+100\hbar/D$ [see the time sequence in Fig.~\ref{fig:sweep} (a)]. Here, we fix the ramp rate to $0.1D^2/\hbar$. 

Figure~\ref{fig:sweep} (b) shows the time-averaged magnetization in the $x$ direction $\hat{S}_{\rm tot}^x\equiv \sum_{j=1}^M\hat{S}_j^x$. We also plot the ground-state magnetization curve obtained by the DMRG calculations. The time-averaged magnetizations do not follow the ground state for both classical and quantum cases. In particular, the time-averaged magnetization of the classical case suddenly changes around $h^x_{\rm fin}\sim 0.25D$ [see Figs.~\ref{fig:sweep} (b) and (d)]. This behavior indicates metastability of the uniformly polarized state, which has been discussed in the context of classical chiral magnets \cite{Togawa2015,Mito2018,Shinozaki2018}. The sudden jump is due to the vanishing of the surface energy barrier, which prevents the chiral solitons from entering the bulk from the boundaries of the system. In contrast to the classical case, the time-averaged magnetization changes smoothly as a function of the magnetic field in the quantum case, as shown in Figs.~\ref{fig:sweep} (b) and (c). We attribute this behavior to macroscopic quantum tunneling between states with different winding numbers due to strong quantum fluctuations in $S=\frac{1}{2}$ systems.

We now discuss the experimental observables. In optical tweezer experiments, the $z$ component of the local spin can be extracted from measured recapture probability after depopulating one of the spin components through a short-lived intermediate state \cite{Scholl2022}. We can also obtain the $x$ and $y$ components using the quantum-state tomography technique by rotating the measurement basis with a microwave \cite{Signoles2021,Scholl2022,Geier2021}. Since all spin components are observables, the above quantum dynamics can be accessible experimentally.

\section{Quantum many-body scar states of the DH model}\label{sec:quantum_many-body_scar}

In this section, we show that the DH model has QMBS states. The QMBS states \cite{Serbyn2021,Papic2022,Moudgalya2022,Chandran2023} have been discussed in the context of thermalization problems in isolated quantum systems \cite{DAlessio2016,Mori2018} and extensively studied theoretically \cite{Choi2019,Ho2019,Lin2019,Schecter2019,Iadecola2020,Chattopadhyay2020,Mark2020,Mark2020_2,ODea2020,Shibata2020,Michailidis2020,Kuno2020,Sugiura2021,Tang2022,Tamura2022,Sanada2023,Omiya2023,Inversen2023,Moudgalya2022a,Moudgalya2023a} and experimentally \cite{Bernien2017,Bluvstein2021,Su2023}. QMBS states refer to special eigenstates in nonintegrable systems that do not satisfy the strong version of the eigenstate thermalization hypothesis (ETH) \cite{Deutsch1991,Srednicki1994,Rigol2008}. If we set the initial condition with a large overlap with a QMBS state, the relaxation to the thermal equilibrium state is slow or does not occur persistently. Under appropriate conditions, the system exhibits persistent oscillations.

In the following, we show that the DH model has the QMBS states based on a restricted spectrum generating algebra (RSGA) \cite{Moudgalya2020}. For completeness, we recall the definition of the RSGA of order $1$ before showing the results of the DH model. 

Let us consider the Hamiltonian $\hat{H}_0$ and the operator $\hat{Q}^{\dagger}$ such that $\hat{Q}^{\dagger}\cket{\psi_0}\not=0$ for some state $\cket{\psi_0}$. We assume that $\hat{H}_0$, $\hat{Q}^{\dagger}$, and $\cket{\psi_0}$ satisfy the following conditions:
\begin{align}
\hat{H}_0\cket{\psi_0}&=E_0\cket{\psi_0},\label{eq:condition1_RSGA1}\\
[\hat{H}_0, \hat{Q}^{\dagger}]\cket{\psi_0}&=\mathcal{E}\hat{Q}^{\dagger}\cket{\psi_0},\label{eq:condition2_RSGA1}\\
[\,[\hat{H}_0,\hat{Q}^{\dagger}],\hat{Q}^{\dagger}]&=0,\label{eq:condition3_RSGA1}
\end{align}
where $E_0$ and $\mathcal{E}$ are real numbers. These relations are called the RSGA of order 1. If the above relations are satisfied, we can show the following relations:
\begin{align}
\hat{H}_0\cket{\psi_n}&=(E_0+n\mathcal{E})\cket{\psi_n},\label{eq:eigenstate_scar_RSGA}\\
\cket{\psi_n}&\equiv (\hat{Q}^{\dagger})^n\cket{\psi_0}\quad \text{or }\cket{\psi_n}=0.\label{eq:results_RSGA}
\end{align}
The state $\cket{\psi_n}$ corresponds to the (unnormalized) QMBS state.

\subsection{Periodic boundary case}\label{subsec:quantum_many-body_scar_PBC}

In this subsection,  we show the existence of the QMBS states of the DH model under  the periodic boundary conditions. The Hamiltonian is given by
\begin{align}
\hat{H}_{\rm DH}^{\rm PBC}&=D\sum_{j=1}^{M}\hat{S}_j^x(\hat{S}_{j-1}^z-\hat{S}_{j+1}^z)-h^x\sum_{j=1}^M\hat{S}_j^z,\label{eq:DH_Hamiltonian_periodic}
\end{align}
where $\hat{S}_{j+M}^{\mu}=\hat{S}_j^{\mu}$ is imposed and we change the basis $(\hat{S}_j^x, \hat{S}_j^y, \hat{S}_j^z)\to (\hat{S}_j^z, \hat{S}_j^x, \hat{S}_j^y)$ to simplify the calculations. Here, we consider only the case of even $M$ for simplicity. A direct calculation shows that the Hamiltonian $\hat{H}_{\rm DH}^{\rm PBC}$ and the following operator and state satisfy the RSGA relations (\ref{eq:condition1_RSGA1}), (\ref{eq:condition2_RSGA1}), and (\ref{eq:condition3_RSGA1}):
\begin{align}
\hat{Q}^{\dagger}&\equiv \sum_{j=1}^M\hat{P}_{j-1}\hat{S}_j^+\hat{P}_{j+1},\label{eq:definition_of_Q_periodic_supple}\\
\cket{\psi_0}&\equiv \cket{\downarrow_1\downarrow_2\ldots\downarrow_M}\equiv \cket{\Downarrow},\label{eq:definition_of_psi0_supple}
\end{align}
where $\hat{P}_j\equiv 1/2-\hat{S}_j^z$ and $\hat{S}_j^{\pm}\equiv \hat{S}_j^x\pm i\hat{S}_j^y$. The QMBS states are given by
\begin{align}
\cket{S_n}&\propto (\hat{Q}^{\dagger})^n\cket{\Downarrow},\;n=1,2,\ldots, M/2,\label{eq:definition_of_scar_state_periodic_supple}
\end{align}
which satisfy
\begin{align}
\hat{H}_{\rm DH}^{\rm PBC}\cket{S_n}&=\left(\frac{1}{2}M-n\right)h^x\cket{S_n}.\label{eq:scar_state_periodic_supple}
\end{align}
The operator $\hat{Q}^{\dagger}$ generates the transition $\cket{\downarrow_{j-1}\downarrow_j\downarrow_{j+1}}\to \cket{\downarrow_{j-1}\uparrow_j\downarrow_{j+1}}$. From this fact, we can write the QMBS states as
\begin{align}
\cket{S_n}&=\frac{1}{n!\sqrt{N(M,n)}}(\hat{Q}^{\dagger})^n\cket{\Downarrow}\notag \\
&=\frac{1}{\sqrt{N(M,n)}}\sideset{}{'}\sum_{\bm{m}}\cket{\bm{m}},\label{eq:explict_form_of_QMBS_supple}\\
N(M,n)&\equiv \frac{M}{M-n}\binom{M-n}{n},\label{eq:Normalization_constant_scar_PBC}
\end{align}
where $\cket{\bm{m}}\equiv \cket{m_1m_2\ldots m_M}$ represents the basis state, $m_i=\uparrow\text{ or }\downarrow$, $\displaystyle{\sideset{}{'}\sum_{\bm{m}}}$ represents the summation over all possible configurations of $\bm{m}$ where $n$ up spins are not adjacent to each other, and $N(M,n)$ is the normalization constant. We also show that the DH model has another series of QMBS states. We define
\begin{align}
\hat{Q}'^{\dagger}&\equiv \sum_{j=1}^M\hat{P}'_{j-1}\hat{S}_j^-\hat{P}'_{j+1},\label{eq:definition_of_Qprime}\\
\hat{P}_j'&\equiv \frac{1}{2}+\hat{S}_j^z,\label{eq:definition_of_projection_operator_down_state}\\
\cket{\psi_0}&\equiv \cket{\uparrow_1\uparrow_2\ldots\uparrow_M}\equiv \cket{\Uparrow}.\label{eq:definition_of_all_up_state}
\end{align}
Similarly, we can show that the above operators and the states satisfy the RSGA relations. The QMBS states are given by
\begin{align}
\cket{S_n'}&=\frac{1}{n!\sqrt{N(M,n)}}(\hat{Q}'^{\dagger})^n\cket{\Uparrow}=\frac{1}{\sqrt{N(M,n)}}\sideset{}{''}\sum_{\bm{m}}\cket{\bm{m}},\label{eq:scar_state_from_all_up_state}
\end{align}
where $\displaystyle{\sideset{}{''}\sum_{\bm{m}}}$ represents the summation over all possible configurations of $\bm{m}$ where $n$ down spins are not adjacent to each other. The state $\cket{S_n'}$ satisfies
\begin{align}
\hat{H}_{\rm DH}^{\rm PBC}\cket{S_n'}=\left(-\frac{M}{2}+n\right)h^x\cket{S_n'}.\label{eq:eigenenergy_Sn'}
\end{align}
These two towers of QMBS states are related to each other as $\hat{C}_x\cket{S_n'}=\cket{S_n}$ through the unitary operator $\hat{C}_x\equiv \prod_{j=1}^M(2\hat{S}_j^x)$. For $n=M/2$, we can easily show $\cket{S_{M/2}}=|S_{M/2}'\rangle=(\cket{\uparrow\downarrow\ldots\uparrow\downarrow}+\cket{\downarrow\uparrow\ldots\downarrow\uparrow})/\sqrt{2}$. We note that the QMBS states (\ref{eq:definition_of_scar_state_periodic_supple}) and (\ref{eq:scar_state_from_all_up_state}) are almost the same as those with a three-spin interaction found in Refs.~\cite{Iadecola2020,Mark2020,Omiya2023}, while our Hamiltonian (\ref{eq:DH_Hamiltonian_periodic}) is different from theirs. Compared to these studies, the DH model has advantages regarding experimental feasibility \cite{footnote_DH}. We also find the pyramid scar states. See the Supplemental Material (SM) for a detailed calculation \cite{supple}.

We now discuss the symmetry of the DH model. We can show that the following operators commute with the Hamiltonian (\ref{eq:DH_Hamiltonian_periodic}):
\begin{align}
\hat{N}_{\rm sol}^{\rm PBC}&\equiv \sum_{j=1}^M\left(\frac{1}{4}-\hat{S}_j^z\hat{S}_{j+1}^z\right),\label{eq:definition_of_soliton_number_operator_supple}\\
\hat{C}&\equiv \hat{\mathcal{I}}\hat{C}_z,\label{eq:definition_of_C_operator}
\end{align}
where $\hat{\mathcal{I}}$ is the space-inversion operator defined by $\hat{\mathcal{I}}\hat{S}_j^{\mu}\hat{\mathcal{I}}=\hat{S}_{M-j+1}^{\mu}$, and $\hat{C}_z\equiv \prod_{j=1}^M(2\hat{S}_j^z)$. The operator $\hat{N}_{\rm sol}^{\rm PBC}$ counts the number of solitons \cite{Kodama_thesis,Kodama2023} as discussed in Sec.~\ref{sec:Ground_state}. For example, we obtain
\begin{align}
\hat{N}_{\rm sol}^{\rm PBC}\cket{\downarrow\downarrow\downarrow\downarrow\uparrow\downarrow\downarrow\downarrow}&=1\cket{\downarrow\downarrow\downarrow\downarrow\uparrow\downarrow\downarrow\downarrow},\label{eq:example_soliton1_supple}\\
\hat{N}_{\rm sol}^{\rm PBC}\cket{\downarrow\uparrow\downarrow\downarrow\uparrow\downarrow\downarrow\downarrow}&=2\cket{\downarrow\uparrow\downarrow\downarrow\uparrow\downarrow\downarrow\downarrow}.\label{eq:example_soliton2_supple}
\end{align}
By direct calculations, we have
\begin{align}
\hat{N}_{\rm sol}^{\rm PBC}\cket{S_n}&=n\cket{S_n},\quad \hat{C}\cket{S_n}=(-1)^n\cket{S_n},\label{eq:symmetry_sector_QMBS_supple}
\end{align}
where we used the relations $[\hat{N}_{\rm sol}^{\rm PBC}, \hat{Q}^{\dagger}]=\hat{Q}^{\dagger}$ and $\hat{C}\hat{Q}^{\dagger}\hat{C}=-\hat{Q}^{\dagger}$. Therefore, the QMBS state $\cket{S_n}$ is in the symmetry sector specified by $(N_{\rm sol}^{\rm PBC}, C)=(n,(-1)^n)$.

\subsection{Open boundary case}\label{subsec:quantum_many-body_scar_OBC}

In this subsection, we consider QMBS states under the open boundary condition, where the Hamiltonian reads
\begin{subequations}  
\begin{align}
&\hat{H}_{\rm DH}^{\rm OBC}\notag \\
&\equiv D\sum_{j=1}^{M-1}(\hat{S}_j^z\hat{S}_{j+1}^x-\hat{S}_j^x\hat{S}_{j+1}^z)-h^x\sum_{j=1}^M\hat{S}_j^z-\frac{D}{2}(\hat{S}_1^x-\hat{S}_M^x)\label{eq: DH-open} \\
&=D\sum_{j=1}^M\hat{S}_j^x(\hat{S}_{j-1}^z-\hat{S}_{j+1}^z)-h^x\sum_{j=1}^M\hat{S}_j^z.\label{eq:DH_plus_edge_magnetic_field}
\end{align}
\end{subequations}
Here, we defined $\hat{S}_0^z=\hat{S}_{M+1}^z=-\frac{1}{2}$. 
The last term in Eq.~\eqref{eq: DH-open} is necessary to define the soliton number operator that commutes with Eq.~\eqref{eq: DH-open}. In  Eq.~\eqref{eq:DH_plus_edge_magnetic_field}, we add the auxiliary sites $j=0$, $M+1$ and restrict the states to those with the down spins at the two sites. This restriction is equivalent to applying a downward magnetic field with infinite strength to the spins at the auxiliary sites.    

Because the Hamiltonian is formally the same as in the periodic case, the QMBS states exist for Eq.~(\ref{eq:DH_plus_edge_magnetic_field}) with a slight modification. We modify the definition of $\hat{Q}^{\dagger}$ and $\hat{P}_j$ as
\begin{align}
\hat{Q}^{\dagger}&\equiv \hat{S}_1^+\hat{P}_2+\sum_{j=2}^{M-1}\hat{P}_{j-1}\hat{S}_j^+\hat{P}_{j+1}+\hat{P}_{M-1}\hat{S}_M^+\notag \\
&= \sum_{j=1}^M\hat{P}_{j-1}\hat{S}_j^+\hat{P}_{j+1},\label{eq:operator_Q_for_open_boundary_case_supple}\\
\hat{P}_j&\equiv \frac{1}{2}-\hat{S}_j^z,\quad \hat{P}_0\equiv 1,\quad \hat{P}_{M+1}\equiv 1.\label{eq:projection_operator_at_virtual_edge_supple}
\end{align}
A direct calculation shows that the Hamiltonian $\hat{H}_{\rm DH}^{\rm OBC}$, the operator $\hat{Q}^{\dagger}$, and the state $\cket{\Downarrow}$ satisfy the RSGA relations (\ref{eq:condition1_RSGA1}), (\ref{eq:condition2_RSGA1}), and (\ref{eq:condition3_RSGA1}).  The QMBS states are obtained by
\begin{align}
\cket{S_n}&\propto(\hat{Q}^{\dagger})^n\cket{\Downarrow},\label{eq:QMBS_expression_supple}
\end{align}
which satisfies
\begin{align}
\hat{H}_{\rm DH}^{\rm OBC}\cket{S_n}&=\left(\frac{1}{2}M-n\right)h^x\cket{S_n}.\label{eq:QMBS_open_boundary_case}
\end{align}
The explicit expression of the QMBS state is given by
\begin{align}
\cket{S_n}&=\frac{1}{\sqrt{\mathcal{N}(M,n)}}\sideset{}{'}\sum_{\bm{m}}\cket{\bm{m}},\label{eq:explict_form_of_QMBS_open_supple}\\
\mathcal{N}(M,n)&\equiv \binom{M-n+1}{n},\label{eq:definition_of_normalization_constant_QMBS}
\end{align}
where the meaning of the summation is the same as the case of the periodic boundary condition. We can also show that the Hamiltonian commutes with $\hat{C}$ [the definition is the same as the periodic boundary case, see Eq.~(\ref{eq:explict_form_of_QMBS_supple}).] and modified soliton number operator:
\begin{align}
&\hat{N}_{\rm sol}^{\rm OBC}\notag \\
&\equiv \sum_{j=1}^{M-1}\left(\frac{1}{4}-\hat{S}_j^z\hat{S}_{j+1}^z\right)+\left(\frac{1}{4}+\frac{1}{2}\hat{S}_1^z\right)+\left(\frac{1}{4}+\frac{1}{2}\hat{S}_M^z\right)\notag \\
&=\sum_{j=0}^{M}\left(\frac{1}{4}-\hat{S}_j^z\hat{S}_{j+1}^z\right).\label{eq:modified_soliton_number_operator}
\end{align}
This operator counts the number of solitons in the extended Hilbert space spanned by $\cket{\downarrow_0;\bm{m};\downarrow_{M+1}}$. The QMBS states are in the symmetry sector specified by $(N_{\rm sol}^{\rm OBC},C)=(n,(-1)^n)$.

\subsection{Level spacing statistics}\label{subsec:level_spacing_statistics}

\begin{figure}[t]
\centering
\includegraphics[width=8.6cm,clip]{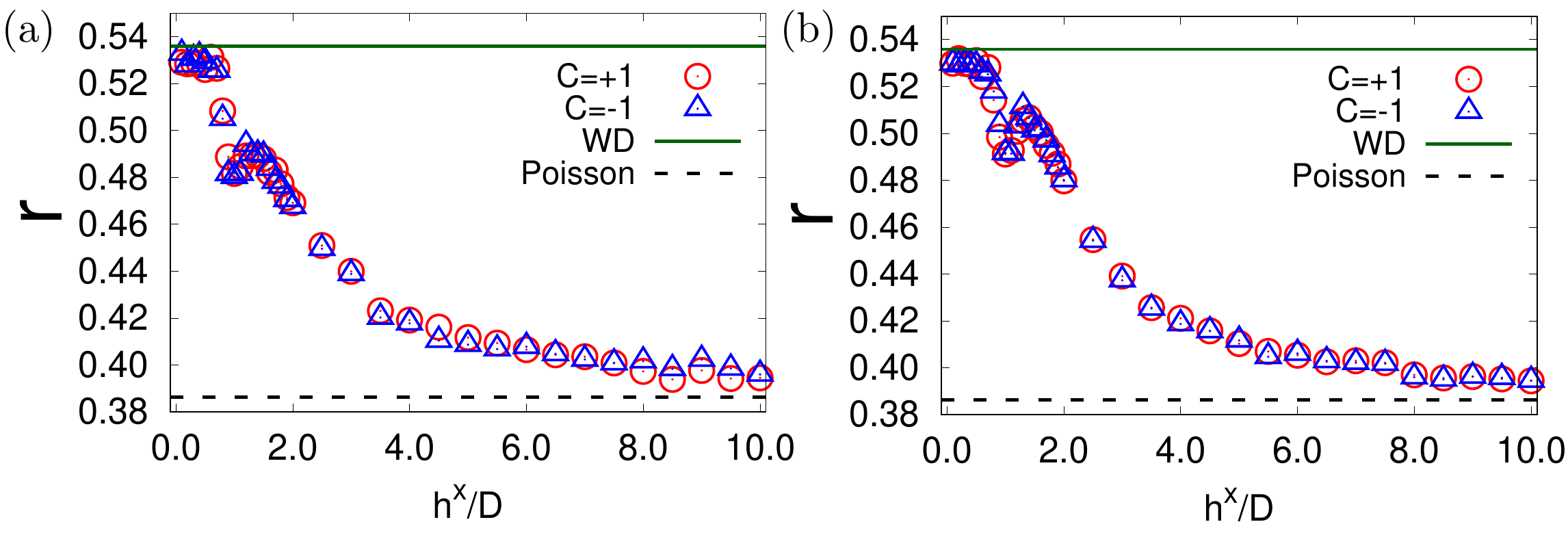}
\caption{$r$ value as a function of the magnetic field. The green solid and black dashed lines represent the $r$ value of the Wigner-Dyson (Gaussian orthogonal ensemble) and Poisson distribution, respectively. (a) DH model without the edge magnetic field for $M=16$. (b) DH model with the edge magnetic field for $M=18$ in $N_{\rm sol}^{\rm OBC}=5$ sector.}
\label{fig:level_spacing_supple}
\vspace{-0.75em}
\end{figure}%
\begin{figure}[t]
\centering
\includegraphics[width=8.6cm,clip]{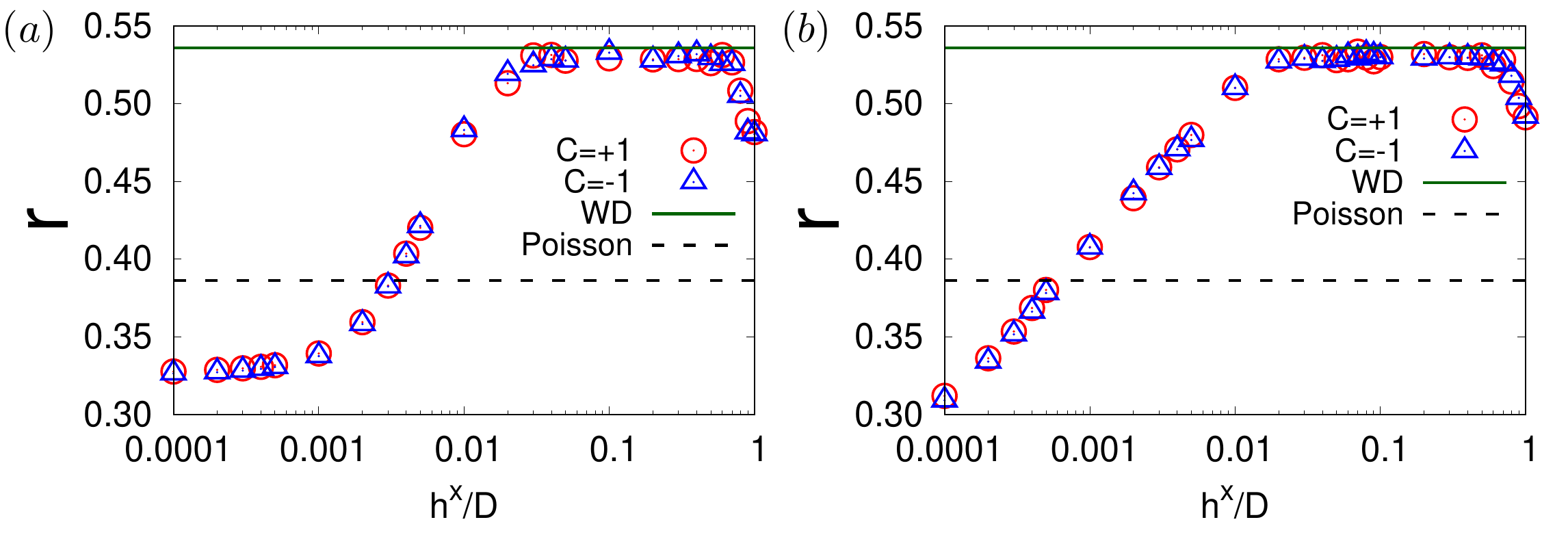}
\caption{$r$ value as a function of the magnetic field in the low-field regime. The green solid and black dashed lines represent the $r$ value of the Wigner-Dyson (Gaussian orthogonal ensemble) and Poisson distribution, respectively. (a) DH model without the edge magnetic field for $M=16$. (b) DH model with the edge magnetic field for $M=18$ in $N_{\rm sol}^{\rm OBC}=5$ sector.}
\label{fig:level_spacing_supple_small_h}
\vspace{-0.75em}
\end{figure}%

In this subsection, we discuss the level spacing statistics. For the states created by $\hat{Q}^{\dagger}$ to be QMBS states, we need to show the  nonintegrablity of the system as well as  a sub-volume law scaling of the entanglement entropy (EE) in QMBS states. We  confirm the former by the mean level spacing ratio \cite{Oganesyan2007,Atas2013} as shown below. Let $E_n$ be the $n$th eigenenergy of the Hamiltonian ($E_{n+1}\ge E_n$) in a specific symmetry sector. We define the $r$ value by $r\equiv \langle\min(r_n, 1/r_n )\rangle$, where $r_n\equiv s_{n+1}/s_n$ and $s_n\equiv E_{n+1}-E_n$.  The symbol $\langle\cdots\rangle$ represents the average of $r_n$. To confirm nonintegrability, we perform the exact diagonalization \cite{Sandvik2010,Jung2020} for the Hamiltonian (\ref{eq:DH_plus_edge_magnetic_field}), which  commutes with the operators $\hat{C}$ and $\hat{N}_{\rm sol}^{\rm OBC}$.

Figure~\ref{fig:level_spacing_supple} (a) shows the results of the magnetic-field dependence of the $r$ value of the DH model without an edge magnetic field. In this case, $\hat{C}$ is conserved, but $\hat{N}_{\rm sol}^{\rm OBC}$ is not. The $r$ value is consistent with the Wigner-Dyson distribution (Gaussian orthogonal ensemble) around $h^x\lesssim D$ ($r\simeq 0.53$). In the high-field regime, the $r$ value approaches the Poisson value ($r\simeq 0.38$), due to the emergent conservation law of $\hat{S}_{\rm tot}^z$ in the high-field regime. We also plot the $r$ value of the DH model with the edge magnetic field in Fig.~\ref{fig:level_spacing_supple} (b), where we calculate the largest symmetry sector. For $M=18$, e.g., $N_{\rm sol}^{\rm OBC}=5$ is the largest. 

Figures \ref{fig:level_spacing_supple_small_h} (a) and (b) represent the $r$ value in the low-field regime. We can see the deviation of the $r$ value from the Wigner-Dyson value around $h^x\sim 0.01D$. This behavior results from the integrability at $h^x=0$; the DH model is equivalent to the {\it XY} model at this point in the absence of the edge magnetic field. The integrability of the {\it XY} model with the edge magnetic field terms was discussed in Refs.~\cite{Bilstein1999,Bilstein2000,Bilstein2000_2}. Regardless of the presence or absence of the edge magnetic field, the $r$ value does not approach the Poisson value. We may attribute this discrepancy to the additional emergent symmetry at $h^x=0$ discussed in Refs.~\cite{Bilstein1999,Bilstein2000,Bilstein2000_2}. 

\subsection{Entanglement entropy and off-diagonal long-range order}\label{subsec:entanglement_entropy_and_ODLRO}
\begin{figure}[t]
\centering
\includegraphics[width=8.6cm,clip]{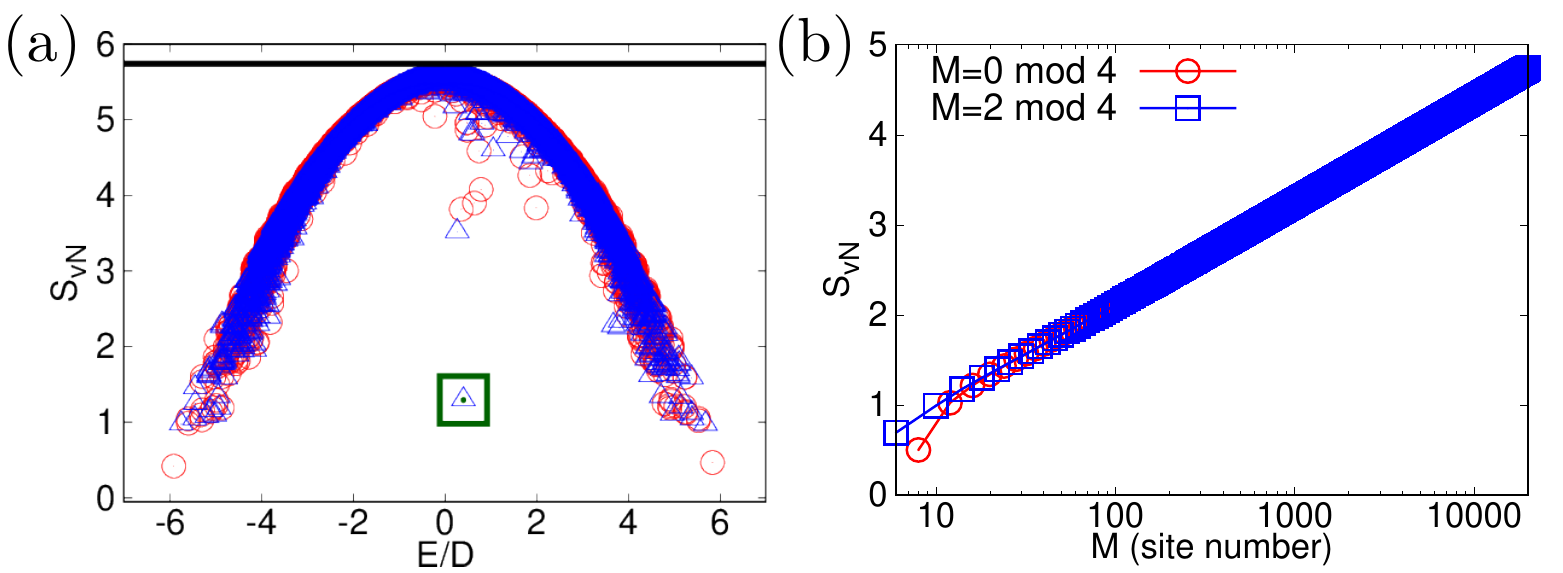}
\caption{(a) Half-chain von Neumann EE as a function of the eigenenergy of Hamiltonian (\ref{eq:DH_plus_edge_magnetic_field}) for $M=18$ and $h^x=0.1D$ in the symmetry sector $N_{\rm sol}^{\rm OBC}=5$ and $C=+1$ (red circle) and $C=-1$ (blue triangle). The black solid line and green square represent the Page value \cite{Page1993} and the EE of the QMBS state $\cket{S_5}$, respectively.  (b) Size dependence of the half-chain von Neumann entanglement entropy of the QMBS state $\cket{S_n}$, where $n=\lfloor M/4\rfloor+2$ with $\lfloor \cdot\rfloor$ being the floor function.
}
\label{fig:scar}
\vspace{-0.75em}
\end{figure}

In this subsection, we discuss the half-chain entanglement entropy of the system under the open boundary condition. Figure \ref{fig:scar} (a) shows the half-chain von Neumann EE as a function of the eigenenergy of the system. We find a QMBS state with a high-energy eigenstate and low EE. One can derive the analytical expression of the EE of the QMBS state (see Appendix~\ref{app:Analytical_EE} for details). We show the size dependence of the EE in Fig.~\ref{fig:scar} (b). The EE obeys the sub-volume law $\sim\ln{M}$.

Then, we discuss the off-diagonal long-range order (ODLRO) \cite{Yang1962} of the QMBS states. In our case, the order parameter is given by $\hat{Q}^{\dagger}$ [see Eq.~(\ref{eq:operator_Q_for_open_boundary_case_supple})]. We can detect ODLRO  by the correlation function $\bra{S_n}\hat{Q}\hat{Q}^{\dagger}\cket{S_n}$ \cite{Tasaki_book}. Using the relation
\begin{align}
\hat{Q}^{\dagger}\cket{S_n}&=\sqrt{\frac{(n+1)(M-2n+1)(M-2n)}{M-n+1}}\cket{S_{n+1}},\label{eq:relation_Qdagger_to_Sn}
\end{align}
we can obtain 
\begin{align}
\frac{1}{M^2}\bra{S_n}\hat{Q}\hat{Q}^{\dagger}\cket{S_n}&=\frac{(n+1)(M-2n+1)(M-2n)}{M^2(M-n+1)}.\label{eq:expectation_value_of_QQdagger}
\end{align}
Here, we fix $n/M=O(1)$. In the thermodynamic limit, Eq.~(\ref{eq:expectation_value_of_QQdagger}) becomes
\begin{align}
&\lim_{M\to\infty,n/M;\text{fixed}}\frac{1}{M^2}\bra{S_n}\hat{Q}\hat{Q}^{\dagger}\cket{S_n}\notag \\
&=\frac{n}{M}\left(1-\frac{n/M}{1-n/M}\right)=O(1).\label{eq:thermodynamic_limit_scar}
\end{align}
Namely,  the QMBS states exhibit ODLRO, which we can interpret as magnon condensation \cite{Iadecola2020}. Equation~\eqref{eq:thermodynamic_limit_scar}  indicates that the DH model violates the ETH because ODLRO is not allowed for ETH-obeying states.

\subsection{Experimental implications}\label{subsec:experimental_implications}

In this subsection, we address how to observe the QMBS states. In the spin-laboratory frame and the original basis, the edge magnetic field term becomes $-(D/2)[\hat{S}_1^x\sin q-\hat{S}_1^y\cos q-\hat{S}_M^x\sin(q M)+\hat{S}_M^y\cos(q M)]$. To realize this term experimentally, we need additional lasers with the same Rabi frequency as the Raman laser and the appropriate phases. Thus, implementing the edge magnetic field in Eq.~(\ref{eq:DH_plus_edge_magnetic_field}) in experiments is technically challenging.  However, we expect to see the effects of the QMBS states without an edge magnetic field, which can be regarded as a perturbation to the Hamiltonian (\ref{eq:DH_plus_edge_magnetic_field}).  Slow relaxation comes about when the initial state has a large overlap with the QMBS states. To verify this, we calculate the quench dynamics starting from the initial state $\cket{\text{xN\'{e}el}}\equiv \cket{+_1-_2+_3-_4\ldots}$. [In this paragraph, we use the original basis used in Hamiltonian (\ref{eq:Hamiltonian_DH_model}).] We use the TEBD method \cite{Vidal2003,Vidal2004} with the second-order Trotter decomposition for real-time evolution. We set the discrete time step to $\Delta t=0.01\hbar/D$, and take the bond dimension to be sufficiently large to reduce the truncation effect. 

\begin{figure}[t]
\centering
\includegraphics[width=8.4cm,clip]{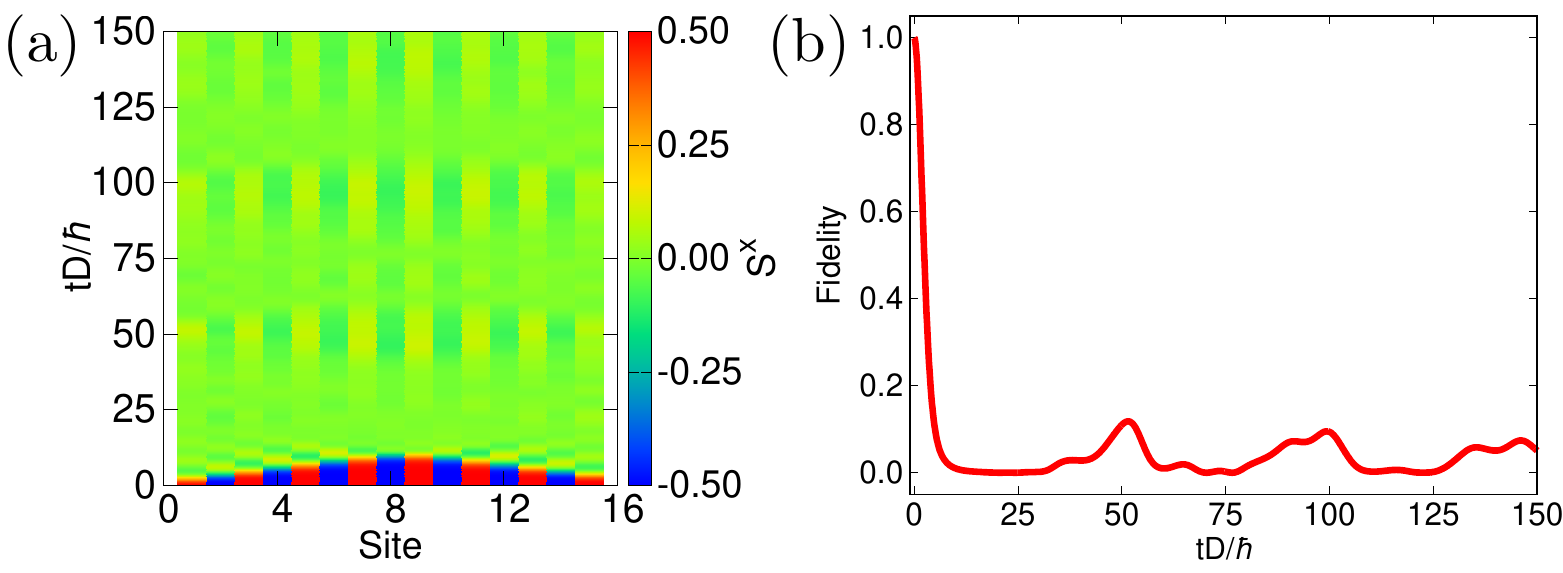}
\caption{
(a) Time evolution of $S^x_j$ for $M=16$ and $h^x=0.1D$. Here, we use the Hamiltonian (\ref{eq:Hamiltonian_DH_model}). (b) Time evolution of the fidelity $|\langle{\text{xN\'{e}el}}|\psi(t)\rangle|^2$ for the same parameters of (a), where $\cket{\psi(t)}$ is the wave function at time $t$.
}
\label{fig:quench_dynamics_xNeel}
\vspace{-0.75em}
\end{figure}

We show the results in Figs.~\ref{fig:quench_dynamics_xNeel} (a) and (b), which exhibit revivals of the xN\'{e}el state around $t\sim 50\hbar/D$, $100\hbar/D$, and $150\hbar/D$.  We attribute this behavior to a large overlap between xN\'{e}el state and  $\cket{S_{M/2}}$. For example, for $M=4$, the QMBS state is given by $\cket{S_2}\propto \cket{+_1-_2-_3+_4}+\cket{+_1-_2+_3-_4}+\cket{-_1+_2-_3+_4}$, which contains the xN\'{e}el state. For general $M$, we can show $|\bracket{\text{xN\'{e}el}}{S_{M/2}}|^2=O(M^{-1})$. This size dependence is unexpected in generic quantum many-body systems. More quantitatively, we can understand the revival behavior as the existence of edge states in the Krylov space. We discuss this point in Appendix~\ref{app:Krylov_subspace}.

\section{Asymptotic quantum many-body scar states of the DH model}\label{sec:asymptotic_scar}

Recently, Gotta {\it et al}. proposed a novel class of nonergodic quantum states, dubbed the AQMBS states \cite{Gotta2023}, which are characterized by the following properties: (i) orthogonality to any exact QMBS state, (ii) small entanglement, and (iii) the energy variance that vanishes in the thermodynamics limit. Due to the property (iii), the relaxation time diverges in the thermodynamics limit. According to Ref.~\cite{Gotta2023}, the AQMBS states can be constructed by deforming the QMBS state. Motivated by Ref.~\cite{Gotta2023}, we construct the AQMBS states in the DH model. While they can be constructed for both periodic and open boundary conditions, we focus on the latter in the following. See the SM for details \cite{supple}. 

The AQMBS states in the DH model are given by
\begin{align}
\cket{AS_n}&\propto \hat{A}^{\dagger}\cket{S_{n-1}},\quad n=1,2,\ldots, M/2,\label{eq:definition_of_asymptotic_scar_state}\\
\hat{A}^{\dagger}&\equiv \sum_{j=1}^Mf_j\hat{P}_{j-1}\hat{S}_j^+\hat{P}_{j+1},\label{eq:definition_of_operator_Adagger}
\end{align}
where $f_j$ is a complex function that satisfies the condition $f_j=-f_{M+1-j}$. This condition ensures $\bracket{AS_n}{S_m}=0$ because $\cket{AS_n}$ and $\cket{S_n}$ have different parity under spatial inversion.
Here, we choose $f_j=\cos[\pi j/(M+1)]$, but other choices are possible (see the SM \cite{supple}).

To verify that the AQMBS states have a small amount of entanglement, we study the half-chain EE of Eq.~(\ref{eq:definition_of_asymptotic_scar_state}), which we can calculate from the matrix product state (MPS) representation \cite{Schollwock2011,Paeckel2019} of Eq.~(\ref{eq:definition_of_asymptotic_scar_state}). Our analytical result shows that the bond dimension of the state (\ref{eq:definition_of_asymptotic_scar_state}) is at most $\chi=n(3n+1)$, which is of the order of $O(M^2)$ because $n$ is at most $O(M)$. The half-chain EE is bounded by $\ln\chi\sim \ln M$. Therefore, the state (\ref{eq:definition_of_asymptotic_scar_state}) obeys a sub-volume law scaling.

\begin{figure}[t]
\centering
\includegraphics[width=8.6cm,clip]{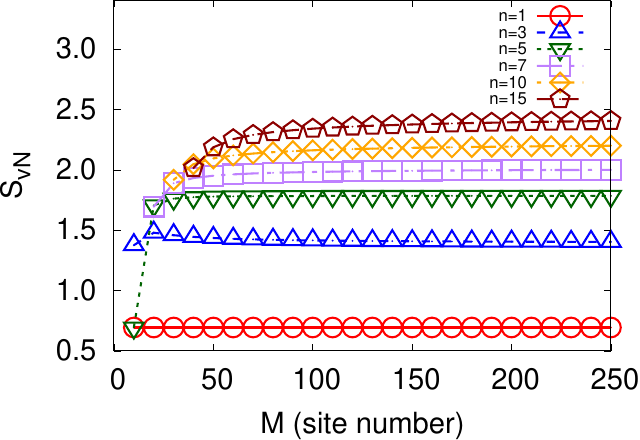}
\caption{System-size dependence of the half-chain EE for the states $\cket{AS_n}$ for several $n$.}
\label{fig:ee_aqmps_obc}
\vspace{-0.75em}
\end{figure}

Figure~\ref{fig:ee_aqmps_obc} shows numerical results of the half-chain EE of the states $\cket{AS_n}$ for several $n$ and system sizes. We see that the system size dependence of the half-chain EE is weak and their values are small compared to the Page value \cite{Page1993}, which is given by $S_{\rm Page}=(M/2)\ln(2)-1/2\simeq 86$ for $M=250$.

\begin{figure}[t]
\centering
\includegraphics[width=8.6cm,clip]{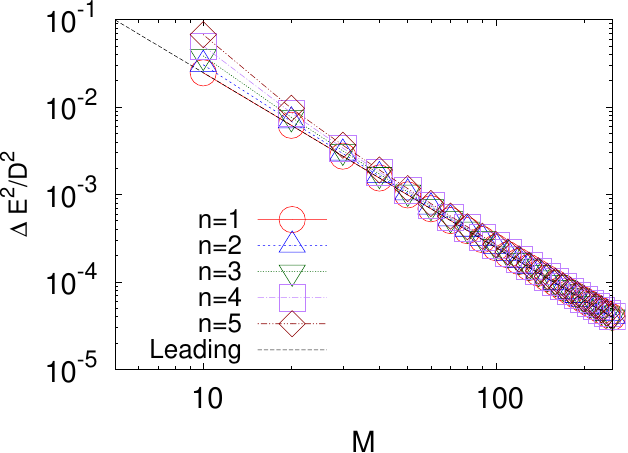}
\caption{The energy variance of the states $\cket{AS_n}$ for several $n$ as a function of the system size $M$. ``Leading'' in the legend means $\Delta E^2/D^2=\pi^2/(4M^2)$.}
\label{fig:asymptotic_scar}
\vspace{-0.75em}
\end{figure}

Next, we estimate the energy variance of the state (\ref{eq:definition_of_asymptotic_scar_state}), which we can write as
\begin{align}
\Delta E^2&\equiv \bra{AS_n}(\hat{H}_{\rm DH}^{\rm OBC})^2\cket{AS_n}-(\bra{AS_n}\hat{H}_{\rm DH}^{\rm OBC}\cket{AS_n})^2\notag \\
&=\frac{\bra{S_{n-1}}[\hat{A}, \hat{H}_{\rm DM}^{\rm OBC}][\hat{H}_{\rm DM}^{\rm OBC}, \hat{A}^{\dagger}] \cket{S_{n-1}}}{\bra{S_{n-1}}\hat{A}\hat{A}^{\dagger}\cket{S_{n-1}}},\label{eq:energy_variance_OBC_case}
\end{align}
where we defined $\hat{H}_{\rm DM}^{\rm OBC}\equiv D\sum_{j=1}^M\hat{S}_j^x(\hat{S}_{j-1}^z-\hat{S}_{j+1}^z)$ and used the fact that $\hat{H}_{\rm DM}^{\rm OBC}\cket{S_{n-1}}=0$ and the state (\ref{eq:definition_of_asymptotic_scar_state}) is an eigenstate of the Zeeman term. For $n=O(1)$, we can exactly evaluate the energy variance as $\Delta E^2=\pi^2D^2/(4M^2)+O(M^{-3})$. We show the numerical results of the energy variance in Fig.~\ref{fig:asymptotic_scar}. For other $n$, we numerically verify that the energy variance goes to zero in the thermodynamic limit (see the SM for details~\cite{supple}.). 

\begin{figure*}[b]
\centering
\includegraphics[width=17cm,clip]{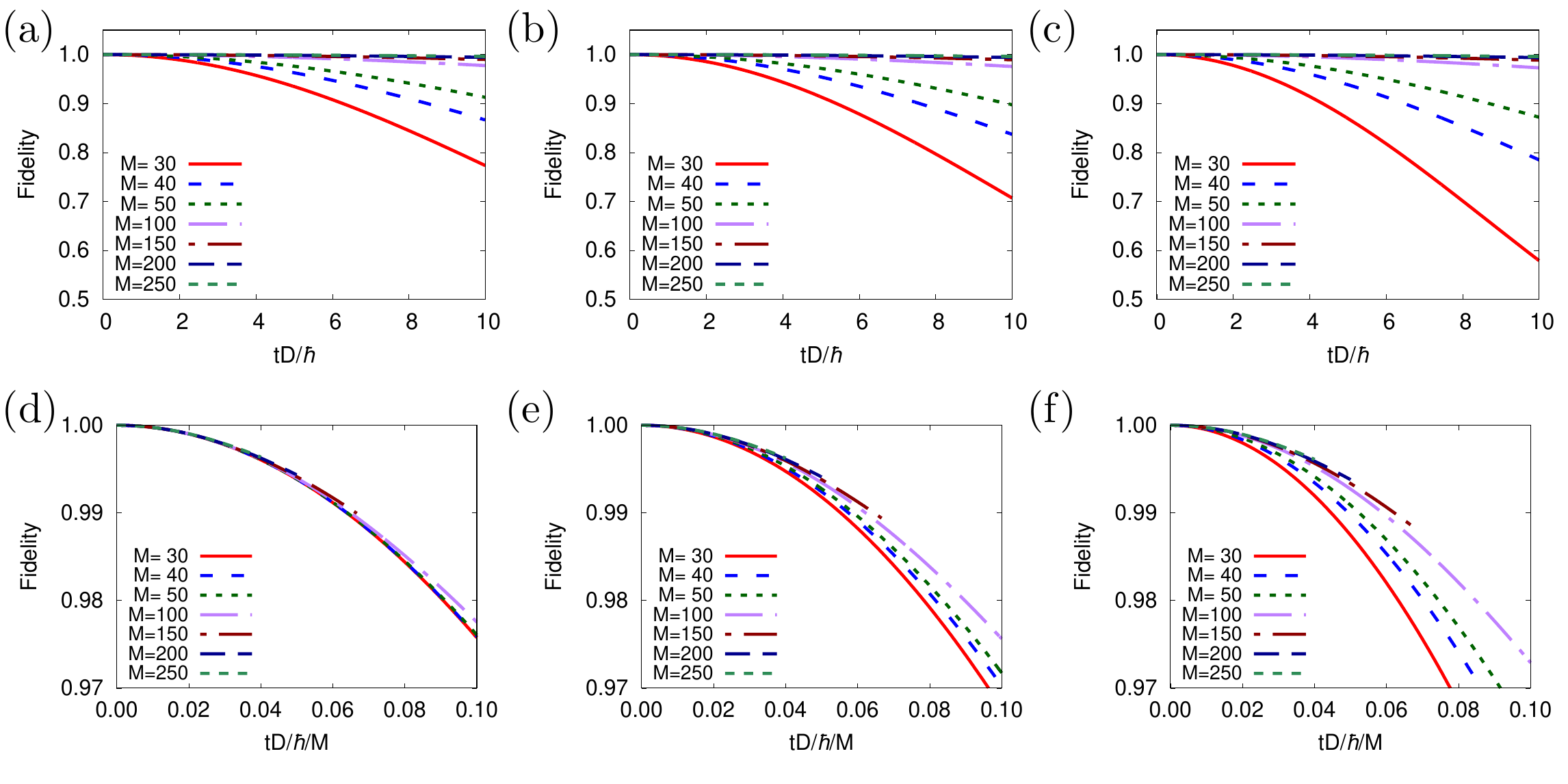}
\caption{Time evolution of the fidelity with the initial condition (a) $\cket{AS_1}$, (b) $\cket{AS_5}$, (c) $\cket{AS_{10}}$. (d)--(f): Rescaled fidelity of (a), (b), and (c), respectively.
}
\label{fig:fidelity_asymptotic_scar_fidelity}
\vspace{-0.75em}
\end{figure*}

Finally, we discuss the dynamics after quench from the state $\cket{AS_n}$. Figures~\ref{fig:fidelity_asymptotic_scar_fidelity}(a)-\ref{fig:fidelity_asymptotic_scar_fidelity}(c) show the fidelity $|\bra{AS_n}e^{-i\hat{H}_{\rm DH}^{\rm OBC}t/\hbar}\cket{AS_n}|^2$ for several system sizes. We use the TEBD method to obtain these results. The discrete time step is set to $\Delta t=0.01\hbar/D$, and the bond dimension is taken to be $\chi=200\sim 310$ so that the truncation error is less than $10^{-10}$. At early times, the fidelity can be approximated as $|\bra{AS_n}\exp(-i\hat{H}_{\rm DH}^{\rm OBC}t/\hbar)\cket{AS_n}|^2\simeq 1-\Delta E^2t^2/\hbar^2$, which implies that the smaller energy variance yields a longer relaxation time. As expected, we see the system size dependence of the fidelity even for $n>1$. We also plot the rescaled fidelity in Figs.~\ref{fig:fidelity_asymptotic_scar_fidelity}(d)-\ref{fig:fidelity_asymptotic_scar_fidelity}(f). We can find a good data collapse by rescaling $t/M$. These results all suggest that the state (\ref{eq:definition_of_asymptotic_scar_state}) is an AQMBS state in the DH model.

\newpage
\section{Summary}\label{sec:summary}
In this paper, we proposed a new method for performing quantum simulations of the $S=\frac{1}{2}$ quantum spin model with the DMI using Rydberg atoms. Our approach is based on the two-photon Raman scheme and unitary transformation, which is feasible with current experimental techniques. One advantage of the Rydberg atom quantum simulator is that it realizes quantum spin models with high tunability. In our scheme, we can easily tune the ratio between the magnitude of the exchange interaction and the DMI in a wide range. This tunability allows us to implement the DH model, which has only the DMI and Zeeman energy. Another advantage is that the heating problem is insignificant compared to the SOC case because the typical timescale is much shorter than the neutral atom case due to the strong interaction between the Rydberg atoms. 

We studied the ground-state properties and nonequilibrium dynamics of the DH model. We showed that the time-averaged magnetization continuously changes as a function of the magnetic field in the fully quantum case, whereas it is discontinuous in the classical case. This behavior may originate from the macroscopic quantum tunneling due to the strong quantum fluctuations.

We have also investigated the nonergodic properties of the DH model. Using the RSGA formalism, we analytically demonstrated the existence of QMBS states in the DH model with open and periodic boundary conditions. Furthermore, we identified a series of AQMBS states in the DH model. These states are orthogonal to all exact QMBS states, have low entanglement, and exhibit slow relaxation due to their small energy variance. 

Our work opens a new avenue for the design of quantum chiral magnets. There are several future directions. One is to extend our scheme to a more complex form of the DMI, which would enable experimental studies of skyrmions in the quantum regime \cite{Ochoa2019}. Another direction is an extension to higher spin systems, allowing us to explore the spin-parity effect in the DH model.

\begin{acknowledgments}
We thank D.~Yamamoto, T.~Nikuni, and S. Moudgalya for their helpful comments. This work was supported by JSPS KAKENHI Grants No.~JP20K14389 (M.K.), No.~JP22H05268 (M.K.), No.~JP22H05267 (T.T.), No.~JP21H05191 (H.K.), No.~JP18K03445 (H.K.), No.~JP23H01093 (H.K.), No.~JP20K03855 (Y.K.), and No.~JP21H01032 (Y.K.).
This research was also supported by Joint Research by the Institute for Molecular Science (IMS program No. 23IMS1101) (Y.K.)
\end{acknowledgments}

\appendix
\clearpage
\begin{widetext}
\section{Derivation of the {\it XXZ} Hamiltonian}\label{app:derivation_XXZ}
In this appendix, we discuss the interaction between Rydberg states $|n_1S_{1/2},m_J=+\frac{1}{2}\rangle\equiv \cket{\downarrow}$ and $|n_2S_{1/2},m_J=+\frac{1}{2}\rangle\equiv \cket{\uparrow}$ with energy $E_{n_1S_{1/2}}$ and $E_{n_2S_{1/2}}$. Since these two states have the same parity, they are not coupled directly via the dipole-dipole interaction. If the atomic distance is large enough, the dipole-dipole interaction can be treated as a perturbation, and the leading term is given by the second-order term \cite{Bijnen_thesis,Whitlock2017}. For simplicity, we consider only the two Rydberg atoms separated by the distance $R$. The Hamiltonian is given by \cite{Weber2017}
\begin{align}
\hat{H}_{\rm two}&=\hat{H}_{\rm Rydberg}+\hat{V}_{\rm dip},\label{eq:interaction_Hamiltonian_two_Rydberg}\\
\hat{H}_{\rm Rydberg}&=\sum_{n,l,J,m_{J}}E_{n,l,J,m_{J}}(\cket{n_,l,J,m_{J}}\bra{n,l,J,m_{J}}\otimes \hat{1}_2)\notag \\
&\quad +\sum_{n',l',J',m_{J}'}E_{n',l',J',m_{J}'}(\hat{1}_1\otimes \cket{n',l',J',m_{J}'}\bra{n',l',J',m_{J}'}),\label{eq:definition_ofH0_two_Rydberg_system}\\
\hat{V}_{\rm dip}&=\frac{1}{4\pi\epsilon_0R^3}\left[\hat{\bm{d}}_1\cdot\hat{\bm{d}}_2-3(\hat{\bm{d}_1}\cdot\tilde{\bm{R}})(\hat{\bm{d}}_2\cdot\tilde{\bm{R}})\right],\label{eq:dipole-dipole_interaction}
\end{align}
where $E_{n,l,J,m_J}$ is the energy level of the Rydberg state specified by $(n,l,J,m_J)$, $n$ is the principal quantum number, $l$ is the orbital angular momentum, $J$ is the total angular momentum, $m_J$ is the magnetic quantum number, $\hat{1}_i$ is the unit operator of the $i$th atom, $\hat{V}_{\rm dip}$ is the dipole-dipole interaction Hamiltonian, $\epsilon_0$ is the electric constant, $\hat{\bm{d}}_i$ is the dipole operator of the $i$th atom, and $\tilde{\bm{R}}\equiv \bm{R}/R$. We assign the basis states as $\cket{1}=\cket{\uparrow\uparrow}$, $\cket{2}=\cket{\uparrow\downarrow}$, $\cket{3}=\cket{\downarrow\uparrow}$, and $\cket{4}=\cket{\downarrow\downarrow}$. The energies of these pair states are defined by
\begin{align}
\mathcal{E}_1&\equiv 2E_{n_2S_{1/2}},\quad \mathcal{E}_2\equiv \mathcal{E}_3\equiv E_{n_2S_{1/2}}+E_{n_1S_{1/2}},\quad \mathcal{E}_4\equiv 2E_{n_1S_{1/2}}.\label{eq:energy_of_pair_state_supple}
\end{align}
Using the standard second-order perturbation theory, we obtain the effective Hamiltonian:
\begin{align}
\hat{H}_{\rm two}^{(2)}&=\sum_{j=1}^4\left(\mathcal{E}_j\cket{j}\bra{j}\right)+\left[J_{11}\cket{1}\bra{1}+J_{22}\cket{2}\bra{2}+J_{33}\cket{3}\bra{3}+J_{44}\cket{4}\bra{4}+\frac{J}{2}(\cket{2}\bra{3}+\cket{3}\bra{2})\right],\label{eq:2nd_order_effective_Hamiltonian_two_Rydberg_atoms}\\
J_{11}&\equiv -\sum_{\substack{n,l,J,m_J,n',l',J',m'_{J'}}}\frac{\bra{1}\hat{V}_{\rm dip}\cket{n,l,J,m_J,n',l',J',m'_{J'}}\bra{n,l,J,m_J,n',l',J',m'_{J'}}\hat{V}_{\rm dip}\cket{1}}{E_{n,l,J,m_J}+E_{n',l',J',m'_{J'}}-\mathcal{E}_{1}}\equiv \frac{C_6(n_2,n_2)}{R^6},\label{eq:definition_of_J11}\\
J_{22}&= J_{33}\equiv-\hspace{-0.8em}\sum_{\substack{n,l,J,m_J,n',l',J',m'_{J'}}}\hspace{-0.8em}\frac{\bra{2}\hat{V}_{\rm dip}\cket{n,l,J,m_J,n',l',J',m'_{J'}}\bra{n,l,J,m_J,n',l',J',m'_{J'}}\hat{V}_{\rm dip}\cket{2}}{E_{n,l,J,m_J}+E_{n',l',J',m'_{J'}}-\mathcal{E}_2}\equiv \frac{C_6(n_1,n_2)}{R^6},\label{eq:definition_of_J22_33}\\
J_{44}&\equiv -\sum_{\substack{n,l,J,m_J,n',l',J',m'_{J'}}}\frac{\bra{4}\hat{V}_{\rm dip}\cket{n,l,J,m_J,n',l',J',m'_{J'}}\bra{n,l,J,m_J,n',l',J',m'_{J'}}\hat{V}_{\rm dip}\cket{4}}{E_{n,l,J,m_J}+E_{n',l',J',m'_{J'}}-\mathcal{E}_4}\equiv \frac{C_6(n_1,n_1)}{R^6},\label{eq:definition_of_J44}\\
\frac{J}{2}&\equiv -\sum_{\substack{n,l,J,m_J, n',l',J',m'_{J'}}}\frac{\bra{2}\hat{V}_{\rm dip}\cket{n,l,J,m_J,n',l',J',m'_{J'}}\bra{n,l,J,m_J,n',l',J',m'_{J'}}\hat{V}_{\rm dip}\cket{3}}{E_{n,l,J,m_J}+E_{n',l',J',m'_{J'}}-\mathcal{E}_{2}}\equiv \frac{\tilde{C}_6(n_1,n_2)}{R^6}.\label{eq:definition_of_Jex}
\end{align}
Because the state $\cket{j} (j=1,2,3,4)$ is an $S$ state, the only nonvanishing intermediate states are $P$ states due to the selection rule. We can write down the effective Hamiltonian in the spin language. The spin operators can be defined as
\begin{align}
\hat{S}_i^+&\equiv \cket{\uparrow_i}\bra{\downarrow_i},\label{eq:definition_of_spin_operators_for_XXZ1}\\
\hat{S}_i^-&\equiv  \cket{\downarrow_i}\bra{\uparrow_i},\label{eq:definition_of_spin_operators_for_XXZ2}\\ 
\hat{S}_i^z&\equiv \frac{1}{2}(\cket{\uparrow_i}\bra{\uparrow_i}-\cket{\downarrow_i}\bra{\downarrow_i}),\label{eq:definition_of_spin_operators_for_XXZ3}\\
\hat{S}_i^x&\equiv (\hat{S}_i^++\hat{S}_i^-)/2,\quad \hat{S}_i^y\equiv (\hat{S}_i^+-\hat{S}_i^-)/(2i).\label{eq:definition_of_spin_operators_for_XXZ4}
\end{align}
From the definitions of the spin operators and Eq.~(\ref{eq:2nd_order_effective_Hamiltonian_two_Rydberg_atoms}), we obtain the {\it XXZ} type Hamiltonian
\begin{align}
\hat{H}_{\rm two}^{(2)}&=\left[E_{n_2S_{1/2}}-E_{n_1S_{1/2}}+\frac{J_{11}-J_{44}}{2}\right](\hat{S}_1^z+\hat{S}_2^z)+J(\hat{S}_1^x\hat{S}_2^x+\hat{S}_1^y\hat{S}_2^y+\delta\hat{S}_1^z\hat{S}_2^z)+\frac{1}{4}(J_{11}+J_{44}+2J_{22}),\label{eq:effective_Hamiltonian_for_two_Rydberg_atoms}\\
\delta&\equiv \frac{J_{11}+J_{44}-2J_{22}}{J}=\frac{C_6(n_1,n_1)+C_6(n_2,n_2)-2C_6(n_1,n_2)}{2\tilde{C}_6(n_1,n_2)}.\label{eq:definition_of_anisotropy_parameters}
\end{align}

We then extend the above results to the case of $M$ atoms. We assume that $M$ Rydberg atoms are arranged in a one-dimensional open chain. In this case, the effective Hamiltonian is given by
\begin{align}
\hat{H}_{\rm multi}^{(2)}&=\frac{1}{2}\sum_{j,k,j\not=k}J_{jk}(\hat{S}_j^x\hat{S}_k^x+\hat{S}_j^y\hat{S}_k^y+\delta\hat{S}_j^z\hat{S}_k^z)+\sum_{j=1}^Mh_j^z\hat{S}_j^z,\label{eq:multi_atom_case_XXZ_Hamiltonian}
\end{align}
where $R_{jk}\equiv d|j-k|$, $J_{jk}\equiv \tilde{C}_6(n_1,n_2)/R_{jk}^6$, and $h_j^z\equiv (E_{n_2S_{1/2}}-E_{n_1S_{1/2}})+\left\{[C_6(n_2,n_2)-C_6(n_1,n_1)]/2\right\}\sum_{k\not=j}1/R_{jk}^6$. In this work, we neglect the second term of $h_j^z$ because this term is small compared with the first term in typical experimental situations.

\section{Derivation of the rotating transverse field term}\label{app:rotating_transverse_field}

In this appendix, we consider the single Rydberg atom at the position $\bm{R}_j\equiv (R_{jx}, R_{jy}, R_{jz})\equiv dj(\cos\theta,0,\sin\theta)$ (see Fig.~\ref{fig:schematic_figure}). We irradiate two linearly polarized lasers with detuning $\hbar\Delta_1=\hbar\omega_1-(E_{n_1S_{1/2}}-E_P)$ and $\hbar\Delta_2=\hbar\omega_2-(E_{n_2S_{1/2}}-E_P)$, where $E_P$ is the energy of the intermediate state $\cket{6P_{3/2}}$. The atom-light interaction Hamiltonian under the rotating wave approximation can be written as
\begin{align}
\hat{H}_{\rm AL}(t)&=E_{n_1S_{1/2}}|n_1S_{1/2}\rangle\langle n_1S_{1/2}|+E_{n_2S_{1/2}}|n_2S_{1/2}\rangle\langle n_2S_{1/2}|+E_P|6P_{3/2}\rangle\langle 6P_{3/2}|\notag \\
&\quad +\frac{\hbar\Omega_1}{2}\left[e^{+i(k_1R_{jx}-\omega_1t)}|n_1S_{1/2}\rangle\langle 6P_{3/2}|+e^{-i(k_1R_{jx}-\omega_1t)}|6P_{3/2}\rangle\langle n_1S_{1/2}|\right]\notag \\
&\quad +\frac{\hbar\Omega_2}{2}\left[e^{+i(k_2R_{jy}-\omega_2t)}|n_2S_{1/2}\rangle\langle 6P_{3/2}|+e^{-i(k_2R_{jy}-\omega_2t)}|6P_{3/2}\rangle\langle n_2S_{1/2}|\right].\label{eq:atom-light_Hamiltonian}
\end{align}
Here, we omit the values of $m_J$. To transform into the rotating frame of the laser field, we define $\hat{H}_0$ as
\begin{align}
\hat{H}_0&=\left(E_{n_1S_{1/2}}-\hbar\tilde{\omega}\right)|n_1S_{1/2}\rangle\langle n_1S_{1/2}|+(E_{n_1S_{1/2}}-\hbar\omega_1-\hbar\tilde{\omega})|6P_{3/2}\rangle\langle 6P_{3/2}|\notag \\
&\quad +(E_{n_1S_{1/2}}-\hbar\omega_1+\hbar\omega_2-\hbar\tilde{\omega})|n_2S_{1/2}\rangle\langle n_2S_{1/2}|,\label{eq:definition_of_H0}\\
\hbar\tilde{\omega}&\equiv \frac{\hbar}{2}(-\Delta_1+\Delta_2).\label{eq:definition_of_tilde_omega}
\end{align}
The Hamiltonian in the rotating frame of the laser field is given by
\begin{align}
\hat{H}_{\rm AL;rot}&\equiv e^{i\hat{H}_0t/\hbar}[\hat{H}_{\rm AL}(t)-\hat{H}_0]e^{-i\hat{H}_0t/\hbar}\notag \\
&=-\frac{\hbar}{2}(\Delta_1-\Delta_2)|n_1S_{1/2}\rangle\langle n_1S_{1/2}|+\frac{\hbar}{2}(\Delta_1+\Delta_2)|6P_{3/2}\rangle\langle 6P_{3/2}|+\frac{\hbar}{2}(\Delta_1-\Delta_2)|n_2S_{1/2}\rangle\langle n_2S_{1/2}|\notag \\
&\quad +\frac{\hbar\Omega_1}{2}\left(e^{+ik_1R_{jx}}|n_1S_{1/2}\rangle\langle 6P_{3/2}|+e^{-ik_1R_{jx}}|6P_{3/2}\rangle\langle n_1S_{1/2}|\right)\notag \\
&\quad +\frac{\hbar\Omega_2}{2}\left(e^{+ik_2R_{jy}}|n_2S_{1/2}\rangle\langle 6P_{3/2}|+e^{-ik_2R_{jy}}|6P_{3/2}\rangle\langle n_2S_{1/2}|\right).\label{eq:atom-light_Hamiltonian_in_rotating_frame}
\end{align}
From now on, we assume that $|\Delta_1-\Delta_2|\ll |\Delta_1|,|\Delta_2|$ and use the standard adiabatic elimination scheme \cite{Steck,Paulisch2014} or Schrieffer-Wolff transformation \cite{CohenTannoudjibook,Bravyi2011}. We then obtain the effective Hamiltonian for $j$th atom:
\begin{align}
\hat{h}_j&=-\hbar\left[\frac{\Delta_1-\Delta_2}{2}+\frac{\Omega_1^2}{4(\Delta_1+\Delta_2)/2}\right]|n_1S_{1/2}\rangle\langle n_1S_{1/2}|+\hbar\left[\frac{\Delta_1-\Delta_2}{2}-\frac{\Omega_2^2}{4(\Delta_1+\Delta_2)/2}\right]|n_2S_{1/2}\rangle\langle n_2S_{1/2}|\notag \\
&\quad -\hbar\frac{\Omega_1\Omega_2}{4(\Delta_1+\Delta_2)/2}e^{+ik_1R_{j x}}|n_1S_{1/2}\rangle\langle n_2S_{1/2}|-\hbar\frac{\Omega_1\Omega_2}{4(\Delta_1+\Delta_2)/2}e^{-ik_1R_{j x}}|n_2S_{1/2}\rangle\langle n_1S_{1/2}|,\label{eq:effective_Hamiltonian_j-th_atom_Raman}
\end{align}
where we used $R_{j y}=0$. In the spin language, the effective Hamiltonian becomes
\begin{align}
\hat{h}_j&=-\hbar\tilde{\Delta}\hat{S}_j^z-\hbar\Omega_{\rm eff}[\cos(k_1dj\cos\theta)\hat{S}_j^x+\sin(k_1dj\cos\theta)\hat{S}_j^y]-\frac{\hbar(\Omega_1^2+\Omega_2^2)}{8\Delta},\label{eq:effective_Hamiltonian_transverse_rotating_field_term_j-th_atom}
\end{align}
where $\tilde{\Delta}= -[\Delta_1-\Delta_2+(\Omega_1^2-\Omega_2^2)/(4\Delta)]$, $\Delta=(\Delta_1+\Delta_2)/2$, and $\Omega_{\rm eff}= \Omega_1\Omega_2/(2\Delta)$. In the main text, we omit the constant term in Eq.~(\ref{eq:effective_Hamiltonian_transverse_rotating_field_term_j-th_atom}).

\section{Relation between the spin-laboratory and spin-rotating frames}\label{app:relation_lab_and_rot_frame}
In this appendix, we discuss the relation between the spin-laboratory and spin-rotating frames. The DMI results from the unitary transformation
\begin{align}
\hat{U}_{\text{s-rot}}&=\prod_{j=1}^Me^{-i\hat{S}_j^zqj},\quad q=k_1d\cos\theta,\label{eq:definition_of_unitary_transformation_supple}
\end{align}
where $k_1$ is the wave number of laser 1, $d$ is the lattice constant, and $\theta$ is the angle between the chain and laser 1 (see Fig.~\ref{fig:schematic_figure}). Under this transformation, the spin operators are transformed to
\begin{align}
\hat{U}_{\text{s-rot}}^{\dagger}\hat{S}_j^x\hat{U}_{\text{s-rot}}&=\hat{S}_j^x\cos(q j)-\hat{S}_j^y\sin(q j),\label{eq:transformation_Sx_supple}\\
\hat{U}_{\text{s-rot}}^{\dagger}\hat{S}_j^y\hat{U}_{\text{s-rot}}&=\hat{S}_j^x\sin(q j)+\hat{S}_j^y\cos(q j),\label{eq:transformation_Sy_supple}\\
\hat{U}_{\text{s-rot}}^{\dagger}\hat{S}_j^z\hat{U}_{\text{s-rot}}&=\hat{S}_j^z.\label{eq:transformation_Sz_supple}
\end{align}
From these expressions, we obtain the Hamiltonian in the spin-rotating frame: $\hat{H}_{\text{s-rot}}=\hat{U}^{\dagger}_{\text{s-rot}}\hat{H}_{\text{s-lab}}\hat{U}_{\text{s-rot}}$. A state in the spin-laboratory frame $\cket{\psi_{\text{s-lab}}}$ is transformed to $\cket{\psi_{\text{s-rot}}}\equiv \hat{U}^{\dagger}_{\text{s-rot}}\cket{\psi_{\text{s-lab}}}$. 

Then, we consider the expectation values of the spin operators in the spin-laboratory and spin-rotating frames. The expectation value in the laboratory frame is defined by
\begin{align}
 \langle\hat{\bm{S}}_j\rangle_{\text{s-lab}}&\equiv \bra{\psi_{\text{s-lab}}}\hat{\bm{S}}_j\cket{\psi_{\text{s-lab}}}.\label{eq:definition_of_spin_expectation_value_in_lab_frome}
\end{align}
The expectation values in the spin-rotating frame are defined by
\begin{align}
\langle\hat{\bm{S}}_j\rangle_{\text{s-rot}}\equiv \bra{\psi_{\text{s-rot}}}\hat{\bm{S}}_j\cket{\psi_{\text{s-rot}}}.\label{eq:definition_of_spin_expectation_value_in_rot_frame}
\end{align}
Using the relation $\cket{\psi_{\text{s-rot}}}=\hat{U}^{\dagger}_{\text{s-rot}}\cket{\psi_{\text{s-lab}}}$, we obtain $\langle\bm{S}_j\rangle_{\text{s-rot}}=\bra{\psi_{\text{s-lab}}}\hat{U}\hat{\bm{S}}_j\hat{U}^{\dagger}\cket{\psi_{\text{s-lab}}}$. Therefore, the relation of the spin expectation value between the spin-laboratory and spin-rotating frames becomes
\begin{align}
\langle\hat{S}_j^x\rangle_{\text{s-rot}}&=\langle\hat{S}_{j}^x\rangle_{\text{s-lab}}\cos(q j)+\langle\hat{S}_j^y\rangle_{\text{s-lab}}\sin(q j),\label{eq:relation_Sx_lab_and_rot_frame_supple}\\
\langle\hat{S}_j^y\rangle_{\text{s-rot}}&=\langle\hat{S}_{j}^x\rangle_{\text{s-lab}}\sin(q j)-\langle\hat{S}_j^y\rangle_{\text{s-lab}}\cos(q j),\label{eq:relation_Sy_lab_and_rot_frame_supple}\\
\langle\hat{S}_j^z\rangle_{\text{s-rot}}&=\langle\hat{S}_j^z\rangle_{\text{s-lab}}.\label{eq:relation_Sz_lab_and_rot_frame_supple}
\end{align}
In the actual experiments, the physical quantities are measured in the spin-laboratory frame. From the above relations, we can obtain the physical quantities in the spin-rotating frame.

We can measure the spin expectation values as follows. 
In typical Rydberg atom experiments with optical tweezers, we can extract $\langle\hat{S}_j^z\rangle_{\text{s-lab}}$  from a measured recapture probability. To obtain $\langle \hat{S}_j^x\rangle_{\text{s-lab}}$ and $\langle \hat{S}_j^y\rangle_{\text{s-lab}}$, the quantum-state tomographic technique is available. That is, we irradiate an appropriate $\pi/2$ pulse to the system before measuring the recapture probability. In our case, we can use a resonant two-photon microwave field as the $\pi/2$ pulse \cite{Signoles2021}. We can obtain $\langle \hat{S}_j^x\rangle_{\text{s-lab}}$ and $\langle \hat{S}_j^y\rangle_{\text{s-lab}}$ in this way. 

For convenience, we summarize the correspondence of typical quantum states and spin expectation values between the spin-laboratory and spin-rotating frames in table \ref{table:correspondence_lab_frame_and_rot_frame}.

\begin{table*}[t]
\centering
\caption{List of the correspondence between the spin-laboratory and spin-rotating frames. $\cket{\uparrow_j}$ and $\cket{\downarrow_j}$ represent the eigenstates of $\hat{S}_j^z$ with eigenvalue $+\frac{1}{2}$ and $-\frac{1}{2}$, respectively.}
\begin{tabular}{cccc}\hline\hline
\multicolumn{2}{c}{Spin-laboratory frame} & \multicolumn{2}{c}{Spin-rotating frame} \\ \hline
\vspace{0.2em} $\cket{\psi_{\text{s-lab}}}$ & $\langle\hat{\bm{S}}_j\rangle_{\text{s-lab}}\equiv \bra{\psi_{\text{s-lab}}}\hat{\bm{S}}_j\cket{\psi_{\text{s-lab}}}$ & $\cket{\psi_{\text{s-rot}}}\equiv \hat{U}^{\dagger}_{\text{s-rot}}\cket{\psi_{\text{s-lab}}}$ & $\langle\hat{\bm{S}}_j\rangle_{\text{s-rot}} \equiv \bra{\psi_{\text{s-rot}}}\hat{\bm{S}}_j\cket{\psi_{\text{s-rot}}}$ \\ \hline
\vspace{0.2em} $\cket{\uparrow_j}$ & $\dfrac{1}{2}(0,0,+1)$ & $\cket{\uparrow_j}$ & $\dfrac{1}{2}(0,0,+1)$ \\
\vspace{0.2em} $\cket{\downarrow_j}$ & $\dfrac{1}{2}(0,0,-1)$ & $\cket{\downarrow_j}$ & $\dfrac{1}{2}(0,0,-1)$ \\ 
\vspace{0.2em} $\dfrac{1}{\sqrt{2}}(\cket{\uparrow_j}\pm\cket{\downarrow_j})$ & $\dfrac{1}{2}(\pm 1,0,0)$ & \hspace{-0.9em}$\dfrac{1}{\sqrt{2}}(e^{+i q j/2}\cket{\uparrow_j}\pm e^{-iqj/2}\cket{\downarrow_j})$ & $\dfrac{1}{2}(\pm\cos qj,\mp\sin qj,0)$ \\ 
\vspace{0.2em} $\dfrac{1}{\sqrt{2}}(e^{-iqj/2}\cket{\uparrow_j} \pm e^{+iqj/2}\cket{\downarrow_j})$ & $\dfrac{1}{2}(\pm \cos qj,\pm \sin qj,0)$ & $\dfrac{1}{\sqrt{2}}(\cket{\uparrow_j}\pm \cket{\downarrow_j})$ & $\dfrac{1}{2}(\pm 1,0,0)$  \\ \hline\hline
\end{tabular}
\label{table:correspondence_lab_frame_and_rot_frame}
\end{table*}

\section{Extension to two-dimensional systems}\label{app:two-dimension}
In this appendix, we discuss an extension of our setup to a two-dimensional system. We consider the setup shown in Fig.~\ref{fig:schematic_figure_2D}. The $j$th atom is located at the lattice site $\bm{R}_j\equiv d(n_{x,j}\bm{e}_x'+n_{z,j}\bm{e}_z')$, where $n_{x,j}$ and $n_{z,j}$ are integers, $\bm{e}_x'\equiv \cos\theta\bm{e}_x+\sin\theta\bm{e}_z$, $\bm{e}'_z\equiv\cos\theta\bm{e}_z-\sin\theta\bm{e}_x$, and $\bm{e}_{x}$ and $\bm{e}_z$ are the unit vectors in the $x$ and $z$ directions, respectively.

\begin{figure}[t]
\centering
\includegraphics[width=13cm,clip]{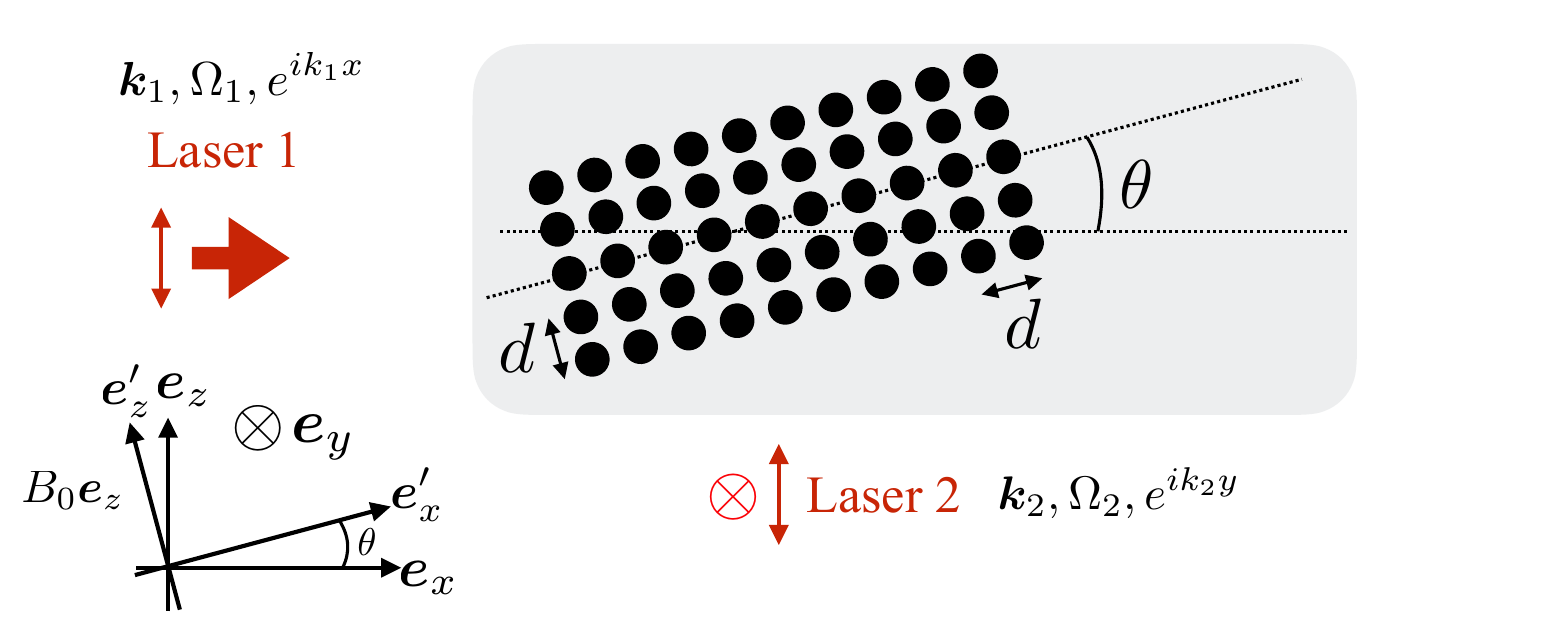}
\caption{Schematics of the setup for the experimental realization of the DMI in a two-dimensional square lattice. The filled black circles represent the position of the Rydberg atoms. The magnetic field is applied along the $z$ axis.}
\label{fig:schematic_figure_2D}
\vspace{-0.75em}
\end{figure}%

In this case, the total Hamiltonian in the spin-laboratory frame is given by
\begin{align}
\hat{H}_{\text{s-lab}}^{2{\rm D}}&=\frac{1}{2}\sum_{\substack{j,k \\ j\ne k}}J_{jk}(\hat{S}_j^x\hat{S}_k^x+\hat{S}_j^y\hat{S}_k^y+\delta\hat{S}_j^z\hat{S}_k^z)-\hbar\Omega_{\rm eff}\sum_j\left[\cos(\bm{k}_1\cdot\bm{R}_j)\hat{S}_j^x+\sin(\bm{k}_1\cdot\bm{R}_j)\hat{S}_j^y\right]-\hbar\tilde{\Delta}\sum_j\hat{S}_j^z,\label{eq:Hamiltonian_laboratory_frame_2D}
\end{align}
where $J_{jk}\equiv \tilde{C}_6(n_1,n_2)/|\bm{R}_j-\bm{R}_k|^6$. Using the unitary operator $\hat{U}_{\text{s-rot}}^{\rm 2D}\equiv \prod_{j}e^{-i\hat{S}_j^z\bm{k}_1\cdot\bm{R}_j}$, we obtain the Hamiltonian in the spin-rotating frame:
\begin{align}
\hat{H}_{\text{s-rot}}^{\rm 2D}&=(\hat{U}^{\rm 2D}_{\text{s-rot}})^{\dagger}\hat{H}_{\text{s-lab}}^{\rm 2D}\hat{U}_{\text{s-rot}}^{\rm 2D}\notag \\
&=\frac{1}{2}\sum_{\substack{j,k \\ j\ne k}}J_{jk}\left\{\cos[\bm{k}_1\cdot(\bm{R}_j-\bm{R}_k)](\hat{S}_j^x\hat{S}_k^x+\hat{S}_j^y\hat{S}_k^y)+\delta\hat{S}_j^z\hat{S}_k^z+\sin[\bm{k}_1\cdot(\bm{R}_j-\bm{R}_k)](\hat{S}_j^x\hat{S}_k^y-\hat{S}_j^y\hat{S}_k^x)\right\}\notag \\
&\quad -\hbar\Omega_{\rm eff}\sum_j\hat{S}_j^x-\hbar\tilde{\Delta}\sum_j\hat{S}_j^z.\label{eq:Hamiltonian_in_rotating_frame_2D}
\end{align}
This result shows that the interaction depends on the directions. For example, when $\theta=0$, the DMI appears only for the $x$ direction.

\section{Analytical expression of the von Neumann entanglement entropy of the quantum many-body scar states}\label{app:Analytical_EE}
The QMBS states of the DH model are essentially the same as those found in Ref.~\cite{Iadecola2020}. In this appendix, we follow their approach and calculate the von Neumann EE of the QMBS states in our model.

We first rewrite Eq.~(\ref{eq:explict_form_of_QMBS_open_supple}) as
\begin{align}
\cket{S_n}=\sum_{\bm{m}_A,\bm{m}_B}M_{\bm{m}_A,\bm{m}_B}^n\cket{\bm{m}_A}\cket{\bm{m}_B},\label{eq:scar_state_splitting_into_subsystems_supple}
\end{align}
where $\bm{m}_{A(B)}$ denotes a spin configuration in the subsystem $A(B)$, $M^n_{\bm{m}_A,\bm{m}_B}$ is a $D_{A}\times D_B$ matrix and $D_{A(B)}$ is the dimension of the subsystem $A$ $(B)$. From the explicit expression of the scar state (\ref{eq:explict_form_of_QMBS_open_supple}), $M^n_{\bm{m}_A,\bm{m}_B}$ becomes zero if spin configurations ($\cket{\bm{m}}=\cket{\bm{m}_A}\cket{\bm{m}_B}$) violate the constraint that up spins are not adjacent to each other. In the following, we consider the half-chain EE, i.e., the sizes of the subsystem $A$ and $B$ are the same. In this case, Eq.~(\ref{eq:scar_state_splitting_into_subsystems_supple}) reduces to
\begin{align}
\cket{S_n}&=\sum_{\bm{m}_A,\bm{m}_B}\sum_{k=0}^{K}M^{n,k}_{\bm{m}_A,\bm{m}_B}\cket{\bm{m}_A}\cket{\bm{m}_B},\label{eq:scar_state_decomposition_number_of_up_states_supple}
\end{align}
where $k$ represents the number of up spins in the subsystem $A$, and $K\equiv\min({n, 1+\lfloor M/4\rfloor}) $. The element of the matrix $M^{n,k}_{\bm{m}_A,\bm{m}_B}$ takes the value $1/\sqrt{\mathcal{N}(M,n)}$ [see Eq.~(\ref{eq:definition_of_normalization_constant_QMBS})] if $\cket{\bm{m}_A}$ contains $k$-up spins and $\cket{\bm{m}_B}$ contains $(n-k)$-up spins and the constraint of the spin configuration is satisfied, otherwise zero.

To calculate the EE, we consider the matrix element of the reduced density matrix $\rho^{n,k}_{\bm{m}_A,\bm{m}_A'}\equiv \sum_{\bm{m}_B}M^{n,k}_{\bm{m}_A,\bm{m}_B}(M^{n,k}_{\bm{m}'_A,\bm{m}_B})^{\dagger}$. This matrix takes the form
\begin{align}
\rho^{n,k}_{\bm{m}_A,\bm{m}_B}&=\frac{1}{\mathcal{N}(M,n)}
\begin{bmatrix}
l_{1,k}\bm{1}_{D_{1,k}\times D_{1,k}} & l_{2,k}\bm{1}_{D_{1,k}\times D_{2,k}}  \\
l_{2,k}\bm{1}_{D_{2,k}\times D_{1,k}} & l_{2,k}\bm{1}_{D_{2,k}\times D_{2,k}}
\end{bmatrix}
,\label{eq:expression_of_matrix_M_M_dagger_supple}\\
D_{1,k}&\equiv \mathcal{N}(M/2-1,k),\label{eq:definition_of_D1k_supple}\\
D_{2,k}&\equiv \mathcal{N}(M/2-2,k-1),\label{eq:definition_of_D2k_supple}\\
l_{1,k}&\equiv \mathcal{N}(M/2,n-k),\label{eq:definition_of_m1k_supple}\\
l_{2,k}&\equiv \mathcal{N}(M/2-1,n-k),\label{eq:definition_of_m2k_supple}
\end{align}
where $\bm{1}_{n\times m}$ is an $n\times m$ matrix whose all entries are equal to 1, $D_{1,k} (D_{2,k})$ is the number of possible configurations of the QMBS state in the subsystem $A$ with $m_{M/2}=\downarrow (\uparrow)$. Similarly, $l_{1,k}(l_{2,k})$ is the number of possible configurations of the QMBS state in the subsystem $B$ with $m_{M/2}=\downarrow (\uparrow)$. The maximum rank of the matrix (\ref{eq:expression_of_matrix_M_M_dagger_supple}) is two, and the matrix (\ref{eq:expression_of_matrix_M_M_dagger_supple}) is real symmetric. Therefore, the number of nonzero eigenvalues is at most two. We can show that an eigenvector of the matrix (\ref{eq:expression_of_matrix_M_M_dagger_supple}) is given by $\bm{x}=[c_1,c_1,\ldots, c_1,c_2,c_2,\ldots, c_2]^{\rm T}$, where the number of $c_1$ and $c_2$ are given by $D_{1,k}$ and $D_{2,k}$, respectively. The nonzero eigenvalue can be obtained by calculating the eigenvalues of the following $2\times 2$ matrix:
\begin{align}
\frac{1}{\mathcal{N}(M,n)}
\begin{bmatrix}
l_{1,k}D_{1,k} & l_{2,k}D_{2,k} \\
l_{2,k}D_{1,k} & l_{2,k}D_{2,k}
\end{bmatrix}
.\label{eq:matrix_independent_one_supple}
\end{align}
The eigenvalues are given by
\begin{align}
\lambda_{k,\pm}&=\frac{1}{2\mathcal{N}(M, n)}\left[l_{1,k}D_{1,k}+l_{2,k}D_{2,k}\pm\sqrt{(l_{1,k}D_{1,k}-l_{2,k}D_{2,k})^2+4D_{1,k}D_{2,k}l_{2,k}^2}\right].\label{eq:eigenvalue_of_M_supple}
\end{align}
The von Neumann EE can be written as
\begin{align}
S_{\rm vN}&=-\sum_{k=0}^{K}\sum_{\eta=\pm}\lambda_{k,\eta}\ln \lambda_{k,\eta}.\label{eq:expression_entanglement_entropy_DH_model_OBC_supple}
\end{align}
We can easily evaluate Eq.~(\ref{eq:expression_entanglement_entropy_DH_model_OBC_supple}) numerically, even for large system sizes. 

\section{Edge state in the Krylov space}\label{app:Krylov_subspace}
In this appendix, we discuss the origin of the slow relaxation dynamics starting from the xN\'{e}el state. To investigate this, we consider the basis in the Krylov subspace, which we define as $\mathcal{K}(\hat{H}, \cket{\psi_0})\equiv {\rm span}(\cket{\psi_0}, \hat{H}\cket{\psi_0},\hat{H}^2\cket{\psi_0},\ldots)$, where $\hat{H}$ is a Hermitian operator and $\cket{\psi_0}$ is a root state. We assume that the root state is normalized to unity: $\bracket{\psi_0}{\psi_0}=1$. Using the standard Lanczos algorithm, we can construct the orthonormal basis $\cket{u_i}$ in the Krylov subspace:
\begin{align}
\hat{H}\cket{u_j}&=\alpha_j\cket{u_j}+\beta_{j-1}\cket{u_{j-1}}+\beta_j\cket{u_{j+1}},\quad (j=1,2,\cdots,N_{\rm K}),\label{eq:recursion_relation_for_Lanczos_supple}\\
\alpha_j&=\bra{u_j}\hat{H}\cket{u_j},\label{eq:recursion_alpha_for_Lanczos_supple}\\
\beta_j&=\sqrt{\left[\bra{u_j}(\hat{H}-\alpha_j)-\beta_{j-1}\bra{u_{j-1}}\right]\left[(\hat{H}-\alpha_j)\cket{u_j}-\beta_{j-1}\cket{u_{j-1}}\right]},\label{eq:recursion_beta_for_Lanczos_supple}
\end{align}
where $\cket{u_0}=0$, $\cket{u_1}=\cket{\psi_0}$, $\beta_0=0$, $\beta_N=0$, and $N_{\rm K}$ is the dimension of the Krylov subspace. Here, $\alpha_j$ and $\beta_j$ are real by definition. The matrix representation of the Hamiltonian in the Krylov basis becomes the tridiagonal form:
\begin{align}
\bra{u_i}\hat{H}\cket{u_j}&=
\begin{bmatrix}
\alpha_1 & \beta_1 &  &  &  \\
\beta_1 & \alpha_2 & \beta_2 &  & \\
 & \beta_2 & \alpha_3 & \beta_3 &   &  \\
 & &\ddots & \hspace{-0.1em}\ddots & \ddots & \\
& & & \beta_{N_{\rm K}-2}&\alpha_{N_{\rm K}-1} & \beta_{N_{\rm K}-1} \\
& & & & \beta_{N_{\rm K}-1}& \alpha_{N_{\rm K}}
\end{bmatrix}
.\label{eq:matrix_representation_Hamiltonian_Krylov_basis_supple}
\end{align}

Here, we set $\hat{H}$ as the DH Hamiltonian [Eq.~(\ref{eq:Hamiltonian_DH_model}) in the main text] and $\cket{\psi_0}=\cket{\text{xN\'{e}el}}$. For the DH model, we can show that $\alpha_i=0$ for $j=1,2,\ldots, N_{\rm K}$. To prove this, the following symmetry is essential:
\begin{align}
\{\hat{\mathcal{I}}\hat{\mathcal{T}}, \hat{H}_{\rm DH}\}=0,\label{eq:symmetry_DH_model}
\end{align}
where $\{\cdot,\cdot\}$ is the anti-commutator, $\hat{\mathcal{I}}$ is the space-inversion operator [see Eq.~(\ref{eq:definition_of_C_operator}) in the main text], and $\hat{\mathcal{T}}\equiv \prod_{j}(-2i\hat{S}_j^y)\hat{K}$ is the time-reversal operator with $\hat{K}$ being the complex-conjugate operator. For even $M$, the following relation holds:
\begin{align}
\hat{\mathcal{I}}\hat{\mathcal{T}}\cket{\text{xN\'{e}el}}&=(-1)^{M/2}\cket{\text{xN\'{e}el}}.\label{eq:parity_xNeel_state}
\end{align}
From Eqs.~(\ref{eq:symmetry_DH_model}) and (\ref{eq:parity_xNeel_state}), we obtain 
\begin{align}
\bra{\text{xN\'{e}el}}\hat{H}^{2k-1}_{\rm DH}\cket{\text{xN\'{e}el}}=0,\quad (k=1,2,\ldots).\label{eq:H_to_2k-1_is_zero_supple}
\end{align}
For $k=1$, this relation is equivalent to $\alpha_1=0$. From $\alpha_1=0$, it follows that $\alpha_2=0$:
\begin{align}
\alpha_2=\bra{u_2}\hat{H}_{\rm DH}\cket{u_2}&=\frac{1}{\beta_1^2}\bra{u_1}\hat{H}^3_{\rm DH}\cket{u_1}=0.\label{eq:proof_alpha2_is_zero_supple}
\end{align} 
We can also show that $\cket{u_2}$ is an eigenstate of $\hat{\mathcal{I}}\hat{\mathcal{T}}$ and has the opposite parity to $\cket{u_1}$:
\begin{align}
\hat{\mathcal{I}}\hat{\mathcal{T}}\cket{u_2}&=-(-1)^{M/2}\cket{u_2}.\label{eq:oposite_parity_u2_supple}
\end{align}
Using these relations, we can show $\alpha_j=0$ by induction. 

\begin{figure}[t]
\centering
\includegraphics[width=14cm,clip]{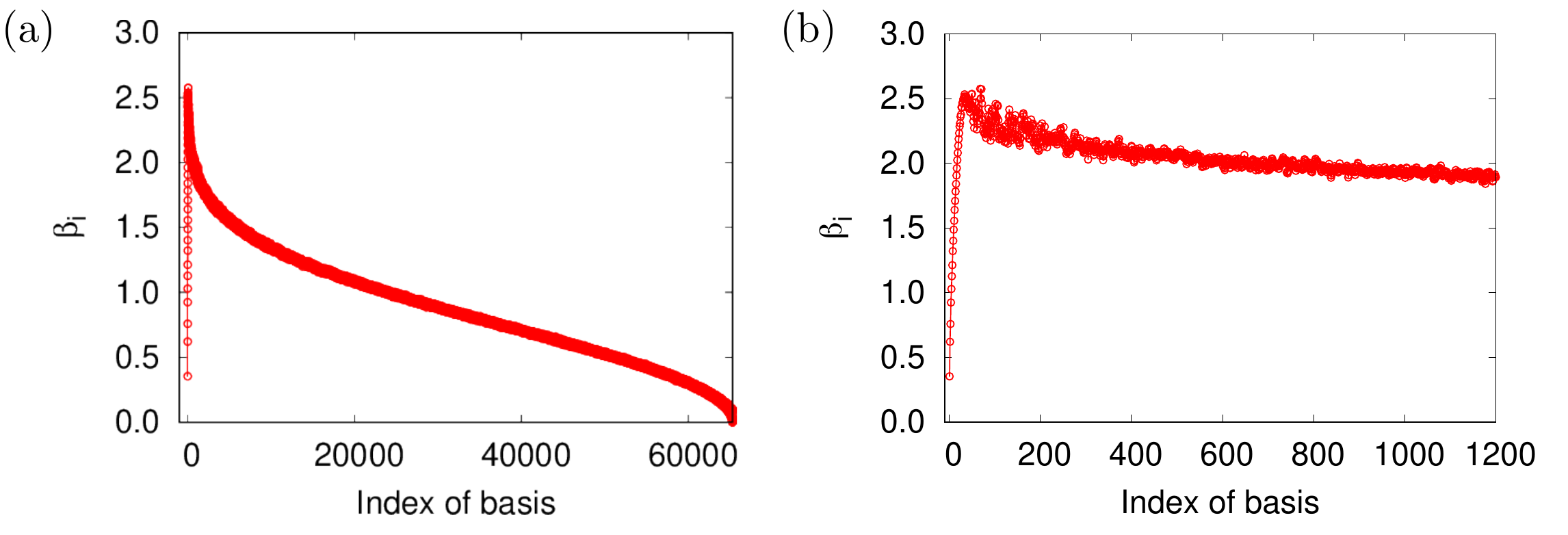}
\caption{(a) Krylov coefficients $\beta_i$ for $M=16$ and $h^x=0.1D$. (b) Magnified view of (a).}
\label{fig:beta_krylov}
\vspace{-0.75em}
\end{figure}%

From the above results, we can write the matrix representation of the DH Hamiltonian as
\begin{align}
H_{\rm DH}&=D
\begin{bmatrix}
0 & \beta_1 &  &  &  \\
\beta_1 & 0 & \beta_2 &  & \\
 & \beta_2 & 0 & \beta_3 &   &  \\
 & &\ddots & \hspace{-0.1em}\ddots & \ddots & \\
& & & \beta_{N_{\rm K}-2}&0 & \beta_{N_{\rm K}-1} \\
& & & & \beta_{N_{\rm K}-1}& 0
\end{bmatrix}
.\label{eq:matrix_representation_DH_Hamiotonian_supple}
\end{align}
We can regard this matrix as a one-dimensional tight-binding Hamiltonian with nonuniform nearest-neighbor hopping amplitude $\beta_j$ (see Fig.~\ref{fig:beta_krylov}). We can easily find that the unitary matrix $\Gamma={\rm diag}(1,-1,1,-1,\ldots)$ satisfies the relation $\{\Gamma, H_{\rm DH}\}=0$. Because all the matrix elements of the matrix $H_{\rm DH}$ are real, we find that this Hamiltonian belongs to the symmetry class BDI in the Altland-Zirnbauer classification \cite{Altland1997}. According to the periodic table of topological insulators and superconductors, edge states are allowed in one-dimensional class BDI systems \cite{Schnyder2008}. 

Here, we numerically diagonalize the Hamiltonian (\ref{eq:matrix_representation_DH_Hamiotonian_supple}). The $j$th eigenvector of the DH Hamiltonian is denoted by $\cket{\Psi^{(j)}}=\sum_ic_i^{(j)}\cket{u_i}$. We plot the overlap between each eigenstate and the xN\'{e}el state (or $\cket{u_1}$) in Fig.~\ref{fig:beta_krylov_edge} (a). We can find an almost zero-energy state with a large overlap with the xN\'{e}el state. We plot the site dependence of the state in Fig.~\ref{fig:beta_krylov_edge} (b). This result clearly shows that the zero-energy eigenstate is localized around $j=1$. We can regard this localized state as a topological edge state. Therefore, we conclude that the slow relaxation dynamics from the xN\'{e}el state can be regarded as a consequence of the topological edge states in the Krylov space. We find similar discussions in the context of the Krylov complexity \cite{Yates2020,Yates2020_2,Rabinovici2021,Rabinovici2022,Trigueros2022, Bhattacharjee2022}.

\begin{figure}[t]
\centering
\includegraphics[width=14cm,clip]{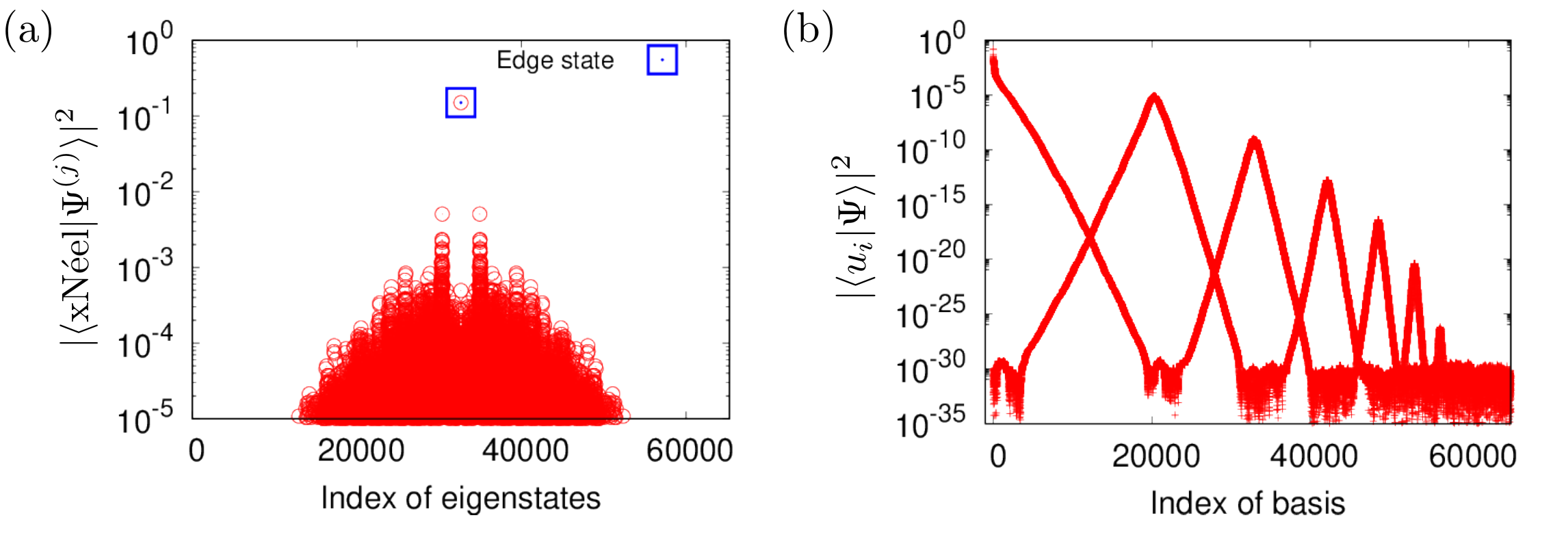}
\caption{(a) Overlap between the xN\'{e}el state and each eigenstate of the Hamiltonian (\ref{eq:matrix_representation_DH_Hamiotonian_supple}) for $M=16$ and $h^x=0.1D$. The blue square represents the state with the largest overlap with the xN\'{e}el state. (b) The site dependence of the wave function of the zero-energy eigenstate is marked by the blue square in (a). }
\label{fig:beta_krylov_edge}
\vspace{-0.75em}
\end{figure}%

\end{widetext}

\begin{thebibliography}{999}
\bibitem{note1}
Here, we introduce the terminology spin-laboratory frame and spin-rotating frame to distinguish the laboratory and rotating frames of the laser field. We note that the Hamiltonian (\ref{eq:effective_Hamiltonian_for_j-th_atom}) is defined in the rotating frame of the laser field and the spin-laboratory frame.
\end{thebibliography}

\newpage
\begin{widetext}
\setcounter{figure}{0}
\setcounter{equation}{0}
\renewcommand{\thefigure}{S\arabic{figure}}
\renewcommand{\theequation}{S\arabic{equation}}

\section*{Supplemental Material for Proposal for simulating quantum spin models with the Dzyaloshinskii-Moriya interaction using Rydberg atoms and the construction of asymptotic quantum many-body scar states}

\section{Pyramid scar states in the DH model}\label{sec:scar_state_DH}

We can show that the DH model with the periodic boundary conditions (PBC) has the pyramid scar states similar to those in Refs.~\cite{sMark2020,sOmiya2023}. The proof is almost the same as those in these references. The pyramid states are defined by
\begin{align}
\cket{S_{n,m}}&\propto (\hat{\mathcal{P}}^{\dagger})^m\cket{S_n},\label{eq:definition_of_Pyramid_state}\\
\hat{\mathcal{P}}^{\dagger}&\equiv \sum_{j=1}^M\sum_{l=1}^{M-2}\hat{P}_{j-1}^{'l}\hat{S}_j^+\hat{P}_{j+1}\notag \\
&=\sum_{j=1}^M\left(\hat{P}'_{j-1}\hat{S}_j^+\hat{P}_{j+1}+\hat{P}'_{j-2}\hat{P}'_{j-1}\hat{S}_j^+\hat{P}_{j+1}+\cdots+\hat{P}'_{j-l}\cdots\hat{P}'_{j-1}\hat{S}_j^+\hat{P}_{j+1}\right),\label{eq:definition_of_Pdagger_for_Pyramid}\\
\hat{P}_{j-1}^{'l}&\equiv \prod_{k=j-l}^{j-1} \hat{P}'_{k}=\hat{P}'_{j-l}\hat{P}_{j-l+1}'\cdots\hat{P}_{j-1}',\label{eq:definition_of_P_prime_l}
\end{align}
where $\cket{S_n}$ is the quantum many-body scar (QMBS) state defined by Eq.~(18) in the main text. The operator $\hat{\mathcal{P}}^{\dagger}$ grows domains to the right. For example, the following transitions occur by the operator $\hat{\mathcal{P}}^{\dagger}$:
\begin{align}
\hat{\mathcal{P}}^{\dagger}\cket{\ldots\downarrow_{j-2}\uparrow_{j-1}\downarrow_j\downarrow_{j+1}\ldots}&=\phantom{2}\cket{\ldots\downarrow_{j-2}\uparrow_{j-1}\uparrow_j\downarrow_{j+1}\ldots},\label{eq:growth_domain_by_Pdagger1}\\
\hat{\mathcal{P}}^{\dagger}\cket{\ldots\downarrow_{j-3}\uparrow_{j-2}\uparrow_{j-1}\downarrow_j\downarrow_{j+1}\ldots}&=2\cket{\ldots\downarrow_{j-3}\uparrow_{j-2}\uparrow_{j-1}\uparrow_j\downarrow_{j+1}\ldots},\label{eq:growth_domain_by_Pdagger2}\\
\hat{\mathcal{P}}^{\dagger}\cket{\ldots\downarrow_{j-4}\uparrow_{j-3}\uparrow_{j-2}\uparrow_{j-1}\downarrow_j\downarrow_{j+1}\ldots}&=3\cket{\ldots\downarrow_{j-4}\uparrow_{j-3}\uparrow_{j-2}\uparrow_{j-1}\uparrow_j\downarrow_{j+1}\ldots}.\label{eq:growth_domain_by_Pdagger3}
\end{align}
The numerical factor $2$ in Eq.~(\ref{eq:growth_domain_by_Pdagger2}) comes from the operator $\hat{P}'_{j-2}\hat{P}'_{j-1}\hat{S}_j^+\hat{P}_{j+1}+\hat{P}'_{j-1}\hat{S}_j^+\hat{P}_{j+1}$. This means that the domain length $l$ gives the numerical factor $l$. Here, we discuss the coefficients of the pyramid state. We introduce the notation $\cket{j_1,j_2,\ldots,j_n}_{(l_1,l_2,\ldots,l_n)}$, where $j_1, j_2,\ldots$ represents the starting positions of the $n$ domains with lengths $l_1,l_2,\ldots,l_n$. Here, we calculate
\begin{align}
(\hat{\mathcal{P}}^{\dagger})^m\cket{j_1,j_2,\ldots,j_n}_{(1,1,\ldots,1)}&=\left(\hat{p}_{j_1}^{\dagger}+\hat{p}_{j_2}^{\dagger}+\ldots+\hat{p}_{j_n}^{\dagger}\right)^m\cket{j_1,j_2,\ldots,j_n}_{(1,1,\ldots,1)},\label{eq:applying_P_to_Sn_state}
\end{align}
where $\hat{p}^{\dagger}_{j_l}$ grows the domain starting at site $j_l$, and the state $\cket{j_1,j_2,\ldots,j_n}_{(1,1,\ldots,1)}$ corresponds to the state containing $\cket{S_n}$. Because the summation over all $l_{j_k}$ is equal to the total number of up spins, we obtain the relation
\begin{align}
\sum_{k=1}^nl_{j_k}=n+m.\label{eq:relation_domain_length_and_number_of_upspins}
\end{align}
Using the multinomial theorem, we obtain
\begin{align}
&\left(\hat{p}_{j_1}^{\dagger}+\hat{p}_{j_2}^{\dagger}+\ldots+\hat{p}_{j_n}^{\dagger}\right)^m\notag \\
&=\sum_{l_1+\ldots+l_n=m+n}\frac{m!}{(l_{j_1}-1)!(l_{j_2}-1)!\ldots(l_{j_n}-1)!}(\hat{p}_{j_1}^{\dagger})^{l_{j_1}-1}\ldots(\hat{p}_{j_n}^{\dagger})^{l_{n_1}-1}.\label{eq:multinomial_theorem_to_hat_P_for_Pyramid}
\end{align}
From the above results, we obtain
\begin{align}
(\hat{\mathcal{P}}^{\dagger})^m\cket{j_1,j_2,\ldots,j_n}_{(1,1,\ldots,1)}&=m!\sum_{l_1+\ldots+l_n=m+n}\cket{j_1,\ldots,j_n}_{(l_1,\ldots,l_n)},\label{eq:proof_of_equal_superposition_Pyramid_state}
\end{align}
where we used the fact that $(\hat{p}_{j_k})^{l_k-1}$ yields the factor $(l_k-1)!$. This result suggests that all states in the pyramids state have equal weight.

Then, we show the relation
\begin{align}
&\hat{H}_{\rm DM}^{\rm PBC}(\hat{\mathcal{P}}^{\dagger})^m\cket{S_{n}}=0,\label{eq:condition_scar_for_pyramid}\\
&\hat{H}_{\rm DM}^{\rm PBC}\equiv D\sum_{j=1}^M\hat{S}_j^x(\hat{S}_{j-1}^z-\hat{S}_{j+1}^z).\label{eq:definition_of_HDM}
\end{align}
To do this, we rewrite the DM interaction term as
\begin{align}
\hat{H}_{\rm DM}^{\rm PBC}&=D\sum_{j=1}^M\hat{S}_j^x(\hat{S}_{j-1}^z-\hat{S}_{j+1}^z)\notag \\
&=\frac{D}{2}\sum_{j=1}^M\left(\cket{\uparrow_{j-1}\uparrow_j\downarrow_{j+1}}\bra{\uparrow_{j-1}\downarrow_j\downarrow_{j+1}}-\cket{\downarrow_{j-1}\uparrow_j\uparrow_{j+1}}\bra{\downarrow_{j-1}\downarrow_j\uparrow_{j+1}}\right)\notag \\
&+\frac{D}{2}\sum_{j=1}^M\left(\cket{\uparrow_{j-1}\downarrow_j\downarrow_{j+1}}\bra{\uparrow_{j-1}\uparrow_j\downarrow_{j+1}}-\cket{\downarrow_{j-1}\downarrow_j\uparrow_{j+1}}\bra{\downarrow_{j-1}\uparrow_j\uparrow_{j+1}}\right)\notag \\
&\equiv \frac{D}{2}\sum_{j=1}^{M}\hat{h}_j^++\frac{D}{2}\sum_{j=1}^M\hat{h}_j^-\notag \\
&\equiv \hat{H}_{\rm DM}^++\hat{H}_{\rm DM}^-,\label{eq:rewrite_DM_term_for_Pyramid_state}
\end{align}
where $\hat{H}_{\rm DM}^{(\pm)}$ increases (decreases) the length of the up domains. Here, we calculate
\begin{align}
\hat{h}^+_{j+l}|\ldots \downarrow_{j-1}\underbrace{\uparrow_j\uparrow_{j+1}\ldots \uparrow_{j+l-1}}_{l-1}\downarrow_{j+l}\downarrow_{j+l+1}\ldots\rangle&=\phantom{-}|\ldots\downarrow_{j-1}\underbrace{\uparrow_j\uparrow_{j+1}\ldots \uparrow_{j+l}}_{l}\downarrow_{j+l+1}\ldots\rangle,\label{eq:domain_l-1_to_l_right}\\
\hat{h}^+_{j}|\ldots\downarrow_{j-1}\downarrow_j\underbrace{\uparrow_{j+1}\uparrow_{j+2}\ldots\uparrow_{j+l}}_{l-1}\downarrow_{j+l+1} \ldots\rangle&=-|\ldots\downarrow_{j-1}\underbrace{\uparrow_j\uparrow_{j+1}\ldots\uparrow_{j+l}}_{l}\downarrow_{j+l+1}\ldots\rangle,\label{eq:domain_l-1_to_l_left}
\end{align}
where we assume $l>1$. We also calculate
\begin{align}
\hat{h}^-_{j+l+1}|\ldots\downarrow_{j-1}\underbrace{\uparrow_j\uparrow_{j+1}\ldots\uparrow_{j+l+1}}_{l+1}\downarrow_{j+l+2} \ldots\rangle&=\phantom{-}|\ldots\downarrow_{j-1}\underbrace{\uparrow_j\ldots\uparrow_{j+l}}_{l}\downarrow_{j+l+1}\downarrow_{j+l+2} \ldots\rangle,\label{eq:domain_l+1_to_l_right}\\
\hat{h}^-_{j-1}|\ldots\downarrow_{j-2}\underbrace{\uparrow_{j-1}\uparrow_j \ldots\uparrow_{j+l}}_{l+1}\downarrow_{j+l+1}\downarrow_{j+l+2}\ldots\rangle&=-|\ldots\downarrow_{j-1}\underbrace{\uparrow_j\ldots\uparrow_{j+l}}_{l}\downarrow_{j+l+1}\downarrow_{j+l+2}\ldots\rangle.\label{eq:domain_l+1_to_l_left}
\end{align}
From these results, we can conclude $\hat{H}_{\rm DM}^{\rm PBC}\cket{S_{n,m}}=0$. Because $\cket{S_{n,m}}$ contains $n+m$-up spins, the pyramid state is an eigenstate of the magnetic-field term $\hat{H}_z\equiv -h^x\sum_{j=1}^M\hat{S}_j^z$. Therefore, we obtain
\begin{align}
\hat{H}_{{\rm DM}}^{\rm PBC}\cket{S_{n,m}}&=-h^x\left(\frac{n+m}{2}-\frac{M-n-m}{2}\right)\cket{S_{n,m}}=-h^x(n+m-M/2)\cket{S_{n,m}}.\label{eq:Pyramid_state_for_Zeeman_term}
\end{align}

\section{Asymptotic quantum many-body scar states in the DH model under the periodic boundary conditions}\label{sec:asymptotic_scar_PBC}
Here, we derive the asymptotic quantum many-body scar (AQMBS) states under the PBC. We define the operator and state
\begin{align}
\hat{Q}^{\dagger}(k)&\equiv \sum_{j=1}^Me^{i k j}\hat{P}_{j-1}\hat{S}_j^+\hat{P}_{j+1},\label{eq:definitino_of_Qdagger_k}\\
\cket{AS_n(k)}&\propto \hat{Q}^{\dagger}(k)\cket{S_{n-1}},\label{eq:definition_of_AQMBS_PBC_case}
\end{align}
where $k\equiv (2\pi/M)Z$ is the crystal momentum, with $Z$ being an integer. Here, we discuss the range of $n$ and $k$. As in the case of the QMBS states, $\cket{AS_n(k)}=0$ for $n>M/2$. When $k=0$, $\cket{AS_n(k=0)}$ reduces to the scar state $\cket{S_n}$ [See Eq.~(18) in the main text.]. For $0<k<\pi$ and $n=M/2$, we can show $\cket{AS_{M/2}(k)}=0$. When $k=\pi$ and $n=M/2$, $\cket{AS_{M/2}(k=\pi)}=(\cket{\uparrow\downarrow\ldots\uparrow\downarrow}-\cket{\downarrow\uparrow\ldots\downarrow\uparrow})/\sqrt{2}$. However, We can easily check that the state $\cket{AS_{M/2}(k=\pi)}$ is an exact eigenstate of the DH Hamiltonian. Thus we will only consider the cases where $0<k\le \pi$ for $n=1,2,\ldots,M/2-1$ and $0<k<\pi$ for $n=M/2$. In these cases, $\bracket{AS_n(k)}{S_m}=0$ holds because the QMBS state and the state (\ref{eq:definition_of_AQMBS_PBC_case}) have different crystal momenta. 

We calculate the energy variance of the state $\cket{AS_n(k)}$, which is given by
\begin{align}
\Delta E^2&\equiv \bra{AS_n(k)}(\hat{H}_{\rm DH}^{\rm PBC})^2\cket{AS_n(k)}-(\bra{AS_n(k)}\hat{H}_{\rm DH}^{\rm PBC}\cket{AS_n(k)})^2\notag \\
&=\frac{\bra{S_{n-1}}[\hat{Q}(k), \hat{H}_{\rm DM}^{\rm PBC}][\hat{H}_{\rm DM}^{\rm PBC}, \hat{Q}^{\dagger}(k)]\cket{S_{n-1}}}{\bra{S_{n-1}}\hat{Q}(k)\hat{Q}^{\dagger}(k)\cket{S_{n-1}}}\equiv \frac{f_{\rm n}(k,n)}{f_{\rm d}(k,n)},\label{eq:energy_variance_expression_PBC}
\end{align}
where $\hat{H}_{\rm DM}^{\rm PBC}$ is defined by Eq.~(\ref{eq:definition_of_HDM}) and we used the fact that $\hat{H}_{\rm DM}^{\rm PBC}\cket{S_{n-1}}=0$ and $\cket{S_{n-1}}$ is an eigenstate of the Zeeman term.

First, we consider the numerator $f_{\rm n}(k,n)$ in Eq.~(\ref{eq:energy_variance_expression_PBC}). To evaluate this, we use the commutation relation
\begin{align}
[\hat{H}^{\rm PBC}_{\rm DM}, \hat{Q}^{\dagger}(k)]&=\frac{D}{2}\sum_je^{i k j}\left[(1-e^{ik})\hat{P}_{j-1}\hat{S}_j^+\hat{S}_{j+1}^+\hat{P}_{j+2}-e^{+ik}\hat{P}_{j-1}'\hat{S}_j^-\hat{S}_{j+1}^+\hat{P}_{j+2}+e^{-ik}\hat{P}_{j-2}\hat{S}_{j-1}^+\hat{S}_j^-\hat{P}_{j+1}'\right].\label{eq:commutation_relation_DM_and_Qdagger_PBC}
\end{align}
Because the QMBS state does not contain the configuration $\cket{\ldots\uparrow\uparrow\ldots}$, the second and third terms in Eq.~(\ref{eq:commutation_relation_DM_and_Qdagger_PBC}) vanish when acting on the QMBS state. From these properties, we obtain
\begin{align}
f_{\rm n}(k,n)&=\frac{D^2}{2}[1-\cos(k)]M\bra{S_{n-1}}\hat{P}_1\hat{P}_2\hat{P}_3\hat{P}_4\cket{S_{n-1}},\label{eq:numerator_PBC}
\end{align}
where we used the translational symmetry. The term $\bra{S_{n-1}}\hat{P}_1\hat{P}_2\hat{P}_3\hat{P}_4\cket{S_{n-1}}$ is called the emptiness formation probability (EFP) \cite{sKorepin1994,sEbler1995}. In this case,  we can analytically calculate the EFP by virtue of the properties of the QMBS. Since the QMBS state has equal coefficients, and the up spins are not adjacent to each other, the EFP is proportional to the total number of configurations by choosing $n-1$ up spins from $M-4$ sites under the constraint. This corresponds to the total number of configurations of the form $|\downarrow\downarrow\downarrow\downarrow\underbrace{\ldots}_{M-4}\rangle$. Therefore, we obtain
\begin{align}
f_{\rm n}(k,n)&=\frac{D^2}{2}[1-\cos(k)]M\dfrac{\displaystyle{\binom{M-4-(n-1)+1}{n-1}}}{\displaystyle{N(M,n-1)}}=\frac{D^2}{2}[1-\cos(k)]\frac{(M-2n)(M-2n+1)(M-2n+2)}{(M-n-1)(M-n)},\label{eq:final_results_numerator_energy_variance_PBC}
\end{align}
where $N(M,n)$ is defined by Eq.~(19) in the main text.

Next, we consider the denominator $f_{\rm d}(k,n)$ in Eq.~(\ref{eq:energy_variance_expression_PBC}). We can rewrite $f_{\rm d}(k,n)$ as
\begin{align}
f_{\rm d}(k,n)&=M\sum_{r=0}^{M-1}e^{ikr}\bra{S_{n-1}}\hat{q}_1\hat{q}^{\dagger}_{1+r}\cket{S_{n-1}}\equiv M\sum_{r=0}^{M-1}e^{ikr}C(r,n-1),\label{eq:denominator_rewrite} \\
\hat{q}_j^{\dagger}&\equiv \hat{P}_{j-1}\hat{S}_j^+\hat{P}_{j+1},\label{eq:definition_of_small_q_PBC}\\
C(r,n)&\equiv \bra{S_{n}}\hat{q}_1\hat{q}^{\dagger}_{1+r}\cket{S_{n}},\quad (0\le r\le M-1).\label{eq:defintion_of_correlation_function}
\end{align}
where we used the translational symmetry of the QMBS state and defined the correlation function $C(r,n-1)$. 

For $r=0$, the correlation function reduces to the EFP:
\begin{align}
C(r=0,n)&=\bra{S_{n}}\hat{P}_1\hat{P}_2\hat{P}_3\cket{S_{n}}=\frac{1}{N(M,n)}\binom{M-3-n+1}{n}=\frac{(M-2n)(M-2n-1)}{M(M-n-1)}.\label{eq:correlation_r=0_term}
\end{align}

For $r=1$ and $M-1$, we find $C(r=1,n)=C(r=M-1,n)=0$ because $\hat{q}_1\hat{q}_2^{\dagger}=0$ and $\hat{q}_1\hat{q}^{\dagger}_M=0$. 

For $r=2$ and $M-2$, we obtain
\begin{align}
C(r=2,n)&=C(r=M-2,n)=\bra{S_{n}}\hat{S}_1^-\hat{P}_2\hat{S}_3^+\hat{P}_4\hat{P}_M\cket{S_n}=\frac{1}{N(M,n)}\binom{M-5-(n-1)+1}{n-1}.\label{eq:correaltion_function_form_conbinatorial_r2}
\end{align}
This corresponds to the total number of configurations $|\uparrow\downarrow\downarrow\downarrow\underbrace{\ldots}_{M-5}\downarrow\rangle$, where $n-1$ up spins are in $\ldots$ region under the constraint. 

In the same manner, for $r=3$ and $M-3$, the correlation function becomes
\begin{align}
C(r=3,n)&=C(r=M-3,n)=\bra{S_n}\hat{S}_1^-\hat{P}_2\hat{P}_3\hat{S}_4^+\hat{P}_5\hat{P}_M\cket{S_n}=\frac{1}{N(M,n)}\binom{M-6-(n-1)+1}{n-1}.\label{eq:correaltion_function_form_conbinatorial_r3}
\end{align}
This corresponds to the total number of configurations $|\uparrow\downarrow\downarrow\downarrow\downarrow\underbrace{\ldots}_{M-6}\downarrow\rangle$.

For $4\le r\le M-4$, we obtain
\begin{align}
C(r,n)&=
\begin{cases}
\vspace{0.3em}\displaystyle{\frac{1}{N(M,n)}\sum_{l=0}^{\lfloor(r-2)/2\rfloor}\binom{r-3-l+1}{l}\binom{M-3-r-(n-1-l)+1}{n-1-l}},\quad \text{for }4\le r\le n,\\
\vspace{0.3em}\displaystyle{\frac{1}{N(M,n)}\sum_{l=0}^{n-1}\binom{r-3-l+1}{l}\binom{M-3-r-(n-1-l)+1}{n-1-l}},\quad \text{for } n+1\le r\le M-(n+1),\\
\displaystyle{\frac{1}{N(M,n)}\sum_{l=n-1-\lfloor(M-r-2)/2\rfloor}^{n-1}\binom{r-3-l+1}{l}\binom{M-3-r-(n-1-l)+1}{n-1-l}},\;\text{for }M-n\le r\le M-4,
\end{cases}
\label{eq:correlation_function_r_dep}
\end{align}
where $\lfloor \cdot\rfloor$ is the floor function. This corresponds to the total number of configurations $|\uparrow\downarrow\underbrace{\ldots}_{r-3}\downarrow\downarrow\downarrow\underbrace{\ldots}_{M-r-3}\downarrow\rangle$. To derive Eq.~(\ref{eq:correlation_function_r_dep}), we used the fact that the maximum number of up spins in $Q$ consecutive sites under the constraint is given by $\lfloor(Q+1)/2\rfloor$. 

Although the above results are exact, the expressions are complicated and not useful for understanding the properties of the energy variance. Here, we derive the asymptotic form of the correlation function $C(r,n)$. In the following, we assume $n=O(1)$ and $Z=O(1)$ for technical reasons. The leading term of the numerator can be easily obtained:
\begin{align}
f_{\rm n}(k,n)=\left[\frac{\pi^2Z^2}{M^2}+O\left(\frac{1}{M^4}\right)\right]\left[M+O(1)\right].\label{eq:order_estimate_numerator_PBC}
\end{align}
The leading terms of the denominator except for $4\le r\le M-4$ are given by
\begin{align}
C(r=0,n)&=\bra{S_{n}}\hat{P}_1\hat{P}_2\hat{P}_3\cket{S_n}=1+O\left(\frac{1}{M}\right),\label{eq:leading_term_r=0_PBC}\\
C(r=2,n)&=C(r=3,n)=C(r=M-3,n)=C(r=M-2,n)=\frac{n}{M}+O\left(\frac{1}{M^2}\right).\label{eq:leading_term_r=2,3_PBC}
\end{align}
Because we assumed $n=O(1)$, the leading terms for $4\le r\le n$ or $M-n\le M-4$ can be obtained by setting $l=0$ and $l=n-1$ in the summation, respectively. These are given by
\begin{align}
C(r,n)&=\frac{n}{M}+O\left(\frac{1}{M^2}\right),\quad \text{for }4\le r\le n\text{ or }M-n\le r\le M-4,\label{eq:leading_term_4<=r<=n_PBC}
\end{align}
To evaluate the leading term for $n+1\le r\le M-(n+1)$, we need a nontrivial identity called Gould's identity \cite{sGould1956}:
\begin{align}
\sum_{k=0}^n\binom{\alpha+\beta k}{k}\binom{\gamma+\beta(n-k)}{n-k}&=\sum_{k=0}^n\binom{\alpha+\epsilon+\beta k}{k}\binom{\gamma-\epsilon+\beta(n-k)}{n-k},\label{eq:Gould_idensity}
\end{align}
where $\alpha,\beta,\gamma,\epsilon$ are complex numbers, and $n$ is a nonnegative integer. If we set $\beta=-1$, we can apply Gould's identity to the second expression of Eq.~(\ref{eq:correlation_function_r_dep}). The correlation function reduces to
\begin{align}
C(n+1\le r\le M-n-1 ,n)&=\frac{1}{N(M,n)}\sum_{l=0}^{n-1}\binom{n-1-l}{l}\binom{M-2n+l-3}{n-1-l}=\frac{n}{M}+O\left(\frac{1}{M^2}\right),\label{eq:correlation_function_intermediate_case}
\end{align}
where we used $n=O(1)$ and set $l=0$ to obtain the leading term. From the above results, the correlation function becomes
\begin{align}
C(r,n)&=
\begin{cases}
\displaystyle{1+O\left(\frac{1}{M}\right),\quad r=0,}\\
\vspace{0.25em}0,\quad r=1,\;M-1,\\
\displaystyle{\frac{n}{M}+O\left(\frac{1}{M^2}\right)},\quad \text{otherwise}.
\end{cases}
\label{eq:leading_term_correlation_function_PBC}
\end{align}
The denominator can be evaluated as
\begin{align}
f_{\rm d}(k,n)&=M+O(1)+M\sum_{r=2}^{M-2}e^{ikr}\left[\frac{n-1}{M}+O\left(\frac{1}{M^2}\right)\right]\notag \\
&=M+O(1)+M\sum_{r=1}^{M}e^{ikr}\left[\frac{n-1}{M}+O\left(\frac{1}{M^2}\right)\right]-[e^{ik1}+e^{ik(M-1)}]\left[n-1+O\left(\frac{1}{M}\right)\right]\notag \\
&=M+O(1),\label{eq:leading_term_of_denominator_PBC}
\end{align}
where we used $\sum_{r=1}^Me^{ikr}=0$. Therefore, the energy variance can be written as
\begin{align}
\Delta E^2&=\frac{\pi^2Z^2D^2}{M^2}+O\left(\frac{1}{M^3}\right),\quad \text{for }n=O(1)\text{ and }Z=O(1).\label{eq:energy_variance_leading_term_for_PBC}
\end{align}
This result means that the energy variance vanishes in the thermodynamics limit when $n=O(1)$ and $Z=O(1)$. Therefore, the state (\ref{eq:definition_of_AQMBS_PBC_case}) satisfies the condition of the AQMBS. 

Here, we show the numerical results of the energy variance for several $n$ and $Z$. Figure \ref{fig:energy_variance_aqmps_pbc}(a) shows the $n$-dependence of the energy variance for several site numbers and $Z=1$. From this result, the energy variance is an increasing function of $n$ for fixed $Z$. Figure \ref{fig:energy_variance_aqmps_pbc}(b) shows the $n$-dependence of the energy variance for several $Z$ for $M=2000$. From these results, the energy variance is maximum at $n=M/2-1$. Figure \ref{fig:energy_variance_aqmps_pbc}(c) shows the energy variance at $n=M/2-1$. We can see that the energy variance goes to zero for large $M$ even in the case of large $n$. Our numerical results suggest that the state (\ref{eq:definition_of_AQMBS_PBC_case}) satisfies the conditions of AQMBS for large $n$ and $Z=O(1)$ cases. In the case of $Z=O(M)$, the state (\ref{eq:definition_of_AQMBS_PBC_case}) does not satisfy the condition of the AQMBS state. In fact, we can obtain the analytical expressions of the energy variance for $n=1$ and $2$:
\begin{align}
\Delta E^2(n=1)&=\frac{D^2}{2}(1-\cos k),\label{eq:energy_variance_n=1_PBC}\\
\Delta E^2(n=2)&=\frac{D^2}{2}\frac{(M-4)[1-\cos(k)]}{M-4-2\cos(k)}.\label{eq:energy_vriance_n=2_PBC}
\end{align}
We can see $\Delta E^2\not=0$ for $k=\pi\;(Z=M/2)$ in the thermodynamic limit. These results mean that $Z=O(1)$ is a necessary condition for the state (\ref{eq:definition_of_AQMBS_PBC_case}) to be an AQMBS.

\begin{figure}[t]
\centering
\includegraphics[width=17cm,clip]{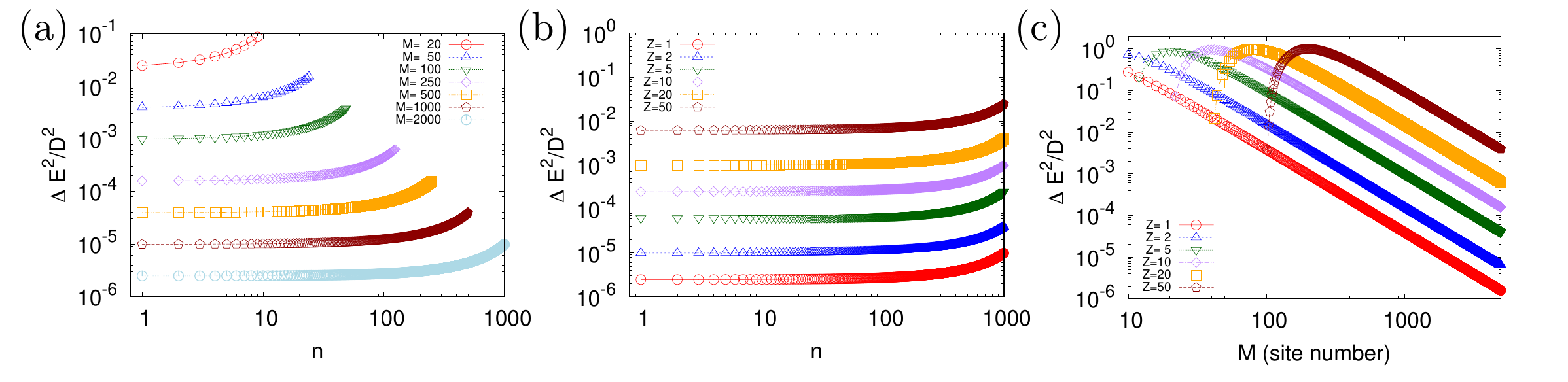}
\caption{(a) $n$ dependence of the energy variance for $Z=1$. (b) $n$ dependence of the energy variance for several $Z$ and $M=2000$. (c) System size dependence of the energy variance for several $Z$ and $n=M/2-1$. }
\label{fig:energy_variance_aqmps_pbc}
\vspace{-0.75em}
\end{figure}%

\section{Asymptotic quantum many-body scar states in the DH model under the open boundary conditions}\label{sec:asymptotic_scar_OBC}
Here, we derive the AQMBS state under open boundary conditions (OBC) with the edge magnetic field. In the OBC case, we will show  that the following states satisfy the conditions of the AQMBS states:
\begin{align}
\cket{AS_n}&\propto\hat{A}^{\dagger}\cket{S_{n-1}},\quad n=1,2,\ldots,M/2,\label{eq:AQMBS_state_OBC}\\
\hat{A}^{\dagger}&\equiv \sum_{j=1}^Mf_j\hat{P}_{j-1}\hat{S}_j^+\hat{P}_{j+1},\label{eq:definition_of_operator_A_for_OBC}
\end{align}
where $\cket{S_{n-1}}$ is the QMBS state with the OBCs [see Eq.~(35) in the main text.], and $f_j$ is a complex function satisfying $f_j=-f_{M+1-j}$, which is sufficient for ensuring the orthonormality $\bracket{AS_n}{S_m}=0$. The QMBS state $\cket{S_m}$ is an eigenstate of the space inversion operator $\hat{\mathcal{I}}$ with eigenvalue $+1$. From the condition $f_j=-f_{M+1-j}$, the operator $\hat{A}^{\dagger}$ has odd parity under spatial inversion. Therefore, we obtain the orthonormality between the QMBS states and $\cket{AS_n}$. Here, we choose $f_j=\cos[\pi j/(M+1)]$. This function is an eigenfunction of the tight-binding model with Dirichlet boundary conditions \cite{sMartin-Delgato1995}. However, this choice is not unique. We will see later that $f_j=\cos[\pi(2j-1)/(2M)]$, an eigenfunction of the tight-binding model with Neumann boundary conditions, also gives qualitatively similar results.

Let us show that the half-chain EE of the state (\ref{eq:AQMBS_state_OBC}) obeys a sub-volume law scaling. The strategy is to write down the matrix product state (MPS) representation of the state (\ref{eq:AQMBS_state_OBC}) and to investigate the scaling of its bond dimension. Here, we construct a matrix product operator (MPO) representation for the operator $\hat{Q}^{\dagger}$ using the technique of the finite state machine \cite{sCrosswhite2008,sMotruk2016,sPaeckel2017}. The results are given by
\begin{align}
\hat{Q}^{\dagger}&\equiv \sum_{\bm{n},\bm{n}'}\hat{Q}^{\dagger}_0(\hat{Q}^{\dagger}_1)^{n_1n_1'}\cdots (\hat{Q}^{\dagger}_M)^{n_Mn_M'}\hat{Q}^{\dagger}_{M+1}\cket{\bm{n}}\bra{\bm{n}'},\label{eq:definition_of_MPO_Qdagger}\\
\hat{Q}^{\dagger}_i&=
\begin{bmatrix}
\hat{1}_i & \hat{P}_i & 0 & 0 \\
0 & 0 & \hat{S}_i^+ & 0 \\
0 & 0 & 0 & \hat{P}_i \\
0 & 0 & 0 & \hat{1}_i
\end{bmatrix}
,\text{ for }1\le i\le M,\quad \hat{Q}^{\dagger}_0=
\begin{bmatrix}
1 & 1 & 0 & 0
\end{bmatrix}
,\quad \quad \hat{Q}^{\dagger}_{M+1}=
\begin{bmatrix}
0 & 0 & 1 & 1
\end{bmatrix}^{\rm T}
,\label{eq:MPO_for_Qdagger_OBC}
\end{align}
where $\hat{Q}_i^{\dagger}$ is an operator-valued matrix and $(\hat{Q}_i^{\dagger})^{n_in_i'}\equiv \bra{n_i}\hat{Q}_i^{\dagger}\cket{n_i'}$. The MPOs of $(\hat{Q}^{\dagger})^2$ and $(\hat{Q}^{\dagger})^3$ are given by
\begin{align}
&\includegraphics[width=7cm,clip]{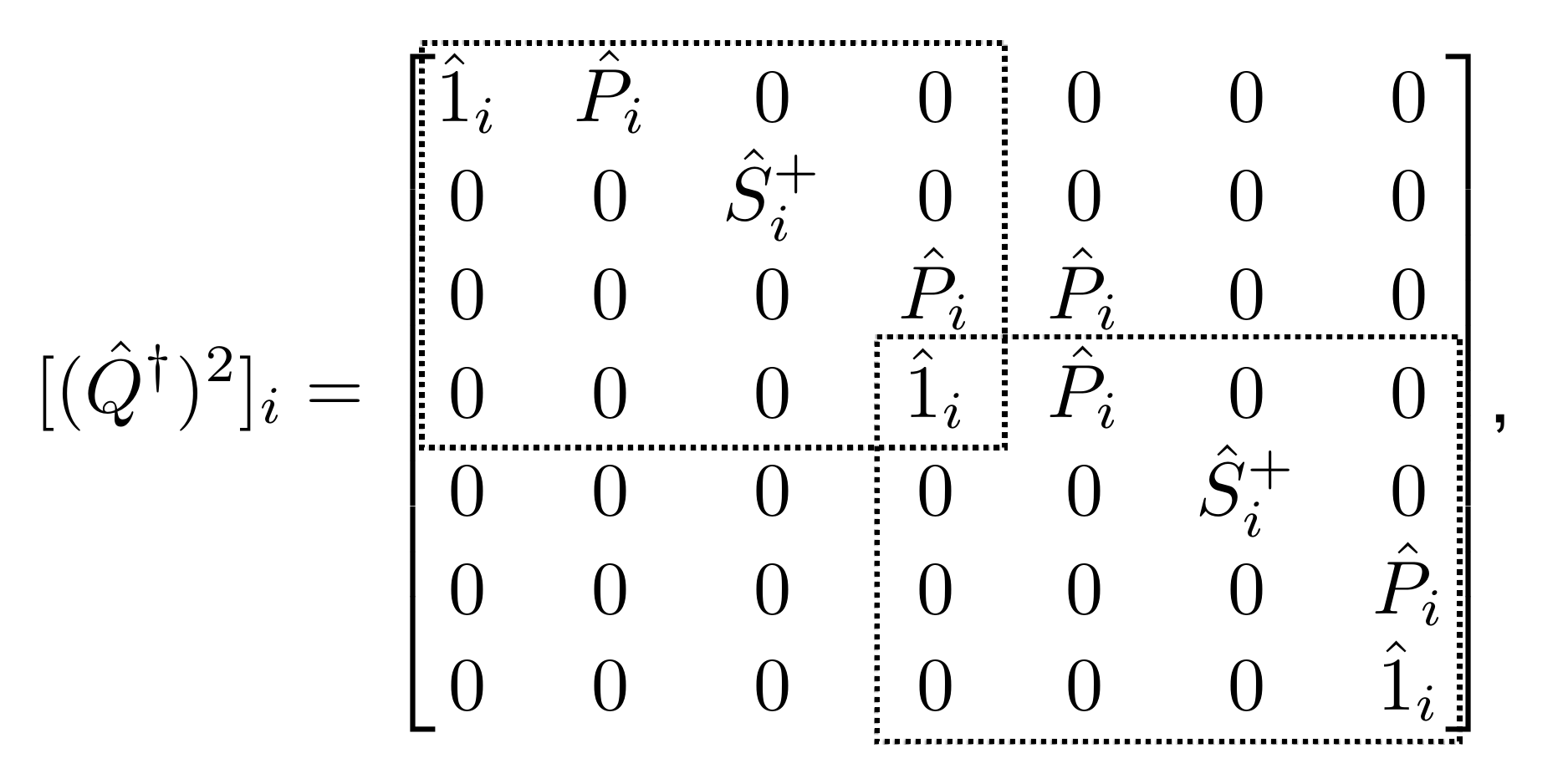}\label{eq:MPO_Qdagger_squared_OBC}\\
\quad [(\hat{Q}^{\dagger})^2]_0&=
\begin{bmatrix}
1 & 1 & 0 & 0 & 0 & 0 & 0
\end{bmatrix}
,\quad [(\hat{Q}^{\dagger})^2]_{M+1}=
\begin{bmatrix}
0 & 0 & 0 & 0 & 0 & 1 & 1
\end{bmatrix}^{\rm T}
,\label{eq:MPO_Qdagger_squared_OBC_edge}
\end{align}
\begin{align}
&\includegraphics[width=9.5cm,clip]{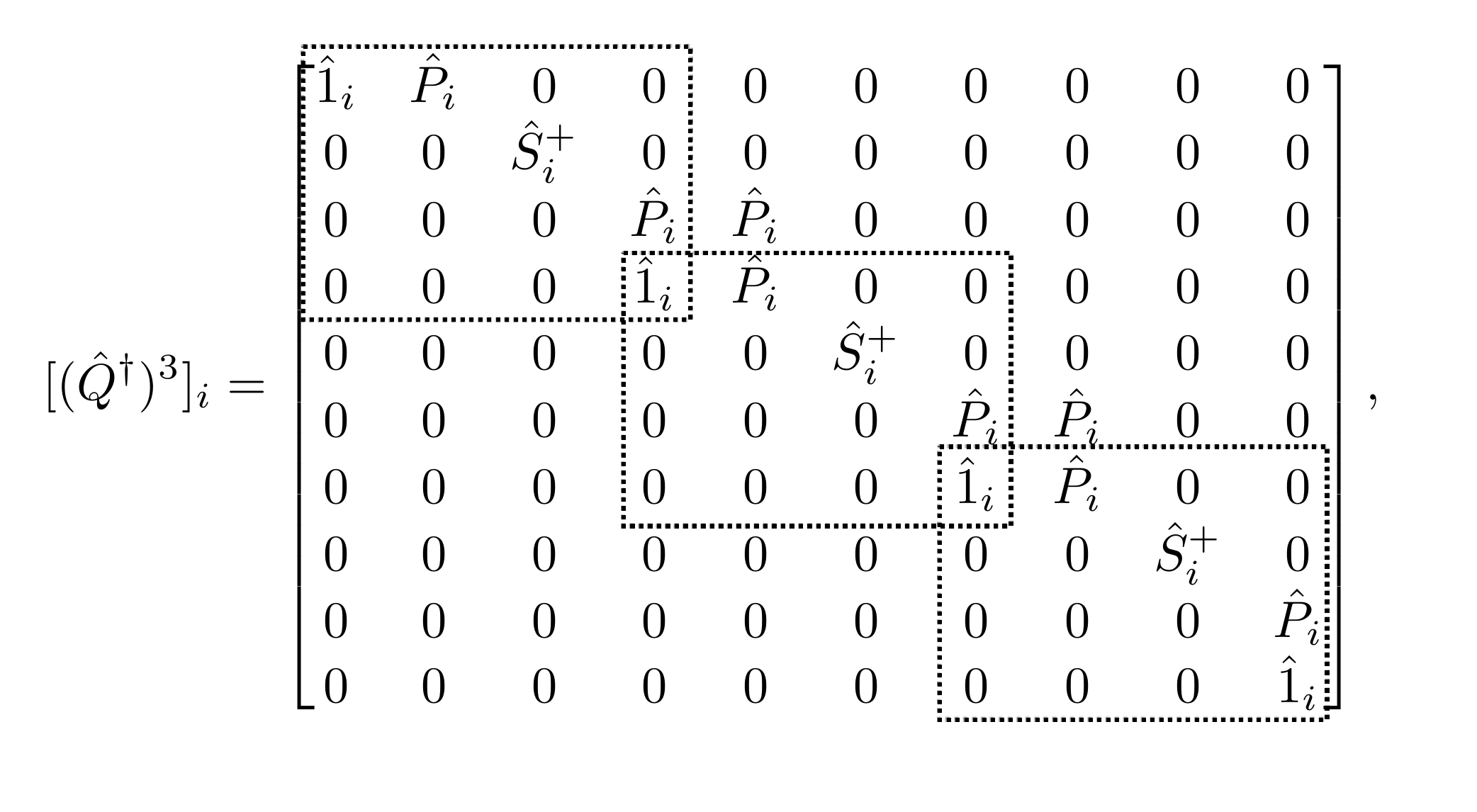}\label{eq:MPS_Qdagger_cubic_OBC}\\
[(\hat{Q}^{\dagger})^3]_0&=
\begin{bmatrix}
1 & 1 & 0 & 0 & 0 & 0 & 0 & 0 & 0 & 0
\end{bmatrix}
,\quad [(\hat{Q}^{\dagger})^3]_{M+1}=
\begin{bmatrix}
0 & 0 & 0 & 0 & 0 & 0 & 0 & 0 & 1 & 1
\end{bmatrix}^{\rm T}
.\label{eq:MPS_Qdagger_cubic_OBC_edge}
\end{align}
From these results, we can see the block structure of the MPOs, which are shown in the dotted black lines. We can obtain the MPO of $[\hat{Q}^{\dagger}]^n$ with the bond dimension $\chi_n=3n+1$:
\begin{align}
\{[(\hat{Q}^{\dagger})^n]_{i}\}_{a,b}&=
\begin{cases}
\vspace{0.5em}\hat{1}_i,\quad \text{for }(a,b)=(3c+1,3c+1),\quad c=0,1,2,\ldots,n,\\
\vspace{0.5em}\hat{P}_i,\quad \text{for }
\begin{cases}
(a,b)=(3c-2,3c-1),\quad c=1,2,\ldots,n,\\
(a,b)=(3c,3c+1),\quad c=1,2,\ldots,n,\\
(a,b)=(3c,3c+2),\quad c=1,2,\ldots,n-1,
\end{cases}
\\
\vspace{0.5em}\hat{S}_i^+,\quad \text{for }(a,b)=(3c-1,3c),\quad c=1,2,\ldots, n,\\
0,\quad \text{otherwise},
\end{cases}
\label{eq:matrix_element_of_Qdagger_to_n}\\
\{[(\hat{Q}^{\dagger})^n]_0\}_{1,b}&=
\begin{cases}
\vspace{0.5em}1,\quad \text{for }b=1,2,\\
\vspace{0.5em}0,\quad \text{otherwise},
\end{cases}
\quad \{[(\hat{Q}^{\dagger})^n]_{M+1}\}_{a,1}=
\begin{cases}
\vspace{0.5em}1,\quad \text{for }a=\chi_n-1,\chi_n,\\
\vspace{0.5em}0,\quad \text{otherwise },
\end{cases}
\label{eq:matrix_element_of_Qdagger_to_n_edge}
\end{align}
where $a,b=1,2,\ldots,\chi_n$. Applying this MPO to $\cket{\Downarrow}$, we obtain the MPS representation of the QMBS $\cket{S_n}$ with the bond dimension $\chi_n$. We note that a similar method is used in Sec.~IV~B of Ref.~\cite{sMoudgalya2018}. Next, we consider the MPS representation of the state (\ref{eq:AQMBS_state_OBC}). The MPO representation of $\hat{A}^{\dagger}$ is given by
\begin{align}
\hat{A}^{\dagger}_i&=
\begin{bmatrix}
\hat{1}_i & \hat{P}_i & 0 & 0 \\
0 & 0 & f_i\hat{S}_i^+ & 0 \\
0 & 0 & 0 & \hat{P}_i \\
0 & 0 & 0 & \hat{1}_i 
\end{bmatrix}
,\quad \hat{A}^{\dagger}_0=
\begin{bmatrix}
1 & 1 & 0 & 0
\end{bmatrix}
,\quad \quad \hat{A}^{\dagger}_{M+1}=
\begin{bmatrix}
0 & 0 & 1 & 1
\end{bmatrix}^{\rm T}
.\label{eq:MPO_representation_Adagger}
\end{align}
For $n=2$, the state (\ref{eq:AQMBS_state_OBC}) reduces to $\cket{AS_2}\propto \hat{A}^{\dagger}\hat{Q}^{\dagger}\cket{\Downarrow}$. The operator $\hat{A}^{\dagger}\hat{Q}^{\dagger}$ can be written as
\begin{align}
\hat{A}^{\dagger}\hat{Q}^{\dagger}&\propto \sum_{j=1}^{M-2}\sum_{k=j+2}^M(\hat{P}_{j-1}f_j\hat{S}_j^+\hat{P}_{j+1}\hat{P}_{k-1}\hat{S}_k^+\hat{P}_{k+1}+\hat{P}_{j-1}\hat{S}_j^+\hat{P}_{j+1}\hat{P}_{k-1}f_k\hat{S}_k^+\hat{P}_{k+1})\equiv \sum_{l=1}^2\hat{F}^l,\label{eq:product_Adagger_and_Qdagger_OBC}\\
\hat{F}^1&\equiv \sum_{j=1}^{M-2}\sum_{k=j+2}^M\hat{P}_{j-1}(f_j\hat{S}_j^+)\hat{P}_{j+1}\hat{P}_{k-1}\hat{S}_k^+\hat{P}_{k+1},\label{eq:definition_of_F1}\\
\hat{F}^2&\equiv \sum_{j=1}^{M-2}\sum_{k=j+2}^M\hat{P}_{j-1}\hat{S}_j^+\hat{P}_{j+1}\hat{P}_{k-1}(f_k\hat{S}_k^+)\hat{P}_{k+1}.\label{eq:definition_of_F2}
\end{align}
From this expression, we can obtain the MPO representation of $\hat{F}^l$ by replacing $S_i^+$ with $(f_i\hat{S}_i^+)$ in the $l$th block of $[(\hat{Q}^{\dagger})^2]_i$, where the $l$th block of $[(\hat{Q}^{\dagger})^2]_i$ is defined by the $l$th $4\times 4$ matrix enclosed by the dotted black line in (\ref{eq:MPO_Qdagger_squared_OBC}). Because the MPO representation of the sum of the two operators is given by the direct sum of the MPOs of each operator \cite{sSchollwock2011}, the MPO representation of $\hat{A}^{\dagger}\hat{Q}^{\dagger}$ becomes
\begin{align}
(\hat{A}^{\dagger}\hat{Q}^{\dagger})_i&=
\left[
\begin{array}{c|c}
(\hat{F}^1)_i & O \\ \hline
O & (\hat{F}^2)_i
\end{array}
\right].\label{eq:MPO_representation_of_AdaggerQdagger}
\end{align}
The bond dimension of $\hat{A}^{\dagger}\hat{Q}^{\dagger}$ is $2(3n+1)$ because the bond dimension of $\hat{F}^l$ is the same as that of $\hat{Q}^{\dagger}$. Generalizing this procedure for $n>2$, we can obtain the MPO representation of the operator $\hat{A}^{\dagger}(\hat{Q}^{\dagger})^{n-1}$, ant its bond dimension is given by $n(3n+1)$. Since the half-chain EE is bounded by the bond dimension, the EE of the state (\ref{eq:AQMBS_state_OBC}) is bounded by $S_{\rm vN}\le \ln[n(3n+1)]\le \ln[O(M^2)]$. Here, we used $n=1,2,\ldots,M/2$. Therefore, the half-chain EE of the state (\ref{eq:AQMBS_state_OBC}) obeys a sub-volume law scaling, which is a condition of the AQMBS. 

Then, we calculate the energy variance of the state (\ref{eq:AQMBS_state_OBC}). In the same manner as the PBC, the expression of the energy variance becomes
\begin{align}
\Delta E^2&=\frac{\bra{S_{n-1}}[\hat{A}, \hat{H}_{\rm DM}^{\rm OBC}][\hat{H}_{\rm DM}^{\rm OBC},\hat{A}^{\dagger}]\cket{S_{n-1}}}{\bra{S_{n-1}}\hat{A}\hat{A}^{\dagger}\cket{S_{n-1}}}\equiv\frac{f_{\rm n}(n)}{f_{\rm d}(n)},\label{eq:energy_variance_OBC}
\end{align}
where $\hat{H}_{\rm DM}^{\rm OBC}\equiv D\sum_{j=1}^M\hat{S}_j^x(\hat{S}_{j-1}^z-\hat{S}_{j+1}^z)$. Using the commutation relation,
\begin{align}
[\hat{H}_{\rm DM}^{\rm OBC}, \hat{A}^{\dagger}]&=\frac{D}{2}\sum_{j=1}^{M-1}(f_j-f_{j+1})\hat{P}_{j-1}\hat{S}_j^+\hat{S}_{j+1}^+\hat{P}_{j+2}+\frac{D}{2}\sum_{j=1}^{M-2}f_j\hat{P}_{j-1}\hat{S}_j^+\hat{S}_{j+1}^-\hat{P}_{j+2}'-\frac{D}{2}\sum_{j=2}^{M-2}f_{j+1}\hat{P}_{j-1}'\hat{S}_j^-\hat{S}_{j+1}^+\hat{P}_{j+2},\label{eq:commutator_HDM_and_A_OBC}
\end{align}
and the properties of the QMBS states, we obtain the numerator $f_{\rm n}(n)$:
\begin{align}
f_{\rm n}(n)&=\frac{D^2}{2}\left[1-\cos\left(\frac{\pi}{M+1}\right)\right]\sum_{j=1}^{M-1}g_j^2E_{n-1}(j),\label{eq:energy_variance_numerator_OBC}\\
g_j&\equiv \sin\left[\frac{\pi(2j+1)}{2(M+1)}\right],\label{eq:definition_of_gj_for_OBC}\\
E_{n-1}(j)&\equiv \bra{S_{n-1}}\hat{P}_{j-1}\hat{P}_{j}\hat{P}_{j+1}\hat{P}_{j+2}\cket{S_{n-1}},\label{eq:definition_of_En-1_j_OBC}
\end{align}
where $E_{n-1}(j)$ is the EFP. We can obtain the explicit expression of Eq.~(\ref{eq:definition_of_En-1_j_OBC}) in a manner similar to the PBC case:
\begin{align}
E_{n-1}(1)&=E_{n-1}(M)=\frac{1}{\mathcal{N}(M,n-1)}\binom{M-3-(n-1)+1}{n-1},\label{eq:EFP_OBC_edge}\\
E_{n-1}(j)&=
\begin{cases}
\vspace{0.5em}\displaystyle{\frac{1}{\mathcal{N}(M,n-1)}\sum_{s=0}^{\lfloor(j-1)/2\rfloor}\binom{j-2-s+1}{s}\binom{M-j-2-(n-1-s)+1}{n-1-s}},\quad 2\le j\le n-1,\\
\vspace{0.5em}\displaystyle{\frac{1}{\mathcal{N}(M,n-1)}\sum_{s=0}^{n-1}\binom{j-2-s+1}{s}\binom{M-j-2-(n-1-s)+1}{n-1-s}},\quad n\le j\le M-n,\\
\vspace{0.5em}\displaystyle{\frac{1}{\mathcal{N}(M,n-1)}\hspace{-0.3em}\sum_{s=n-\lfloor(M-j+1)/2\rfloor}^{n-1}\hspace{-0.6em}\binom{j-2-s+1}{s}\hspace{-0.3em}\binom{M-j-2-(n-1-s)+1}{n-1-s}}, M-n+1\le j\le M-1.\\
\end{cases}
\label{eq:EFP_OBC_not_edge}
\end{align}
Here, $\mathcal{N}(M,n)$ is defined by Eq.~(36) in the main text, and Eqs~(\ref{eq:EFP_OBC_edge}) and (\ref{eq:EFP_OBC_not_edge}) correspond to the total number of configurations $|\downarrow\downarrow\downarrow\underbrace{\ldots}_{M-3}\rangle$ and $|\underbrace{\ldots}_{j-2}\downarrow\downarrow\downarrow\downarrow\underbrace{\ldots}_{M-j-2}\rangle$, respectively. 

The denominator of the energy variance $f_{\rm d}(n)$ can be rewritten as
\begin{align}
f_{\rm d}(n)&=\sum_{j=1}^M\sum_{l=1}^Mf_jf_l\bra{S_{n-1}}\hat{q}_j\hat{q}_l^{\dagger}\cket{S_{n-1}}\equiv \sum_{j=1}^M\sum_{l=1}^Mf_jf_lC(j,l,n-1),\label{eq:denominator_energy_variance_OBC}\\
\hat{q}_j^{\dagger}&\equiv \hat{P}_{j-1}\hat{S}_j^+\hat{P}_{j+1},\label{eq:definition_of_small_q_for_OBC}\\
C(j,l,n)&\equiv \bra{S_n}\hat{q}_j\hat{q}_l^{\dagger}\cket{S_n}.\label{eq:definition_of_correlation_function}
\end{align}
In the case of OBC, the proof is more involved than in the PBC case due to the lack of translational symmetry. Since the QMBS state is an eigenstate of the space-inversion operator $\hat{\mathcal{I}}$, we can show $C(j,l,n)=C(l,j,n)$. Thus, its suffices to consider the case $C(j,j+r,n)$\; $(0\le r\le M-j)$.

For $r=0$, the operator becomes $\hat{q}_j\hat{q}_j^{\dagger}=\hat{P}_{j-1}\hat{P}_j\hat{P}_{j+1}$. We obtain
\begin{align}
C(1,1,n)&=C(M,M,n)=\bra{S_n}\hat{P}_1\hat{P}_2\cket{S_n}=\frac{1}{\mathcal{N}(M,n)}\binom{M-2-n+1}{n},\label{eq:correlation_r=0_j=1_M_OBC}\\
C(j,j,n)&=\bra{S_n}\hat{P}_{j-1}\hat{P}_j\hat{P}_{j+1}\cket{S_n}\notag \\
&=
\begin{cases}
\vspace{0.5em}\displaystyle{\frac{1}{\mathcal{N}(M,n)}\sum_{s=0}^{\lfloor(j-1)/2\rfloor}\binom{j-2-s+1}{s}\binom{M-j-1-(n-s)+1}{n-s} },\quad 2\le j\le n,\\
\vspace{0.5em}\displaystyle{\frac{1}{\mathcal{N}(M,n)}\sum_{s=0}^{n}\binom{j-2-s+1}{s}\binom{M-j-1-(n-s)+1}{n-s} },\quad n+1\le j\le M-(n+1),\\
\vspace{0.5em}\displaystyle{\frac{1}{\mathcal{N}(M,n)}\sum_{s=n-\lfloor(M-j)/2\rfloor}^{n}\binom{j-2-s+1}{s}\binom{M-j-1-(n-s)+1}{n-s} },\;M-n\le j\le M-1.\\
\end{cases}
\label{eq:correlation_r=0_2<=j_M-1_OBC}
\end{align}
Equations (\ref{eq:correlation_r=0_j=1_M_OBC}) and (\ref{eq:correlation_r=0_2<=j_M-1_OBC}) correspond to the total number of configurations $|\downarrow\downarrow\underbrace{\ldots}_{M-2}\rangle$ and $|\underbrace{\ldots}_{j-2}\downarrow\downarrow\downarrow\underbrace{\ldots}_{M-j-2}\rangle$, respectively.

For $r=1$, the correlation function reduces to $C(j,j+1,n)=0$ because $\hat{q}_j\hat{q}_{j+1}^{\dagger}=0$. 

For $r=2$, we obtain
\begin{align}
C(1,3,n)&=C(M-2,M,n)=\bra{S_n}\hat{S}_1^-\hat{P}_2\hat{S}_3^+\hat{P}_4\cket{S_n}=\frac{1}{\mathcal{N}(M,n)}\binom{M-4-(n-1)+1}{n-1},\label{eq:correlation_r=2_OBC}\\
C(j,j+2,n)&=\bra{S_n}\hat{P}_{j-1}\hat{S}_j^-\hat{P}_{j+1}\hat{S}_{j+2}^+\hat{P}_{j+3}\cket{S_n}\notag \\
&=
\begin{cases}
\vspace{0.5em}\displaystyle{\frac{1}{\mathcal{N}(M,n)}\sum_{s=0}^{\lfloor(j-1)/2\rfloor}\binom{j-2-s+1}{s}\binom{M-j-3-(n-1-s)+1}{n-1-s} },\quad 2\le j\le n-1,\\
\vspace{0.5em}\displaystyle{\frac{1}{\mathcal{N}(M,n)}\sum_{s=0}^{n-1}\binom{j-2-s+1}{s}\binom{M-j-3-(n-1-s)+1}{n-1-s} },\quad n\le j\le M-(n+1),\\
\vspace{0.5em}\displaystyle{\frac{1}{\mathcal{N}(M,n)}\sum_{s=n-\lfloor(M-j)/2\rfloor}^{n-1}\binom{j-2-s+1}{s}\binom{M-j-3-(n-1-s)+1}{n-1-s} },\;M-n\le j\le M-3.\\
\end{cases}
\label{eq:correlation_r=2_2<=j_OBC}
\end{align}
Equations (\ref{eq:correlation_r=2_OBC}) and (\ref{eq:correlation_r=2_2<=j_OBC}) correspond to the total number of configurations $|\uparrow\downarrow\downarrow\downarrow\underbrace{\ldots}_{M-4}\rangle$ and $|\underbrace{\dots}_{j-2}\downarrow\uparrow\downarrow\downarrow\downarrow\underbrace{\ldots}_{M-j-3}\rangle$, respectively.

For $r=3$, we obtain
\begin{align}
C(1,4,n)&=C(M-3,M,n)=\bra{S_n}\hat{S}_1^-\hat{P}_2\hat{P}_3\hat{S}_4^+\hat{P}_5\cket{S_n}=\frac{1}{\mathcal{N}(M,n)}\binom{M-5-(n-1)+1}{n-1},\label{eq:correlation_r=3_j=1_M_OBC}
\end{align}
\begin{align}
&C(j,j+3,n)\notag \\
&=\bra{S_n}\hat{P}_{j-1}\hat{S}_j^-\hat{P}_{j+1}\hat{P}_{j+2}\hat{S}_{j+3}^+\hat{P}_{j+4}\cket{S_n}\notag \\
&=
\begin{cases}
\vspace{0.5em}\displaystyle{\frac{1}{\mathcal{N}(M,n)}\sum_{s=0}^{\lfloor(j-1)/2\rfloor}\binom{j-2-s+1}{s}\binom{M-j-4-(n-1-s)+1}{n-1-s} },\quad 2\le j\le n-1,\\
\vspace{0.5em}\displaystyle{\frac{1}{\mathcal{N}(M,n)}\sum_{s=0}^{n-1}\binom{j-2-s+1}{s}\binom{M-j-4-(n-1-s)+1}{n-1-s} },\quad n\le j\le M-(n+2),\\
\vspace{0.5em}\displaystyle{\frac{1}{\mathcal{N}(M,n)}\sum_{s=n-\lfloor(M-j-1)/2\rfloor}^{n-1}\binom{j-2-s+1}{s}\binom{M-j-4-(n-1-s)+1}{n-1-s} },\;M-(n+1)\le j\le M-4.\\
\end{cases}
\label{eq:correlation_r=3_2<=j_OBC}
\end{align}
Equations (\ref{eq:correlation_r=3_j=1_M_OBC}) and (\ref{eq:correlation_r=3_2<=j_OBC}) correspond to the total number of configurations $|\uparrow\downarrow\downarrow\downarrow\downarrow\underbrace{\ldots}_{M-5}\rangle$ and $|\underbrace{\ldots}_{j-2}\downarrow\uparrow\downarrow\downarrow\downarrow\downarrow\underbrace{\ldots}_{M-j-4}\rangle$, respectively.

For $j=1$ and $r=M-1$, we obtain
\begin{align}
C(1,M,n)&=\bra{S_n}\hat{S}_1^-\hat{P}_2\hat{P}_{M-1}\hat{S}_M^+\cket{S_n}=\frac{1}{\mathcal{N}(M,n)}\binom{M-4-(n-1)+1}{n-1}.\label{eq:correlation_4<=r_j=1_OBC}
\end{align}
Here, it corresponds to the total number of configurations $|\uparrow\downarrow\underbrace{\ldots}_{M-4}\downarrow\downarrow\rangle$. For $j=1$ and $4\le r\le M-2$, we obtain
\begin{align}
&C(1,1+r,n)\notag \\
&=C(M-r,M,n)=\bra{S_n}\hat{S}_1^-\hat{P}_2\hat{P}_r\hat{S}_{r+1}^+\hat{P}_{r+2}\cket{S_n}\notag \\
&=
\begin{cases}
\vspace{0.5em}\displaystyle{\frac{1}{\mathcal{N}(M,n)}\sum_{s=0}^{\lfloor(r-2)/2\rfloor}\binom{r-3-s+1}{s}\binom{M-r-2-(n-1-s)+1}{n-1-s} },\quad 4\le r\le n,\\
\vspace{0.5em}\displaystyle{\frac{1}{\mathcal{N}(M,n)}\sum_{s=0}^{n-1}\binom{r-3-s+1}{s}\binom{M-r-2-(n-1-s)+1}{n-1-s} },\quad n+1\le r\le M-n,\\
\displaystyle{\frac{1}{\mathcal{N}(M,n)}\sum_{s=n-\lfloor(M+1-r)/2\rfloor}^{n-1}\binom{r-3-s+1}{s}\binom{M-r-2-(n-1-s)+1}{n-1-s} },\;M-n+1\le j\le M-2.\\
\end{cases}
\label{eq:correlation_j=1_4<=r_OBC}
\end{align}
Here, Eq.~(\ref{eq:correlation_j=1_4<=r_OBC}) corresponds to the total number of configurations $|\uparrow\downarrow\underbrace{\ldots}_{r-3}\downarrow\downarrow\downarrow\underbrace{\ldots}_{M-r-2}\rangle$. For $4\le r\le M-r$ and $2\le j\le M-r-1$, we obtain
\begin{align}
&C(j,j+r,n)\notag \\
&=\bra{S_n}\hat{P}_{j-1}\hat{S}_j^-\hat{P}_{j+1}\hat{P}_{j+r-1}\hat{S}_{j+r}^+\hat{P}_{j+r+1}\cket{S_n}\notag \\
&=\frac{1}{\mathcal{N}(M,n)}\sum_{p=0}^{p_{\rm max}}\sum_{q=q_{\rm min}(p)}^{q_{\rm max}(p)}\binom{j-2-p+1}{p}\binom{r-3-q+1}{q}\binom{M-j-r-1-(n-1-p-q)+1}{n-1-p-q},\label{eq:correlation_function_j_j+r_OBC}
\end{align}
where $p_{\rm max}$, $q_{\rm max}(p)$, and $q_{\rm min}(p)$ are defined by
\begin{align}
p_{\rm max}&=
\begin{cases}
\lfloor j/2\rfloor,\quad j\le n-1,\\
n-1,\quad n\le j,
\end{cases}
\label{eq:definition_of_pmax}\\
q_{\rm max}(p)&=
\begin{cases}
\lfloor (r-1)/2\rfloor,\quad r\le n-p,\\
n-1-p,\quad n-p+1\le r,
\end{cases}
\label{eq:definition_of_qmax}\\
q_{\rm min}(p)&=
\begin{cases}
0,\quad j+r\le M-(n-1-p),\\
n-1-p-\lfloor(M-j-r)/2\rfloor,\quad M-(n-1-p)+1\le j+r.
\end{cases}
\label{eq:definition_of_qmin}
\end{align}
Equation (\ref{eq:correlation_function_j_j+r_OBC}) corresponds to the total number of configurations $|\underbrace{\ldots}_{j-2}\downarrow\uparrow\downarrow\underbrace{\ldots}_{r-3}\downarrow\downarrow\downarrow\underbrace{\ldots\ldots}_{M-j-r-1}\rangle$.

From the above calculations, we obtain the expressions of the correlation function $C(j,l,n)$ for all patterns. However, as in the case of PBC, the expressions are complicated. Here, we also evaluate the leading order contribution to the energy variance under the assumption $n=O(1)$. 

Here, we consider the numerator of the energy variance. From Gould's identity (\ref{eq:Gould_idensity}), the leading order of the EFP is given by
\begin{align}
E_{n-1}(j)&=1+O\left(\frac{1}{M}\right).\label{eq:leading_order_of_EFP_OBC}
\end{align}
Using the relation
\begin{align}
\sum_{j=1}^{M-1}g_j^2&=\frac{M}{2}+O(1),\label{eq:sumation_of_gj_squared_expansion}
\end{align}
we can evaluate the numerator as
\begin{align}
f_{\rm n}(n)&=\left[\frac{\pi^2D^2}{4M^2}+O\left(\frac{1}{M^4}\right)\right]\left[\frac{M}{2}+O(1)\right].\label{eq:leading_order_of_numerator}
\end{align}

Then, we consider the denominator. Similarly to the PBC case, we can obtain the leading term of the correlation function using Gould's identity and the assumption $n=O(1)$ except Eq.~(\ref{eq:correlation_function_j_j+r_OBC}). To obtain the leading term of Eq.~(\ref{eq:correlation_function_j_j+r_OBC}), we need an  extension of Gould's identity:
\begin{align}
\sum_{i=0}^n\sum_{j=0}^{n-i}\binom{p+\beta i}{i}\binom{q+\beta j}{j}\binom{s+\beta(n-i-j)}{n-i-j}=\sum_{i=0}^n\sum_{j=0}^{n-i}\binom{p+y_1+\beta i}{i}\binom{q+y_2+\beta j}{j}\binom{s+y_3+\beta(n-i-j)}{n-i-j},\label{eq:extended_version_of_Gould's_identity}
\end{align}
where $p, q, s, \beta, y_1, y_2, y_3$ are complex numbers, $n$ is a nonnegative integer, and $y_1+y_2+y_3=0$. The proof is as follows: According to Ref.~\cite{sGould1956}, the following identity holds:
\begin{align}
f(\alpha,\beta,x)&\equiv \frac{x^{\alpha+1}}{(1-\beta)x+\beta}=\sum_{k=0}^{\infty}\binom{\alpha+\beta k}{k}z^k,\quad z\equiv \frac{x-1}{x^{\beta}}.\label{eq:key_identity_from_Gould}
\end{align}
For $y_1+y_2+y_3=0$, we can easily find 
\begin{align}
f(p,\beta,x)f(q,\beta, x)f(s,\beta,x)=f(p+y_1,\beta,x)f(q+y_2,\beta, x)f(s+y_3,\beta,x).\label{eq:proof_for_extended_Gould_totyu1}
\end{align} 
Using Eq.~(\ref{eq:proof_for_extended_Gould_totyu1}) and the relation
\begin{align}
\sum_{i=0}^{\infty}\sum_{j=0}^{\infty}\sum_{k=0}^{\infty}h_{i,j,k}&=\sum_{n=0}^{\infty}\sum_{i=0}^n\sum_{j=0}^{n-i}h_{i,j,n-i-j},\label{eq:triple_summation_relation}
\end{align}
we obtain the desired result (\ref{eq:extended_version_of_Gould's_identity}). 

From the above results, the leading term of the correlation function becomes
\begin{align}
C(j,l,n-1)&=
\begin{cases}
\vspace{0.5em}\displaystyle{1+O\left(\frac{1}{M}\right),\quad j=l,}\\
\vspace{0.5em}0,\quad |j-l|=1,\\
\displaystyle{\frac{n-1}{M}+O\left(\frac{1}{M^2}\right)},\quad \text{otherwise}.
\end{cases}
\label{eq:leading_term_of_correlation_function_OBC}
\end{align}
Thus we have
\begin{align}
f_{\rm d}(n)&=\sum_{j=1}^Mf_j^2\left[1+O\left(\frac{1}{M}\right)\right]+\sum_{j,l,j\not=l,|j-l|\not=1}f_jf_l\times O\left(\frac{1}{M}\right)\notag \\
&=\left[\frac{M}{2}+O(1)\right]+\left[\left(\sum_{j=1}^Mf_j\right)^2-\sum_{j=1}^Mf_j^2-2\sum_{j=1}^{M-1}f_jf_{j+1}\right]\times O\left(\frac{1}{M}\right)\notag \\
&=\frac{M}{2}+O(1),\label{eq:leading_term_of_denominator}
\end{align}
where we used $\sum_{j=1}^Mf_j^2=(M-1)/2$, $\sum_{j=1}^Mf_j=0$, $\sum_{j=1}^Mf_j^2=O(M)$, and $\sum_{j=1}^{M-1}f_jf_{j+1}=O(M)$.

From the above results, we obtain the energy variance
\begin{align}
\Delta E^2&=\frac{\pi^2D^2}{4M^2}+O\left(\frac{1}{M^3}\right),\quad \text{for }n=O(1).\label{eq:energy_variance_leading_term_OBC}
\end{align}
This result means $\Delta E^2\to 0$ in the thermodynamics limit. Therefore, the states $\cket{AS_n}$ satisfy the conditions of the AQMBS at least when $n=O(1)$. For larger $n$, our numerical results show that the energy variance goes to zero in the thermodynamic limit [see Fig.~\ref{fig:energy_variance_obc_several_n}(a)]. As in the case of the PBC case, the energy variance is an increasing function of $n$. In Fig.~\ref{fig:energy_variance_obc_several_n}(b), we show the system size dependence of the energy variance for $n=M/2$. This result means that the state $\cket{AS_n}$ satisfies the condition of the AQMBS even in large $n$ cases.

\begin{figure}[t]
\centering
\includegraphics[width=17cm,clip]{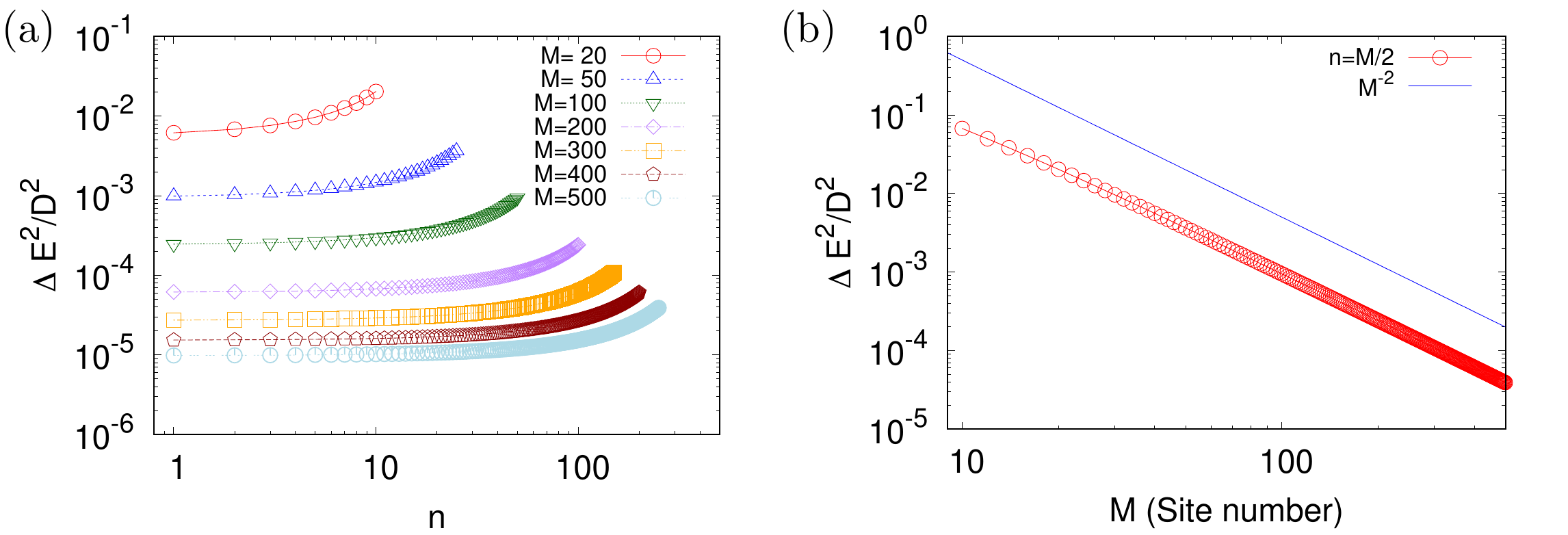}
\caption{(a) $n$ dependence of the energy variance for several system sizes. (b) System size dependence of the energy variance for $n=M/2$. The solid blue line represents $M^{-2}$ for a guide to the eye.
}
\label{fig:energy_variance_obc_several_n}
\vspace{-0.75em}
\end{figure}%

Finally, we remark on the choice of $f_j$. When we choose $f_j=\cos[\pi(2j-1)/(2M)]$, we can show that the leading term of the energy variance is the same as Eq.~(\ref{eq:energy_variance_leading_term_OBC}).


\end{widetext}
\end{document}